\documentclass[twocolumn]{aastex62}

\usepackage{natbib}

\begin{document}

\title{Quantifying Variability of YSOs in the Mid-IR Over Six Years with NEOWISE}

\correspondingauthor{Jeong-Eun Lee}

\author{Wooseok Park}
\affil{School of Space Research, Kyung Hee University,
1732, Deogyeong-daero, Giheung-gu, Yongin-si, Gyeonggi-do 17104, Korea\\ :\href{mailto:jeongeun.lee@khu.ac.kr}{jeongeun.lee@khu.ac.kr}}

\author[0000-0003-3119-2087]{Jeong-Eun Lee}
\affil{School of Space Research, Kyung Hee University,
1732, Deogyeong-daero, Giheung-gu, Yongin-si, Gyeonggi-do 17104, Korea\\ :\href{mailto:jeongeun.lee@khu.ac.kr}{jeongeun.lee@khu.ac.kr}}

\author{Carlos Contreras Pe{\~n}a}
\affil{Centre for Astrophysics Research, University of Hertfordshire, Hatfield AL10 9AB, UK}
\affil{School of Physics, Astrophysics Group, University of Exeter, Stocker Road, Exeter EX4 4QL, UK}

\author[0000-0002-6773-459X]{Doug Johnstone}
\affil{NRC Herzberg Astronomy and Astrophysics, 5071 West Saanich Road, Victoria, BC, V9E 2E7, Canada}
\affil{Department of Physics and Astronomy, University of Victoria, 3800 Finnerty Road, Elliot Building, Victoria, BC, V8P 5C2, Canada}

\author{Gregory Herczeg}
\affil{Kavli Institute for Astronomy and Astrophysics, Peking University, Yiheyuan 5, Haidian Qu, 100871 Beijing, China}
\affil{Department of Astronomy, Peking University, Yiheyuan 5, Haidian Qu, 100871 Beijing, China}

\author{Sieun Lee}
\affil{School of Space Research, Kyung Hee University,
1732, Deogyeong-daero, Giheung-gu, Yongin-si, Gyeonggi-do 17104, Korea\\ :\href{mailto:jeongeun.lee@khu.ac.kr}{jeongeun.lee@khu.ac.kr}}

\author{Seonjae Lee}
\affil{Department of Physics and Astronomy, Seoul National University, 1 Gwanak-ro, Gwanak-gu, Seoul 08826, Korea}

\author{Anupam Bhardwaj}
\affil{Korea Astronomy and Space Science Institute (KASI), 776 Daedeokdae-ro, Yuseong-gu, Daejeon 34055, Korea}

\author{Gerald H.\ Moriarty-Schieven}
\affil{NRC Herzberg Astronomy and Astrophysics, 5071 West Saanich Road, Victoria, BC, V9E 2E7, Canada}

\begin{abstract}
Variability in young stellar objects (YSOs) can be caused by various time-dependent phenomena associated with star formation, including accretion rates, geometric changes in the circumstellar disks, stochastic hydromagnetic interactions between stellar surfaces and inner disk edges, reconnections within the stellar magnetosphere, and hot/cold spots on stellar surfaces. We uncover and characterize $\sim$1700 variables from a sample of $\sim$5400 YSOs in nearby low-mass star-forming regions using mid-IR light curves obtained from the 6.5-years NEOWISE All Sky Survey. The mid-IR variability traces a wide range of dynamical, physical, and geometrical phenomenon.
We classify six types of YSO mid-IR variability based on their light curves: secular variability (\textit{Linear, Curved, Periodic}) and stochastic variability (\textit{Burst, Drop, Irregular}). YSOs in earlier evolutionary stages have higher fractions of variables and higher amplitudes for the variability, with the recurrence timescale of FUor-type outbursts (defined here as $\Delta$W1 or $\Delta$W2 $>1$ mag followed by inspection of candidates) of $\sim$1000 years in the early embedded protostellar phase.
Known eruptive young stars and subluminous objects show fractions of variables similar to the fraction ($\sim55\%$) found in typical protostars,  
suggesting that these two distinct types are not distinct in variability over the 6.5-year timescale. 
Along with brightness variability, we also find a diverse range of secular color variations, which can be attributed to a competitive interplay between the variable accretion luminosity of the central source and the variable extinction by material associated with the accretion process.

\vspace{10mm}
\end{abstract}

%\vspace{10\baselineskip}

\vspace{10mm}

\section{Introduction} \label{sec:intro} 

The rich variability of young stellar objects has been characterized since before the objects had been identified as young \citep{1945Joy}. Recent and ongoing experiments in time-domain astronomy are providing an unbiased evaluation of this variability, usually at optical wavelengths (e.g., ASAS-SN, \citealt{shappee14}; ZTF, \citealt{bellm14}; TESS,  \citealt{ricker14}; WASP, \citealt{rigon17}, and many others).
 Time domain experiments at IR-to-mm wavelengths are more rare but complementary to optical surveys \citep[e.g.][]{carpenter01,2017Lucas,herczeg17}. These longer wavelength emission surveys probe heavily extincted and colder objects as well as different physical processes, often using dust as the physics laboratory.
         
The variability of young stellar objects, revealed primarily by the optical surveys, comes in many flavors, each tracing different physics and locations in the system \citep[e.g.][]{1994Herbst,hillenbrand15}.  Long accretion bursts, tracing instabilities in the disk, last for years or even decades and are seen as enduring enhancements in the broadband luminosity \citep[e.g.][]{covey11,contreras17,johnstone18}. Shorter accretion bursts, likely driven by magnetospheric instabilities, typically last for hours and are seen as temporary enhancements in broadband luminosity \citep[e.g.][]{alencar10,venuti15,2017Cody}.   
Reconnections in the stellar magnetosphere and possibly the star-disk interaction region leads to brief enhancements at optical wavelengths, with counterparts from X-rays through radio wavelengths \citep[e.g.][]{flaccomio12,tofflemire17,mairs19}.  Periodic signals in light curves on stellar rotation periods trace  magnetic spots on the stellar surface  \citep{grankin08,lanza16,gully17,sergison20}.  Changes in the disk scale height (or other changes in the disk morphology) can last for days or decades and are detectable from long-lasting drops in optical and near-IR emission, perhaps with brightening at other wavelengths \citep[e.g.][]{natta97,bouvier07,2013Bouvier,2015Rodriguez}.  Each of these processes has a broad range of potential timescales, with signatures in the light curves that depend sensitively on wavelength.

While these physical descriptions have been developed primarily from optical surveys, long-wavelength time-domain surveys play a particularly important role in probing mass aggregation in young stellar objects.  The youngest protostars, deeply embedded in dusty envelopes, are visible only at longer wavelengths during the main stages of stellar growth \citep[e.g.][]{kospal07,safron15,hunter18,syliu18}.  At mm wavelengths, any variability in the brightness of the dust envelope is a consequence of temporal changes in the protostellar (accretion) luminosity \citep[e.g.][]{johnstone13,macfarlane19,baeketal20,pena20}.  Because the  envelope dust acts as a bolometer radiating with a temperature equilibrating absorption and emission of radiation, mm wavelength observations are straightforward to interpret but less sensitive than the mid-IR and far-IR to changes in bolometric luminosity; small changes are challenging to detect \citep{mairs17, johnstone18}.

Monitoring protostars in the mid-IR is potentially powerful for evaluating protostellar accretion variability, as well as the rich tapestry of other physics and morphological changes -- i.e.\ both a tool and a trouble. The mid-IR time domain for young stellar objects opened up with the YSOVAR program \citep{morales11,2014Stauffer,2014Cody,wolk18} during the extended mission of the \textit {Spitzer Space Telescope}. 
YSOVAR has shown that, similar to the observed behaviour at optical and near-IR wavelengths, variability in YSOs is also common in the mid-IR.  Embedded YSOs display larger amplitudes than more evolved sources \citep{wolk18} and the variability at younger stages occurs over longer timescales \citep{2014Gunther}.  The program has also revealed the complexity of YSO variability, as in many cases the physical mechanisms driving variability in the mid-IR do not lead to or are uncorrelated with optical/near-IR changes \citep{2014Cody}.  
A comparison between {\it Spitzer} and {\it WISE} mid-IR photometry 
%\textit{Spitzer} and \textit{WISE} photometry \citep{scholz13} 
yielded several variable protostars, suggesting that large outbursts may be more common in the youngest phases of stellar assembly than at the end stages of accretion (\citealt{scholz13} and \citealt{fischer19}, see comparisons with outburst frequencies for older phases by \citealt{hillenbrand15} and \citet{contreras19}.

The extension of the \textit{WISE} mission (NEOWISE) provides all-sky photometric monitoring at 3--5 $\mu$m, with epochs every six months and a time baseline of a decade between the first \textit{WISE} and most recent NEOWISE epoch.  The NEOWISE monitoring has been used to find two large outbursts of protostars  \citep{kun19,lucas20}, characterize how an instability moves through the disk prior to an optical outburst \citep{2018Hillenbrand,leeyh20}, evaluate how mid-IR changes correlate with luminosity changes \citep{pena20}, and identify disk height variations as the cause of a prominent fade of AA Tau \citep{covey21}.  Beyond protostars, NEOWISE variability has been a useful probe of other objects, including contact binaries \citep{petrosky20}, white dwarfs \citep{wang19}, Wolf-Rayet stars \citep{williams19}, tidal disruption events \citep{jiang21}, and quasars and AGN \citep{wang20,sheng20}.

In this paper, we systematically evaluate mid-IR variability of $\sim$5400 known, nearby young stellar objects (YSOs) ranging from Class 0 through Class III, over the 6.5-year span of NEOWISE imaging and for a few cases including the 15-year span reaching back to \textit{Spitzer} and \textit{WISE} observations. Section \ref{sec:wise} describes the \textit{WISE} YSO samples used in this paper. Section \ref{sec:methods} details our analysis of stochastic and secular variability, while Section \ref{sec:classification} defines the six types of variables revealed.  In Section \ref{sec:discussion} we consider 
variability across evolutionary stages, mechanisms for variability, extrema in the context of episodic accretion, secular colour changes, and long-term variability.

\section{\textit{WISE/NEOWISE} yso samples} \label{sec:wise}

\begin{deluxetable*}{cccccc}
\tablecaption{YSO Catalogues and Classifications\label{tab:table1}}
\tablenum{1}
%%%%% table 1 %%%%%%
%% The values (usually only l,r and c) in the last part of
%% \begin{deluxetable}{} command tell LaTeX how many columns
%% there are and how to align them.
%% Keep a portrait orientation
%% Over-ride the default font size
%% Use Default (12pt)
%% Use \tablewidth{?pt} to over-ride the default table width.
%% If you are unhappy with the default look at the end of the
%% *.log file to see what the default was set at before adjusting
%% this value.
%% This is the title of the table.
%% This command over-rides LaTeX's natural table count
%% and replaces it with this number.  LaTeX will increment 
%% all other tables after this table based on this number
%% The \tablehead gives provides the column headers.  It
%% is currently set up so that the column labels are on the
%% top line and the units surrounded by ()s are in the 
%% bottom line.  You may add more header information by writing
%% another line between these lines. For each column that requries
%% extra information be sure to include a \colhead{text} command
%% and remember to end any extra lines with \\ and include the 
%% correct number of &s.
\tablehead{\colhead{} & \colhead{Region} & \colhead{Class 0/I [P]\tablenotemark{a} } & \colhead{Class II [D]} & \colhead{Class III+Evolved [PMS+E]} & \colhead{Total}}

%% All data must appear between the \startdata and \enddata commands
\startdata
\cite{megeath12} &  Orion A/B  &  319 (478)\tablenotemark{b}  &  2160 (2991) &   - &  2479 (3469) \\
\\
% \cite{stutz13} &  &  &  &  &\\
\cite{dunham15}  & Aquila  &  105 (148)  &   275 (330)  &  742 (841)  &   1122 (1319)  \\
    & Auriga/CMC  & 35 (43) & 67 (73) & 17 (17)  & 119 (133) \\
    & Cepheus  & 16 (29) & 50 (61)  &  12 (13) & 78 (103)  \\
    & Chamaeleon  & 5 (12)  &  57 (81) & 17 (23) & 79 (116) \\
    & Corona Australis  &  5 (15) & 17 (22) & 13 (17)  & 35 (54)   \\
    & IC5146  & 25 (38)  & 66 (79) & 14 (15) & 105 (132) \\
    & Lupus &  12 (13) & 53 (58) & 84 (111) & 149 (182) \\
    & Musca  & 1 (1)  & 1 (1) &  5 (11)&  7 (13) \\
    & Ophiuchus  & 57 (74)  & 167 (177) & 42 (51) & 266 (302) \\
    & Perseus  & 79 (111)  & 225 (235) & 35 (39) & 339 (385)  \\
    & Serpens  & 42 (52) & 118 (131)  & 37 (44)  & 197 (227)  \\
\\
\cite{esplin19}  &  Taurus  & 34 (45)  &  203 (238) &  186 (209)\tablenotemark{c} &  423 (492) \\
\hline
Total  &  & 735 (1059)  &  3459 (4477)  &  1204 (1391)  &  5398 (6927) \\
\enddata
\tablenotetext{a}{The P classification also includes flat-spectrum YSOs.}
\tablenotetext{b}{Numbers in front are for the \textit{WISE} samples satisfying our selection criteria. Numbers in parentheses include all YSOs from the adopted catalogues.}
\tablenotetext{c}{As mentioned in the text, for Taurus this classification includes only bona fide Class III YSOs, uncontaminated by AGBs. }

%% Include any \tablenotetext{key}{text}, \tablerefs{ref list},
%% or \tablecomments{text} between the \enddata and 
%% \end{deluxetable} commands
%% No \tablecomments indicated
%% No \tablerefs indicated

\end{deluxetable*}

The \textit{Wide-field Infrared Survey Explorer} (\textit{WISE}) is a NASA Explorer mission to obtain the most comprehensive full mid-IR sky survey \citep{wright10}. \textit{WISE} surveyed the entire sky in four bands, W1 (3.4 $\mu$m), W2 (4.6 $\mu$m), W3 (12 $\mu$m), and W4 (22 $\mu$m), with the spatial resolutions of 6.1”, 6.4”, 6.5”, and 12”, respectively, from January to September 2010. 

After the depletion of hydrogen from the cryostat, \textit{WISE} operated using only the short wavelength bands, W1 and W2. The survey continued as the NEOWISE Post-Cryogenic Mission \citep{mainzer11} for an additional four months, until \textit{WISE} went into hibernation in February 2011. In September 2013,  \textit{WISE} was reactivated and has performed observations in W1 and W2 as NEOWISE-reactivation mission (NEOWISE-R, \citealp{mainzer14}) with the primary purpose to explore the Near-Earth Objects. NEOWISE-R is still operating and the latest released data set consists of 6.5-year W1 and W2 photometric observations.
 
To ensure a comprehensive list of nearby protostellar sources, our \textit{WISE} samples are collected from 20 star-forming regions on the Gould Belt based on various YSO catalogues: \cite{megeath12} for the Orion A and B regions, \cite{esplin19} for the Taurus region, and  \cite{dunham15} for 18 additional star-forming regions, which were covered by the Spitzer Legacy projects ``Cores to Disks'' \citep{evans09} and ``Gould Belt'' \citep{dunham15}. We initially identified 6927 potential YSOs from the NEOWISE archival data using the above catalogues and reclassified these sources as Class 0/I (protostar [P]), Class II (disk [D]), and Class III+Evolved (pre-main sequence+evolved [PMS+E]) to unify the individual classification systems adopted by the different catalogues.

Classification of the evolutionary stages of YSOs is generally consistent among catalogues with minor variation. \cite{megeath12} divided YSOs into two classes using color thresholds: Class 0/I (protostar) and Class II (disk). They additionally classified protostellar candidates: red candidate protostars and faint candidate protostars (see Table 3 of \citealp{megeath12}). Red candidate protostars have no detection by \textit{Spitzer} at 4.5 $\mu$m, 5.8 $\mu$m, or 8 $\mu$m but are bright at 24 $\mu$m, $M_{24} < 7$ mag. These candidates are excluded from our analyses because they are generally too faint in NEOWISE W1 and W2. Faint candidate protostars are faint at 24 $\mu$m, $M_{24} > 7$ mag, but satisfy the color criteria for protostars at shorter wavelengths, and thus, are included as Class 0/I in our analyses.

\cite{dunham15} used extinction corrected spectral indexes to classify YSOs as Class 0/I (protostar), flat-spectrum, Class II (disk), and Class III (evolved\footnote{\cite{dunham15} use the classification `evolved' to refer to the oldest YSOs; however, in this paper we use the term exclusively to refer to AGB contaminants.}). 
There is, however, a possibility to misclassify the background/foreground Asymptotic Giant Branch stars (AGBs) as Class III due to the similar infrared excess between YSOs and AGBs \citep{lee2021agb}. Therefore, in our analysis we reclassify Dunham's Class III YSOs as Class III+Evolved (PMS+E) considering the possible contamination (discussed further in Appendix \ref{appendix:agb}).

The YSO classification in Taurus is more accurate compared to the other two catalogues because \cite{esplin19} combined  color-magnitude diagrams, proper motions, and spectral analyses, to classify YSOs as Class 0 through III. In contrast to the other catalogues, \cite{esplin19} divided Class 0 and I YSOs, but we combine these into Class 0/I for consistency. The authors also subdivided Class II YSOs with the evolutionary stages of the disks; however, we classify all disk sources as Class II in this work.

Using the source coordinates in the catalogues above, we extracted NEOWISE photometric information for each YSO. To ensure that all the photometric data extracted at a given set of coordinates are for the same source, we set a 3" radius criterion; a NEOWISE measurement is considered as the target if its coordinates are located within 3" from the known YSO coordinates.
From the set of NEOWISE raw exposures, we calculate the mean and standard deviation of the distances of the NEOWISE measurements from the known YSO coordinates and consider only those NEOWISE exposures located within 2 sigma from the mean distance in our analyses. We further limit the standard deviation of distance to be less than 0.3", since faint or saturated sources commonly have a large dispersion in position. The W2 data are primarily used to search for variability since YSOs are generally faint in W1. 

We use several additional criteria to confirm the \textit{WISE} sources and construct a high-quality data set for our variability analyses of YSOs as listed below:
\begin{itemize}
\item Sample targets should have been detected in more than 5 epochs in W2;

\item Sample targets must have the standard deviation of distance from the known YSO coordinates smaller than 0.3"; 

\item Samples need to have the mean W2 uncertainty smaller than 0.2 magnitude. 
\end{itemize} 

% \vspace{5mm}

\begin{figure}[t] %ht
 % \epsscale{2}
        
\centering
%\subfigure{
{
\includegraphics[width=1\linewidth]{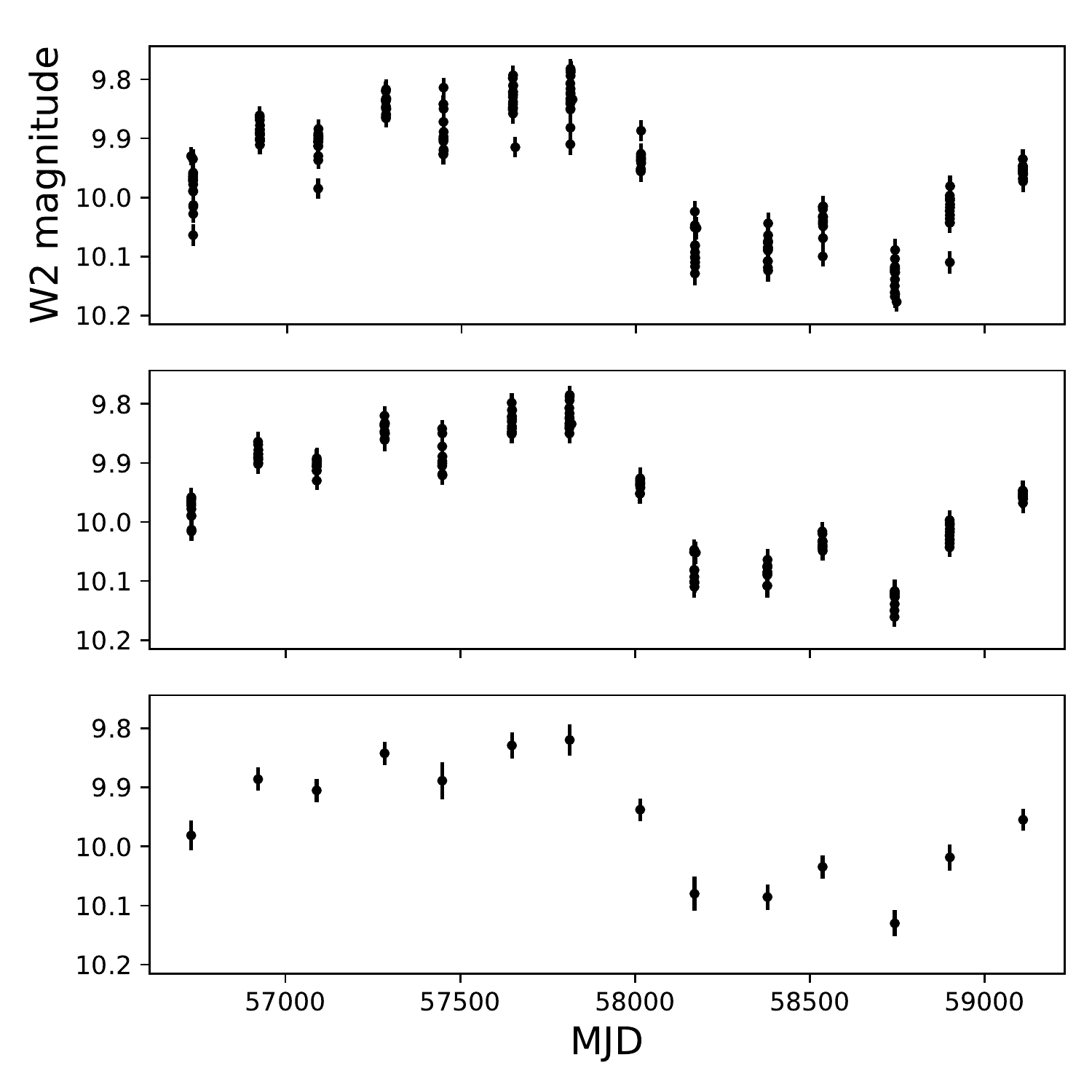}}

\caption{An example of averaging raw exposures of 2MASS J05423983-0921460. (Top) First, raw exposures satisfying the distance criterion are collected. (Middle) Upper and lower 15\% outliers in each epoch are excluded. (Bottom) Finally, the remaining 70\% of exposures are averaged to produce the final light curve and its uncertainty, to be used for our analyses.}     
\label{fig:avg_data}
\end{figure}

NEOWISE is providing all-sky survey photometric data at 3.4 (W1) and 4.6 (W2) $\mu$m every six months \citep{cutri15}, and we utilize 6.5-years of observations. There are thus 14 epochs available for our exploration of YSO variability. Furthermore, each epoch consists of a 10-20 exposures for a given YSO, typically covering less than a few days. We average these exposures within each epoch to provide regularly sampled light curves every six months. In this averaging process, for a given YSO, we use only the middle 70\% of exposures in the range of magnitude to exclude the upper and lower 15\% outliers. Next, we adopt the mean MJD and mean magnitude for the YSO in the epoch. 
The measurement error is calculated by adding, in quadrature, the mean error and the standard deviation (in magnitudes) of the exposures in each epoch. Figure \ref{fig:avg_data} presents an example of the original, outlier removed, and epoch-averaged light curves for a single YSO. Finally, in total, 5398 YSOs satisfied our criteria. Epoch-averaged quantities are provided in Table \ref{tab:table6} for all these sources. 

\begin{figure}[t]
        
\centering
%\subfigure{

{\includegraphics[trim={0.3cm 0.1cm 0.3cm 0.0cm},clip,width=0.98\columnwidth]
{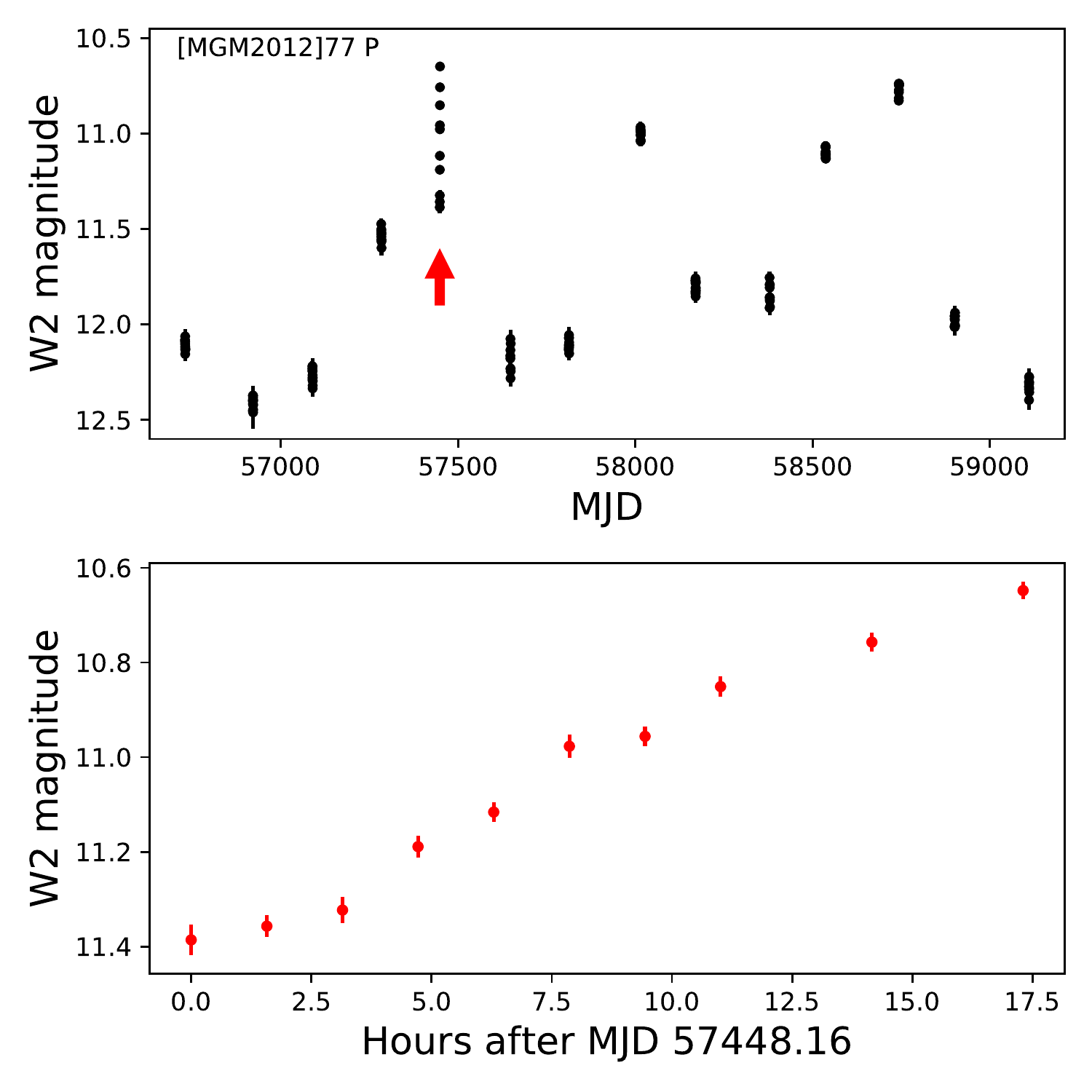} }
\caption{Observed variability on a one day timescale. The upper panel shows the light curve of [MGM2012]77 (P) over 6.5 years and marks the epoch of interest. The lower panel shows a focus on the fifth observing block, March 2016, where exposures show a large dispersion and monotonic brightening in time. The time coverage of the observing block is about one day and the cadence between exposures is about 2 hours.  
\label{fig:short_var}}
\end{figure}

Although we use the averaged magnitude for each observing block, the exposures within observing blocks can have observable time variability within a few days. 
Figure \ref{fig:short_var} shows an example of such short-time variability for the protostar [MGM2012]77 within an observing block, where the cadence of exposures is $\sim$2 hours. In the fifth observing block, March 2016, the W2 light curve sharply increases by 1 magnitude over just one day. In our analysis, for epochs with such short-time variability, the uncertainty of the mean magnitude is estimated to be large. We will report our detailed analyses of short-time variability in a separate paper.

\begin{figure}[t] %ht
 % \epsscale{2}
\centering
 %\subfigure{

{\includegraphics[trim={0.4cm 0.1cm 0.3cm 0.3cm},clip,width=0.98\columnwidth]{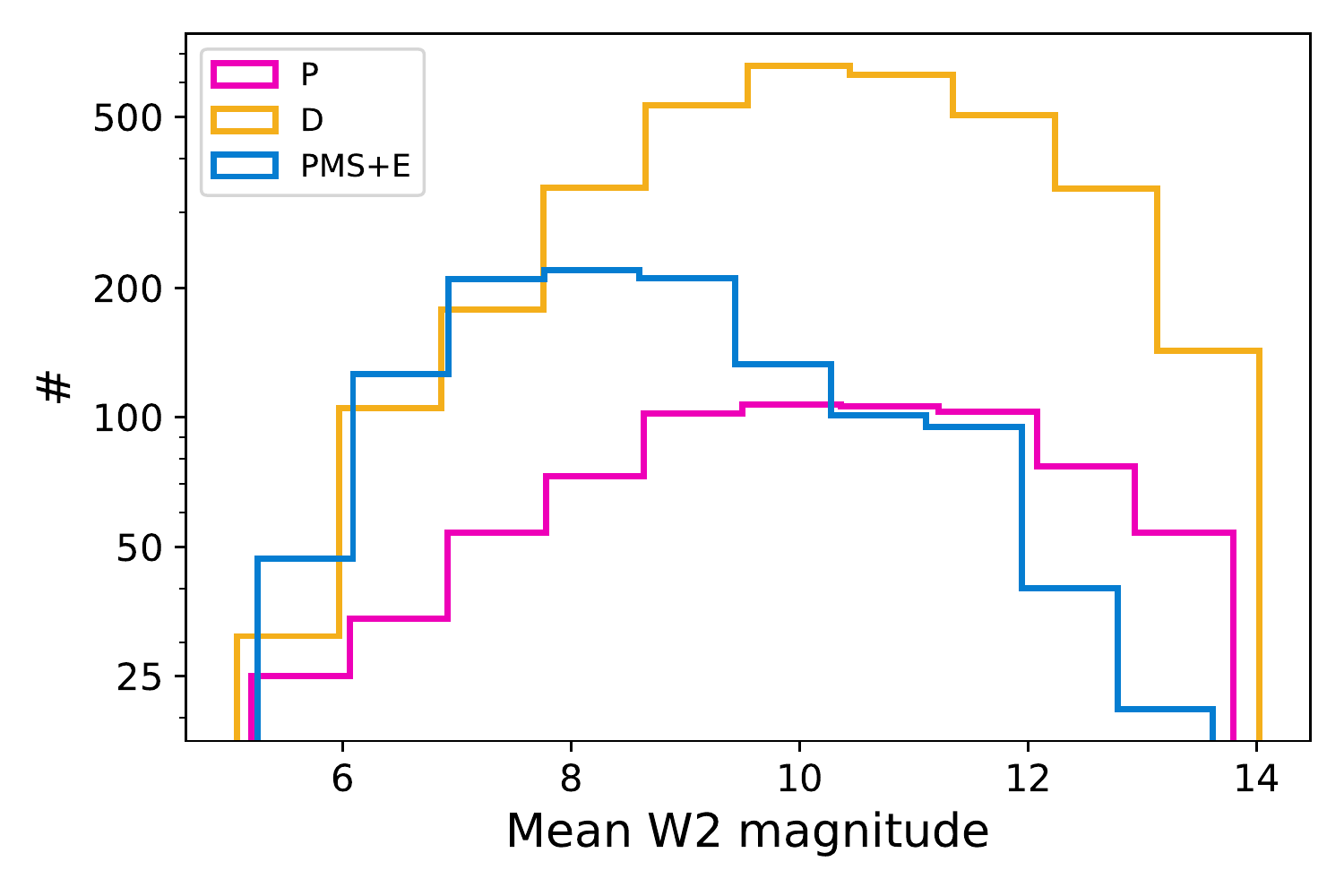}}
\caption{The distribution of mean W2 magnitude of our samples. Colors denote different evolutionary stages: magenta for Class 0/I (P), yellow for Class II (D), and blue for Class III+Evolved (PMS+E).    
\label{fig:meanw2}}
\end{figure}

Figure \ref{fig:meanw2} shows the distribution of the mean W2 magnitude depending on the evolutionary stages of YSOs. The peaks of the distributions for Class 0/I (P) and Class II (D) are located at $\sim$10 magnitude with tails toward brighter sources while the peak for Class III+Evolved (PMS+E) is located at $\sim$8 magnitude with a broad tail toward fainter sources. More than half, 64.1\%, of our sample sources are disks, while $\sim$13.6\% and $\sim$22.3\% are protostars and PMS+E, respectively (see Table \ref{tab:table1}).

% averaged lightcurve plot

% - Secondary : original datapoints(with outlier removed) to show hourly change of brightness 

% \vspace{10mm}
 
% \end{enumerate}

%%%%% table 1 %%%%%%
%% The values (usually only l,r and c) in the last part of
%% \begin{deluxetable}{} command tell LaTeX how many columns
%% there are and how to align them.
\begin{deluxetable*}{ccccc}

%% Keep a portrait orientation

%% Over-ride the default font size
%% Use Default (12pt)

%% Use \tablewidth{?pt} to over-ride the default table width.
%% If you are unhappy with the default look at the end of the
%% *.log file to see what the default was set at before adjusting
%% this value.

%% This is the title of the table.
\tablecaption{Variable Type by YSO Classification\label{tab:table2}}

%% This command over-rides LaTeX's natural table count
%% and replaces it with this number.  LaTeX will increment 
%% all other tables after this table based on this number
\tablenum{2}

%% The \tablehead gives provides the column headers.  It
%% is currently set up so that the column labels are on the
%% top line and the units surrounded by ()s are in the 
%% bottom line.  You may add more header information by writing
%% another line between these lines. For each column that requries
%% extra information be sure to include a \colhead{text} command
%% and remember to end any extra lines with \\ and include the 
%% correct number of &s.
\tablehead{\colhead{} & \colhead{Class 0/I [P]} & \colhead{Class II [D]} & \colhead{Class III+Evolved [PMS+E]} & \colhead{Total}}

%% All data must appear between the \startdata and \enddata commands
\startdata
% Type I (Linear)  & 50 (31)\tablenotemark{a}  & 62 (34)   & 14 (3)    & 126 (68)  \\
\textit{Linear} & 37 (5.0)\tablenotemark{a}  & 31 (0.9)    & 9 (0.7)  & 77\\
\textit{Curved}  & 103 (14.0) & 183 (5.3)  & 27 (2.2)  & 313    \\%number
\textit{Periodic} & 6 (0.8)  & 31 (0.9)   & 81 (6.7) & 118      \\
\textit{Burst}  & 13 (1.8)   & 117 (3.4) & 7 (0.6)  & 137    \\
\textit{Drop} & 0 (0)   & 27 (0.8)   & 7 (0.6)      & 34   \\
\textit{Irregular}\tablenotemark{b} & 244 (33.2) & 757 (21.9) & 54 (4.5)  & 1055  \\
\hline
Total & 403 (54.8) & 1146 (33.1) & 185 (15.4)  & 1734  \\
\enddata

\tablenotetext{a}{Numbers in front are the count of variables for each type, while numbers in parentheses are the fractions (\%) of variables relative to the total \textit{WISE} samples in each evolutionary stage (see Table \ref{tab:table1}). }
\tablenotetext{b}{Sources with SD/$\sigma$ $>3$ but not classified as any specific type of variability.
}

\end{deluxetable*}
%%%%%%%%%%%%%%%%%%%%		

\section{Methods}

\label{sec:methods}
We adopt similar methods used by \cite{johnstone18} and Lee et al. (submitted) to search for variable YSOs: (1) the standard deviation of fluxes in a given light curve, (2) a periodogram analysis for any periodic variation within a light curve, and (3) a linear least square fitting for the linear trend of a rising or declining light curve. We describe each method in this section. %Note that all analyses are done in the flux domain, but light curves are presented in the magnitude domain.

\subsection{Standard Deviation}
\label{subsec:sd}

Following the standard deviation analysis by \cite{johnstone18} and \cite{pena20} at submm wavelengths, we first convert the NEOWISE W2 magnitude into flux, and then measure the standard deviation over the light curve of a given YSO. To isolate variable YSOs, we divide this standard deviation (SD) by the mean flux uncertainty ($\sigma$), which is the mean of errors calculated for individual epochs as described in the previous section. \cite{johnstone18} used SD/$\sigma$ as an indicator of stochastic variability. Here, we define also $\Delta$W2 (Max-Min) as the difference between the maximum and minimum \textit{magnitudes} of each source to measure the fractional flux change between the faintest and brightest epochs. 

\begin{figure}[t] %ht
% \epsscale{0.1}

\centering
{
\includegraphics[trim={1cm 0.6cm 1cm 0.6cm},clip,width=0.98\columnwidth]
{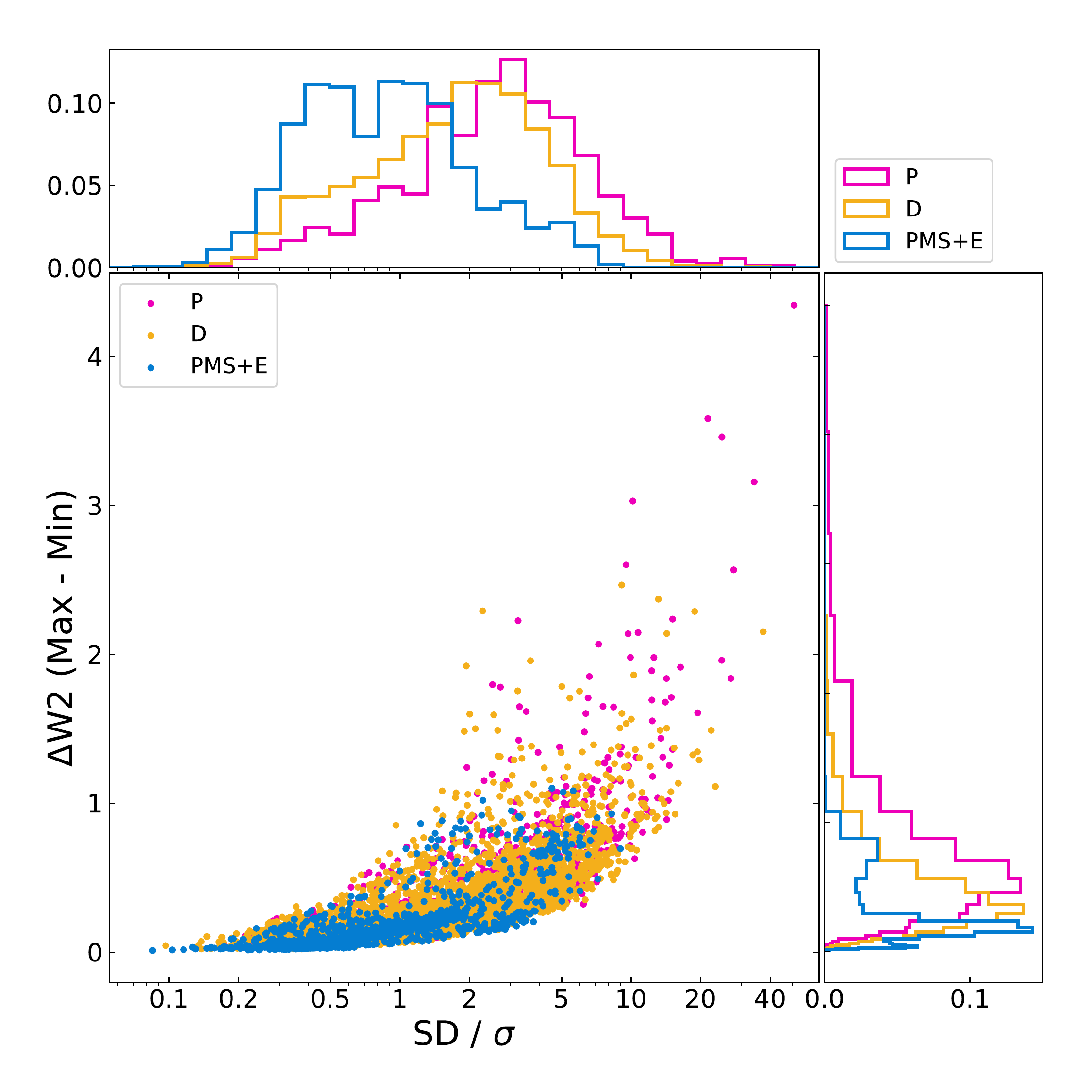}
}
\caption{The fractional flux change between the maximum and minimum phases ($\Delta$W2) as a function of stochasticity (SD/$\sigma$) for all 5398 YSOs selected by the criteria in Section \ref{sec:wise}. SD is the standard deviation of fluxes for a given light curve, and $\sigma$ is the mean flux uncertainty. Colors are the same as those in Figure \ref{fig:meanw2}.}
\label{fig:yso_sc_h}
\end{figure}

Figure \ref{fig:yso_sc_h} shows $\Delta$W2 (Max-Min) versus SD/$\sigma$ for all 5398 YSOs selected by the criteria in Section \ref{sec:wise}
(Table \ref{tab:table7} provides the derived variability measures for each source). 
%It is clearly seen that 
Targets with large SD/$\sigma$ have large $\Delta$W2. The histograms at the top and right sides  of Figure \ref{fig:yso_sc_h} present probability distributions of SD/$\sigma$ and $\Delta$W2, respectively, for YSOs in three different evolutionary stages as marked with different colors. YSOs in an earlier evolutionary stage have greater variability, that is, larger SD/$\sigma$ and larger $\Delta$W2. 
%Note that 
Sources in the category of PMS+E have three peaks; the two peaks with large probabilities are located at small SD/$\sigma$ and $\Delta$W2 while a peak with a lower probability is located at relatively large SD/$\sigma$ and $\Delta$W2. The PMS+E sources associated with this third small peak in the distribution function are discussed in Appendix \ref{appendix:agb}.

\subsection{Lomb-Scargle Periodogram} \label{subsec:lsp}

\begin{figure}[t] %ht
\centering
{
\includegraphics[trim={0.3cm 0.5cm 0.0cm 0.5cm},clip,width=0.98\columnwidth]{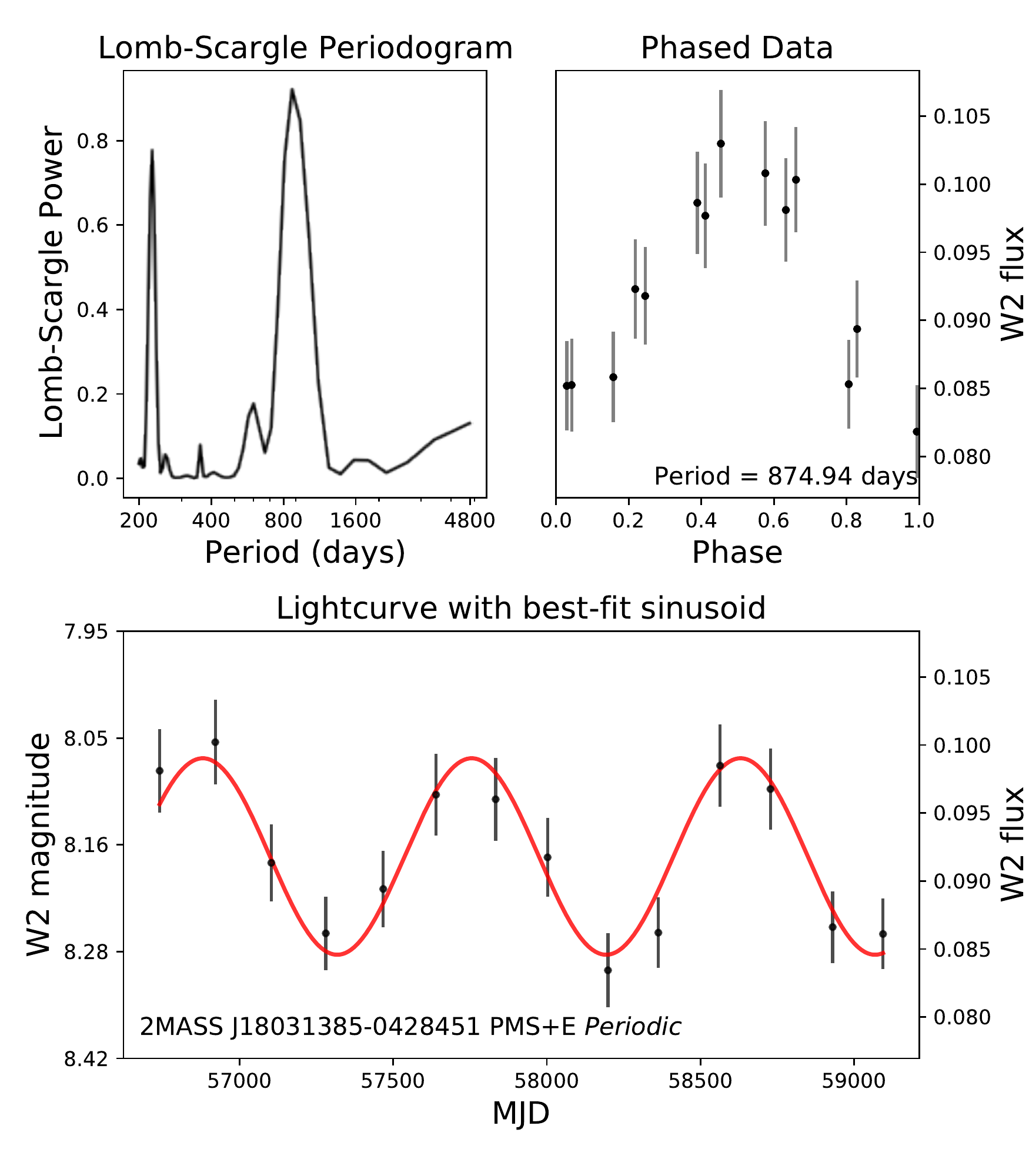}
}
\caption{Lomb-Scargle periodogram analysis of a PMS+E source. The periodogram power spectrum (top left panel) shows the maximum power at a period of about 875 days. The FAP$_{\rm LSP}$ of this period is 6.7$\times10^{-5}$. The phase diagram and the light curve fit by the period are presented in the upper right and bottom panels, respectively.}
\label{fig:lsp}
\end{figure}

The Lomb-Scargle periodogram (LSP, \citealp{lomb76,scargle89}) is a well-known method to detect periodicity from unevenly sampled time-series data. This method is applied to the NEOWISE light curves of our samples using {\tt\string LombScargle} from {\tt\string python} package {\tt\string astropy}. An example of the LSP analysis on an individual source from our sample is presented in Figure \ref{fig:lsp}. Since the NEOWISE survey has been undertaken with a 6-month cadence, periodic variations with period shorter than 6 months cannot be extracted. Therefore, we set the minimum detectable period as 200 days. Furthermore, light curves with periods longer than 1200 days cover at most two full phases in our analysis because the total duration of NEOWISE monitoring is about 2400 days (6.5 years). As a result, we cannot validate decisively the long periodic, $>$ 1200 days, variability. Typically, two full periods of the light curve are needed to confirm and quantify the parameters of the periodicity. Periods longer than 1200 days, however, certainly manifest an increasing or decreasing trend in brightness. Therefore, the LSP analysis is still useful to find long timescale variability although it is difficult to tell whether the variability would be actually periodic or not, and whether the approximated period and amplitude are appropriate.

The false alarm probability, FAP$_{\rm LSP}$, quantifies the uncertainty of a particular LSP peak (see the upper left panel of Figure \ref{fig:lsp}) by quantifying the probability of a false peak due to random errors 
(see \citealp{vanderplas18} for details).
Our false alarm probabilities are assessed using the method developed by \cite{baluev08} to derive
an analytic upper limit of the FAP based on extreme value statistics, taking into account
that the false alarm likelihood increases with the number of 
independent frequencies analysed for each source.
Here we slightly modify Baluev's FAP to determine the false alarm probability of obtaining the found period \textit{or longer}, rather than summing over all periods within the range checked. 
For the majority of YSOs, periods are significantly longer than the half-year cadence of the NEOWISE survey, and thus the unmodified Baluev method systematically overestimates the FAP for these sources. 
We emphasize that for these best-fit long periods we are only estimating a timescale for the observed variability and not implying that the variability is necessarily  repetitive.

%%%%% table 1 %%%%%%
%% The values (usually only l,r and c) in the last part of
%% \begin{deluxetable}{} command tell LaTeX how many columns
%% there are and how to align them.
\begin{deluxetable*}{ccccc}

%% Keep a portrait orientation

%% Over-ride the default font size
%% Use Default (12pt)

%% Use \tablewidth{?pt} to over-ride the default table width.
%% If you are unhappy with the default look at the end of the
%% *.log file to see what the default was set at before adjusting
%% this value.

%% This is the title of the table.
\tablecaption{Combined Variable Types of YSOs\label{tab:table3}}

%% This command over-rides LaTeX's natural table count
%% and replaces it with this number.  LaTeX will increment 
%% all other tables after this table based on this number
\tablenum{3}

%% The \tablehead gives provides the column headers.  It
%% is currently set up so that the column labels are on the
%% top line and the units surrounded by ()s are in the 
%% bottom line.  You may add more header information by writing
%% another line between these lines. For each column that requries
%% extra information be sure to include a \colhead{text} command
%% and remember to end any extra lines with \\ and include the 
%% correct number of &s.
\tablehead{\colhead{} & \colhead{Class 0/I [P]} & \colhead{Class II [D]} & \colhead{Class III+Evolved [PMS+E]} & \colhead{Total}}

%% All data must appear between the \startdata and \enddata commands
\startdata
\textit{Curved} + \textit{Burst} & 2 & 5 &  0 & 7  \\
\textit{Periodic} + \textit{Burst} & 0 & 2 & 1  & 3   \\
\textit{Linear} + \textit{Irregular} & 4 & 8 & 0  & 12  \\
\textit{Curved} + \textit{Irregular} & 25 & 40  & 0  & 65  \\
\textit{Periodic} + \textit{Irregular} & 1 & 1  & 0  & 2  \\
\hline
Total & 32 & 56 & 1  & 89 \\
\enddata
% bur 21
% 114
% 9
% 144

% dim
% 8
% 29
% 8
% 45

% Linear, Curved, Periodic, Burst, Drop, Irregular

%208 648 60 916

%\tablenotetext{a}{Numbers in front are the numbers of variables for each type, and numbers in parentheses are the fractions (\%) of given types of variables relative to the total WISE samples in each evolutionary stage (see Table \ref{tab:table1}). }
%\tablenotetext{b}{Sources with SD/MU $>$ 3 but not classified as any specific type of variability.}
% \tablecomments{}
% Various
% }

%% Include any \tablenotetext{key}{text}, \tablerefs{ref list},
%% or \tablecomments{text} between the \enddata and 
%% \end{deluxetable} commands

%% No \tablecomments indicated

%% No \tablerefs indicated

\end{deluxetable*}
%%%%%%%%%%%%%%%%%%%%

\subsection{Linear Least-Square Fitting} 

We also adopt the linear least square fitting (Lin) to find a linear trend of increasing or decreasing fluxes, which are often fitted by LSP with a very long period. We define the linear FAP, FAP$_{\rm Lin}$ hereafter, with the same formulation as Baluev's FAP to estimate the likelihood of the determined best-fit linear slope (see also Lee et al. submitted).

Throughout the rest of our analysis we consider a source to be robustly fit by a linear slope when FAP$_{\rm Lin}\,<\,10^{-4}$. This threshold ensures that only the best linear fits are represented. For the LSP analysis, we utilize a somewhat lower threshold, FAP$_{\rm LSP}\,<\,10^{-2}$, as we desire to explore the broad range of periods and amplitudes recovered. This lower threshold results in a few false positives within our LSP sample; however, we have checked to ensure that these false positives result only in a small contamination fraction (see Section \ref{sec:classification}).

\begin{figure}[t] %ht
 \centering
{\includegraphics[trim={0.2cm 0.0cm 0.3cm 0.3cm},clip,width=0.98\columnwidth]{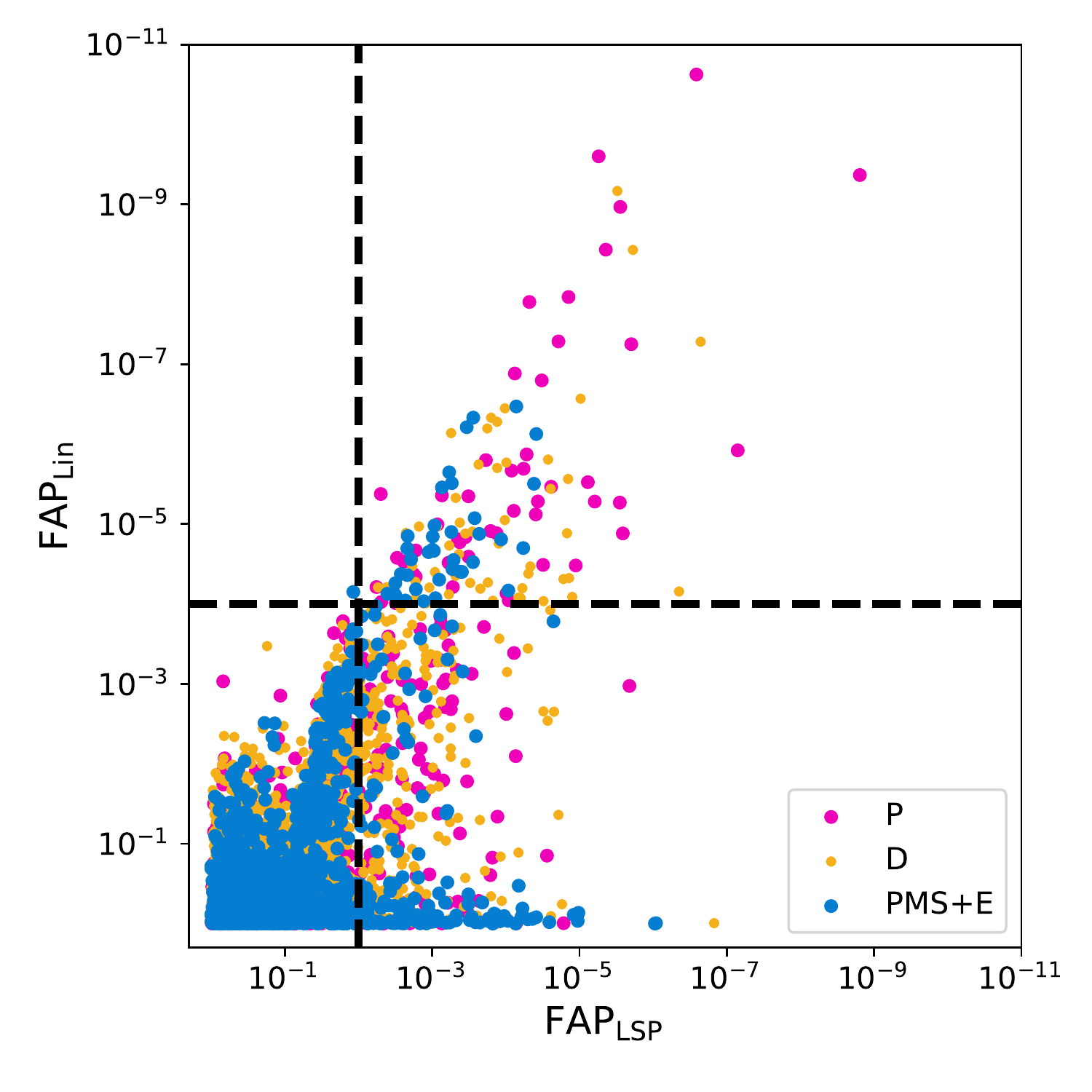} 
}
\caption{Comparison of the linear and periodic false alarm probabilities for all sources in our sample. The horizontal dashed line indicates $10^{-4}$ for FAP$_{\rm Lin}$ and the vertical dashed line indicates $10^{-2}$ for FAP$_{\rm LSP}$.
\label{fig:fap_linb}}
\end{figure}

In Figure \ref{fig:fap_linb}, we consider both the linear and periodic false alarm probabilities for the best fits to all YSOs. The secular variable sources are divided into two zones at FAP$_{\rm LSP}\,<\,10^{-2}$: (1) the upper right region where sources have low FAPs for both LSP and Lin  and (2) the lower right region where sources have high FAP$_{\rm Lin}$ but low FAP$_{\rm LSP}$. Sources with long periods, and thus, predominantly monotonic trends over the timescale covered by NEOWISE are located in the upper right region of Figure \ref{fig:fap_linb}, while sources with periods shorter than 4800 days are located in the lower right region.

 \begin{figure}[t] %ht
\epsscale{2}
{\includegraphics[trim={0.3cm 0.3cm 0.3cm 0.5cm},clip,width=0.98\columnwidth]{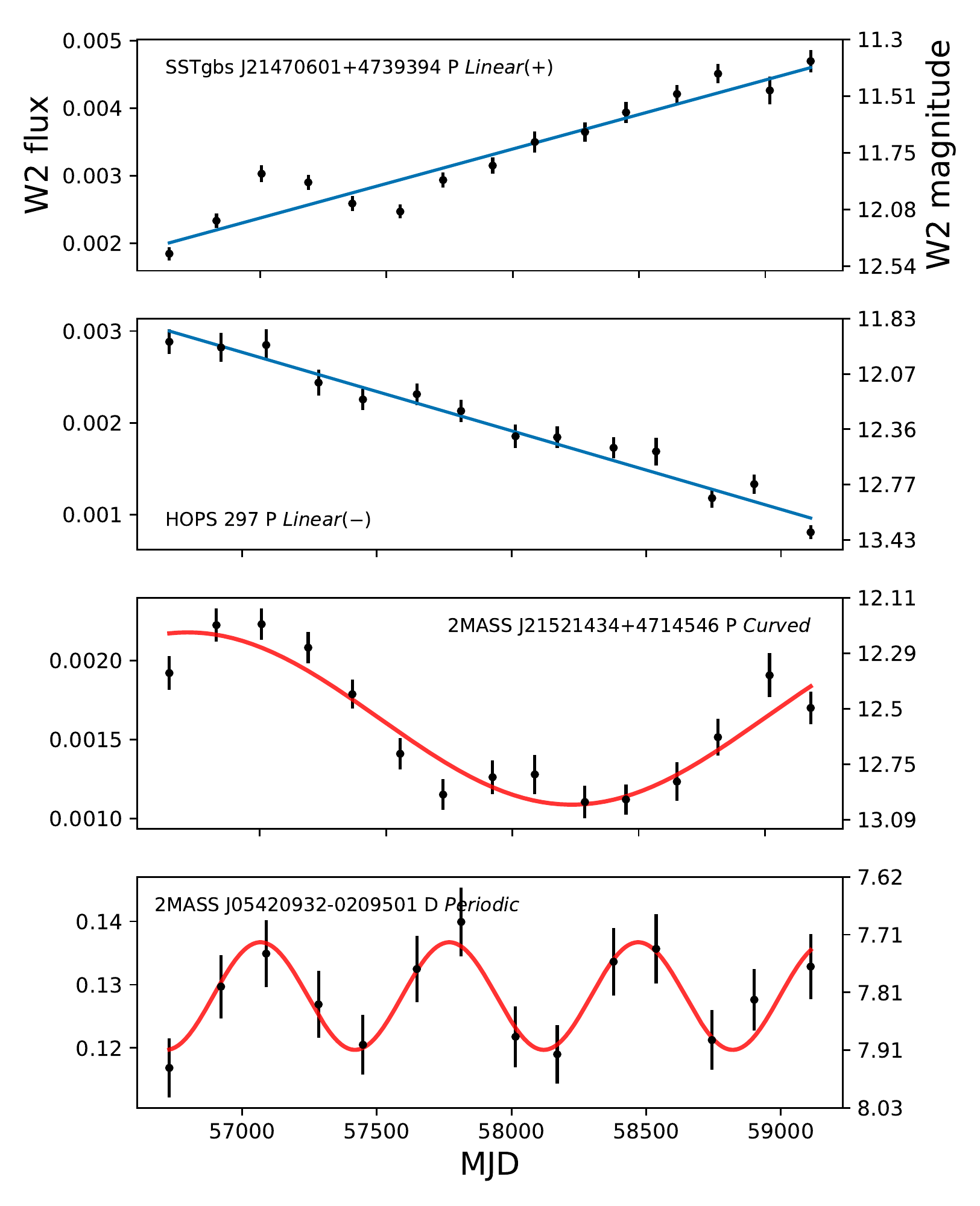}         
}
\centering

\caption{Representative light curves of different types of secular variability. The upper two panels show example light curves of $Linear$ variables while the lower two panels present example light curves of $Curved$ and $Periodic$ variables, respectively. The scale of y-axis shows both flux in Jy (left) and magnitude (right). Color-lines depict the best-fit results by the Lin (blue) and LSP (red).  
\label{fig:secular_ex}}
\end{figure}

\begin{figure*}[t] %ht
\centering
{\includegraphics[trim={0.3cm 0.4cm 0.3cm 0.3cm},clip,width=0.98\linewidth]{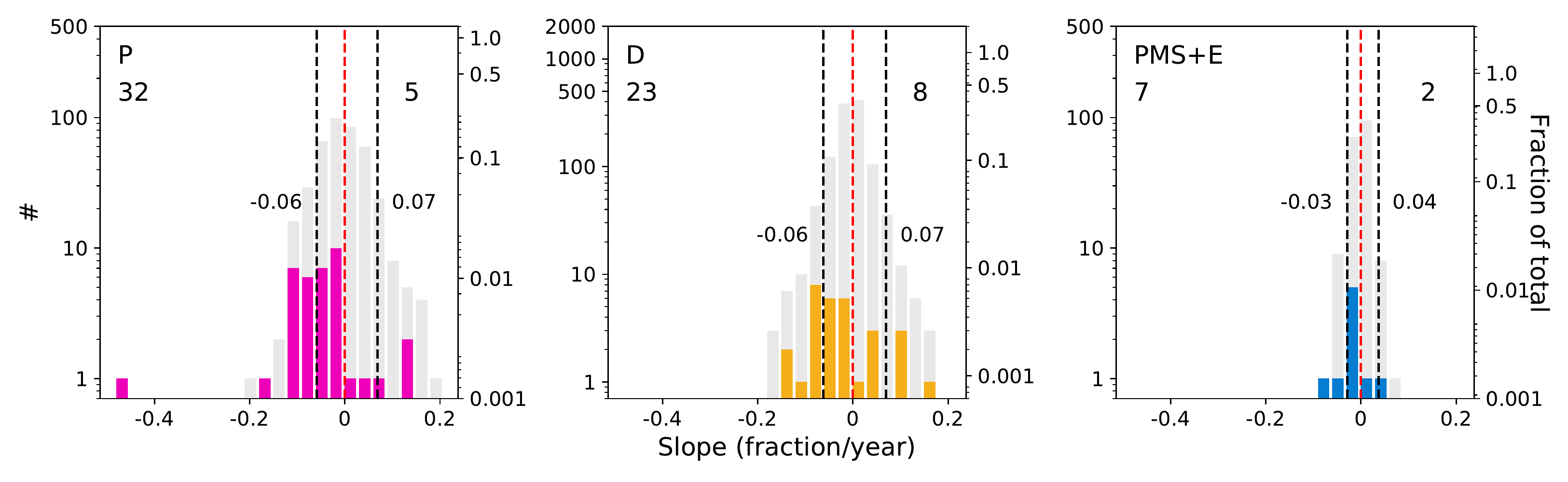}
 }
\caption{The distribution of the measured slopes of linear trends for \textit{Linear} variables. The locations of these \textit{Linear} variables in our diagnostic measure plot, Figure \ref{fig:sec_plot}, are marked by orange triangles. The evolutionary stage of YSOs are indicated on the top left corner of each plot as P, D, and PMS+E. For each plot, the left and the right y-axis show the number of sources and the fractional number relative to the total number, respectively. Note that the y-axis is in log units. Grey histograms show the entire samples while color histograms show the objects with FAP$_{\rm Lin}< 10^{-4}$ (\textit{Linear}). The red dashed vertical lines mark the zero point. The numbers of \textit{Linear}($+$) and \textit{Linear}($-$) sources in each evolutionary stage are shown on the right and left side of each plot. The black vertical dashed lines and the adjacent numbers indicate the median negative and positive slopes.}  
\label{fig:slope}
\end{figure*}

\section{YSO Variability Classification}\label{sec:classification}

In this section, we classify the variability types of YSOs based on the methods described in Section \ref{sec:methods}. First, we divide variability into secular and stochastic and define three different types for each:
%for secular and stochastic variability 
\textit{Linear}, \textit{Curved}, and \textit{Periodic} for secular and \textit{Burst}, \textit{Drop}, and \textit{Irregular} for stochastic. Here we use ``secular" for the regular trends that can be described by simple functions and ``stochastic" for apparently random trends, which cannot be described by simple functions.
For a small subset of variables, the light curves are best explained as a combination of types.

We aim for our classification system to be conservative and thus place strong thresholds on the variability criteria. As such, some sources which are still variable by eye will fall outside our net and not be included here.
The robustness of the variability classification scheme as additional epochs are added to the light curves is presented in Appendix B.

Typically light curves of variables have large standard deviations relative to the flux uncertainty, so the criterion of SD/$\sigma\,>\,3$ can be used to identify variables in general. Of the 5398 candidate young stellar objects in our sample, 1409 satisfy this condition of variability.  However, secular variability can be found even from light curves with low standard deviations because of the underlying regular patterns. In addition, stochastic variability can happen at only one epoch, which does not produce a high standard deviation over the full light curve. Thus, we set the first criterion of $\Delta$W2/$\sigma$(W2)$>3$ in magnitude domain to search for variability; $\sigma$(W2) is the mean uncertainty of W2 magnitudes for a given source. 

With this criterion, for perfect sinusoidal light curves, periodic variability with an amplitude greater than $\sigma$(W2) by a factor of 1.5 will be detected, while the stochastic variability with even one burst or drop event greater than 3$\times\sigma$(W2) will be detected. Within our sample of 5398 candidate YSOs, 3894 satisfy the criterion for $\Delta$W2/$\sigma$(W2)$>3$.
%{\bf needs parallel:  how many are stochastic?}
We apply our methods, which are explained in the previous section, to these 3894 NEOWISE samples to find actual variables despite some having SD/$\sigma<3$.  
We describe each type of secular and stochastic variability below. The number of sources for each variability type for a given evolutionary stage is summarized in Table \ref{tab:table2}. 

\subsection{Secular Variability
\label{subsec:sec_var}}

We applied both Lomb-Scargle Periodogram (LSP) and linear least-square fitting (Lin) methods to isolate secular variables. %Actually, \textcolor{blue}{{\bf all}} 
Secular variables identified by linear least-square fitting are also identified by LSP as variables with long periods. Therefore, LSP alone is enough to isolate secular variables. However, we also apply the linear least-square fitting method to constrain more quantitatively the variability of targets with periods much longer than the time coverage by NEOWISE. 
Roughly 29\% of variables identified in this study are secular variables (Table \ref{tab:table2}), with clear differences seen across evolutionary stage (Class 0/I [P] 36\%, Class II [D] 21\%, Class III + Evolved [PMS+E] 63\%).

\subsubsection{\textit{Linear}: linearly increasing $(+)$ or decreasing $(-)$ light curves}
\label{subsubsec:lin}

We apply the linear least-square fitting analysis to all potential variable targets and classify those with FAP$_{\rm Lin}\,<\,10^{-4}$ as the type, \textit{Linear} (Figure \ref{fig:fap_linb} and Figure \ref{fig:secular_ex}). \textit{Linear} is further subdivided into \textit{Linear}($+$) for positive slopes (increasing light curves, the top panel in Figure \ref{fig:secular_ex}) and \textit{Linear}($-$) for negative slopes (decreasing light curves, the second panel in Figure \ref{fig:secular_ex}). We derive the fractional slopes of the light curves of \textit{Linear} sources, adopting the Equation (1) of \cite{pena20} with a modification; the flux of the first epoch ($f_{0}$) was replaced by the median flux in our calculation. 
Histograms showing the distribution of slopes by evolutionary stage are shown in Figure \ref{fig:slope}; more sources have negative slopes (32) than positive slopes (5) especially for protostars, indicative of a longer decaying timescale than the bursting timescale if the YSO variability is caused by the accretion of the circumstellar material.

\textit{Linear} sources are also identified by LSP as sources with periods
%As a result, using LSP, it is also possible to classify the targets with periods 
longer than 4800 days, since the time coverage by NEOWISE is not long enough to test whether these sources are periodic variables.
%For these variables with long periods, NEOWISE covers only small segments of their light curves where fluxes linearly increase or decrease. 
These targets are located at the upper right region in Figure \ref{fig:fap_linb} with low FAP$_{\rm LSP}$ and FAP$_{\rm Lin}$. Therefore, LSP alone would be good enough to isolate this linear trend, but the amplitude of variability extracted by LSP is possibly overestimated if the variability is not truly periodic. As a result, linear least-square fitting provides a more reliable quantity of variability for this type. 

\subsubsection{\textit{Curved}: curved light curves}

The light curve in the third panel from the top of Figure \ref{fig:secular_ex} shows the {\it Curved} type, with a larger FAP$_{\rm Lin}$ and a lower FAP$_{\rm LSP}$ than our criteria.
However, 
although the periodicity of a target is found by LSP with a low FAP, it is difficult to classify \textit{decisively} the target as \textit{a periodic variable} if the period is not shorter than 1200 days. That is, two periodic cycles at least must be covered by NEOWISE in order to be classified as periodic variables. Targets with periods somewhat longer than 1200 days show curved light curves.
%, which are certainly not linear. 
Therefore, we classify the sources with FAP$_{\rm LSP}\,<\,10^{-2}$ and periods between 1200 days and 4800 days as \textit{Curved}.
%The third panel from the top of Figure \ref{fig:secular_ex} shows a representative light curve of \textit{Curved} variables.
Some variables classified currently as \textit{Curved} could be classified as \textit{Periodic} if they are observed longer in the future. 
%The \textit{Curved} variables are marked with green open circles in Figure \ref{fig:sec_plot}. 

\subsubsection{\textit{Periodic}: periodic light curves}

\textit{Periodic} variables are defined as periodic light curves with periods shorter than 1200 days and FAP$_{\rm LSP}\,<\,10^{-2}$. An example \textit{Periodic} light curve is presented in the bottom panel in Figure \ref{fig:secular_ex}, where the overlaid red line shows the sinusoidal function found by LSP. Many of the \textit{Periodic} and \textit{Curved} sources are not perfectly fit by symmetric sinusoidal functions, likely due to the different heating and cooling timescales \citep[see, e.g., discussion in][]{leeyh20} as well as interspersed stochastic events on top of the secular periodic variability (see also Section \ref{subsec:var_comb_types}). 
%The \textit{Periodic} variables are marked with blue open circles in Figure \ref{fig:sec_plot}. 
Note that the NEOWISE light curves are not appropriate for identification of secular variability on timescales shorter than 6 months due to the cadence of the NEOWISE survey.

\begin{figure}[h] %ht
\epsscale{2}
\centering
{\includegraphics[trim={0.1cm 0.4cm 0.5cm 0.1cm},clip,width=0.98\columnwidth]{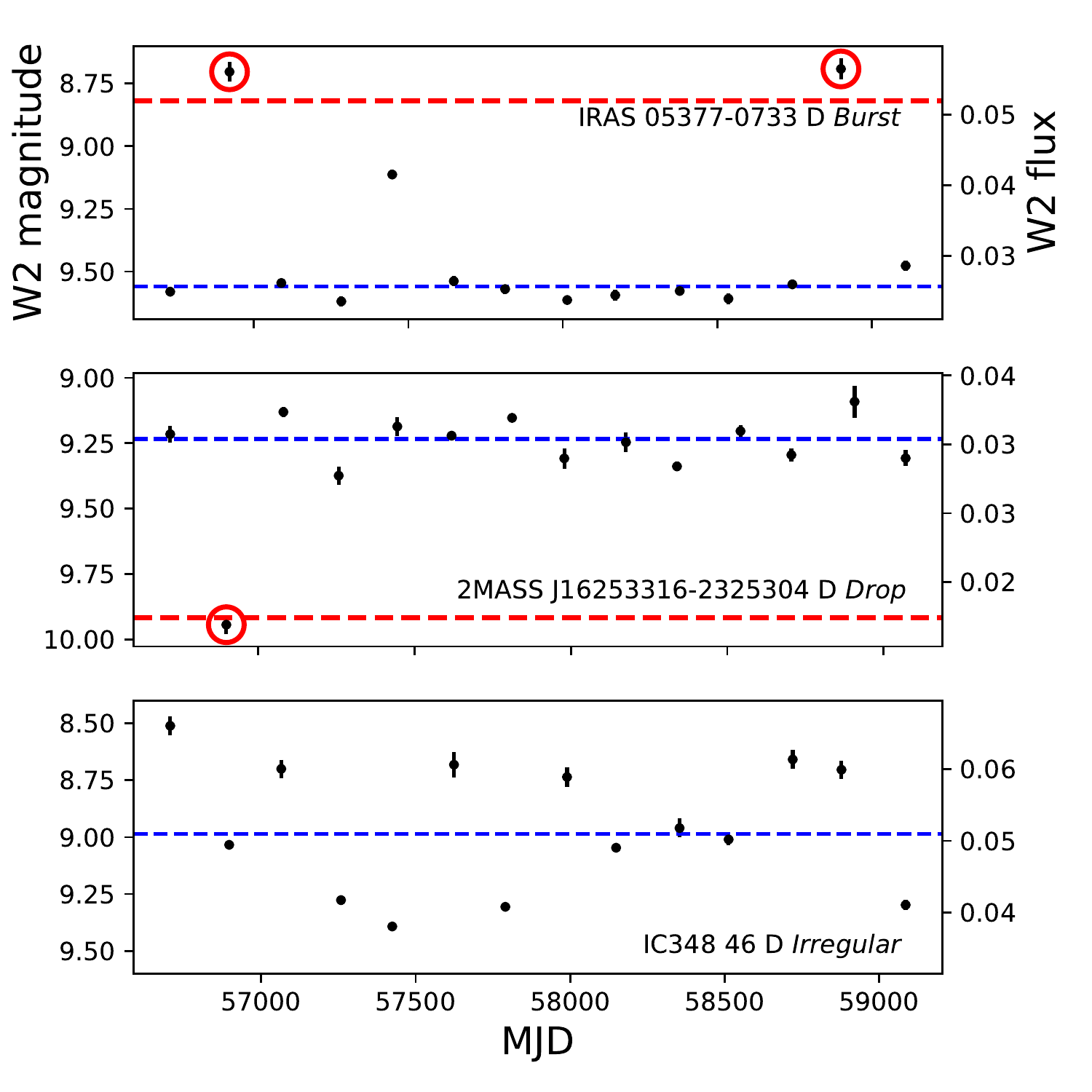}         
}

\caption{Representative light curves of different types of stochastic variability. The dashed blue line indicates the median magnitude. The criterion for \textit{Burst} is 
(median magnitude - minimum magnitude) $>$ $0.8\times$$\Delta$W2. For \textit{Drop}, the criterion is (maximum magnitude - median magnitude) $>$ $0.8\times$$\Delta$W2. For \textit{Burst} and \textit{Drop} sources in the top and middle plots, respectively, these criteria are indicated by dashed red lines. The burst and drop events are marked by red circles.
\label{fig:stoch_ex}}
\end{figure}

\begin{figure}[h] %ht
\epsscale{2}
\centering
{\includegraphics[trim={0.3cm 0.1cm 0.3cm 0.0cm},clip,width=0.98\columnwidth]{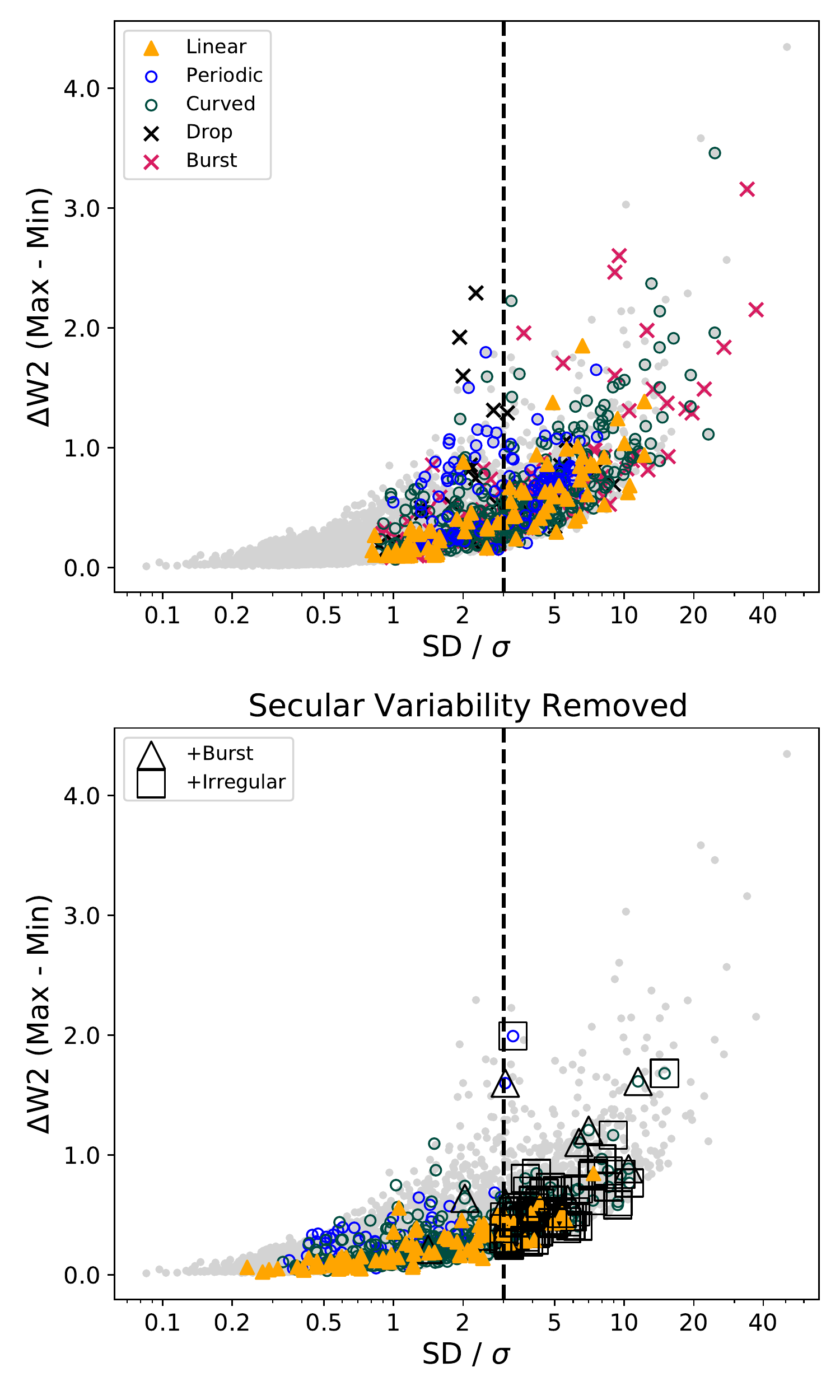}         }
\caption{(Upper) The same as Figure \ref{fig:yso_sc_h} except that variable types are marked with color symbols. The vertical line indicates SD/$\sigma$ of 3, which is generally adopted as the criterion for \textit{irregular} variables. 
(Lower) Recalculation of the plotted values after removing secular trends, if any. The open triangles and squares denote the combined variability types classified from the residual light curves.
\label{fig:sec_plot}}
\end{figure}

\begin{figure}[t] %ht
\epsscale{2}
\centering
{\includegraphics[trim={0.0cm 0.3cm 0.0cm 0.0cm},clip,width=0.98\columnwidth]{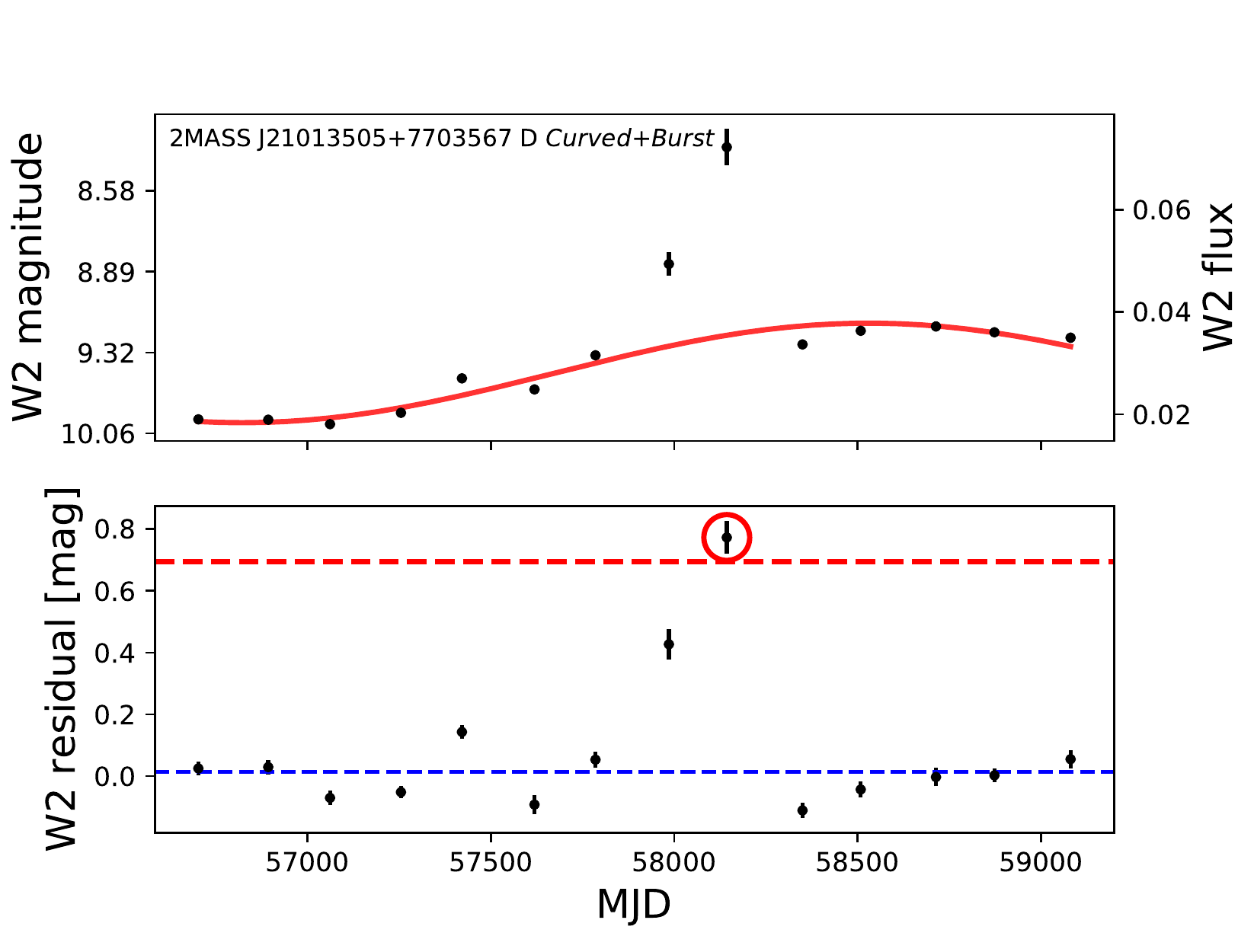}         
}
\caption{A representative light curve of combined types. 2MASS J21013505+7703567 (D) is classified as a combined variable, \textit{Curved+Burst}, with a \textit{Burst} event during the ninth epoch combined with a \textit{Curved} light curve. The top panel shows the secular \textit{Curved} variability, as presented by a best-fit sinusoidal function in the red solid line. This best-fit sinusoid is subtracted from the original light curve to make the residual light curve presented in the bottom panel. From this residual light curve, the stochastic variability is investigated, and classified as \textit{Burst}. The blue and red dashed lines, and the red circle in the bottom panel are the same as those in Figure \ref{fig:stoch_ex}.} 
\label{fig:combined_ex}
\end{figure}

\subsection{Stochastic Variables}\label{subsec:stoch_var}

We define all variability without any regular pattern as stochastic variability and divide further into three types: \textit{Burst}, \textit{Drop}, and \textit{Irregular}. As described below, \textit{Burst} and \textit{Drop} are identified by sudden brightening and dimming only in a few epochs (i.e. with short timescales) over the 6.5-years light curve, while \textit{Irregular} is identified by the random distribution of brightness with a high standard deviation.
Out of 3894 sources with $\Delta$W2/$\sigma$(W2) $>3$, we identified 508 secular variables.
Stochastic variability was searched from the remaining 3386 samples.
1226 samples are classified as stochastic, and these are $\sim$71\% of the entire variables identified in this study.

\subsubsection{\textit{Burst}: light curves with brightness enhancements}

\textit{Burst} variables have brightness enhancements at a few epochs with stable fluxes over the rest of the epochs. As mentioned above, since flux enhancements at a few epochs do not increase the standard deviation largely, we adopt $\Delta$W2/$\sigma$(W2) $>3$ as the first criterion for \textit{Burst} and \textit{Drop}. An additional constraint is necessary to identify targets with brightening events only over a few epochs. For this, we utilize $\Delta$W2 combined with the median and minimum magnitudes in the magnitude domain; a target is classified as \textit{Burst} if the target satisfies the conditions of ({\it median magnitude -- minimum magnitude}) $>0.8\times\Delta$W2.
The top panel of Figure \ref{fig:stoch_ex} shows a representative example light curve for \textit{Burst} variables. In total, 137 \textit{Burst} variables are identified, representing about 11\% of the stochastic sources (Table \ref{tab:table2}).

\subsubsection{\textit{Drop}: light curves with brightness decrements}

The light curves of \textit{Drop} variables show the opposite trend to those of \textit{Burst}; magnitude dips appear at a few epochs. The criterion for \textit{Drop} is the same as those of \textit{Burst}, except for the replacement of minimum magnitude with maximum magnitude: ({\it maximum magnitude - median magnitude})$>0.8\times\Delta$W2. An example light curve for the \textit{Drop} variables is presented in the middle panel of Figure \ref{fig:stoch_ex}. Only 34 variables are classified as \textit{Drop}, or about a quarter the number of \textit{Burst} variables (Table \ref{tab:table2}). The drops could be caused by short timescale extinction events, probably due to the geometric effect of disks (see Section \ref{subsec:mech_var}). 

\subsubsection{\textit{Irregular}: irregular light curves}

After identifying five different types of variables from \textit{Linear} to \textit{Drop} from all 3894 targets with \linebreak $\Delta $W2/$\sigma$(W2) $>3$, the number of remaining targets is 3215. The light curves of these remaining targets look random. To identify true variables with \textit{Irregular} light curves, we finally adopted the general condition for variability by adding, SD/$\sigma>3$. An example light curve for \textit{Irregular} is presented in the bottom panel of Figure \ref{fig:stoch_ex}.
Almost two thirds, $\sim$61\%, of the entire variable sample are \textit{Irregular} (Table \ref{tab:table2}).

\subsection{Variables with Combined Types}\label{subsec:var_comb_types}

The light curves of some secular variables show stochastic variability on top of their secular trends. Therefore, we subtracted the secular trends from their light curves to identify the stochasticity interspersed over the secular trends. Figure \ref{fig:sec_plot} compares the distribution of $\Delta$W2 against SD/$\sigma$ before (top) and after (bottom) subtracting the secular trends found by LSP or Lin, if any. 
We applied the criteria of stochastic variability (Section \ref{subsec:stoch_var}) to the residual light curves of previously classified secular variables (Section \ref{subsec:sec_var}) and found 89 variables in total with combined types.
Therefore, the combined types are a subset of secular variables and listed in Table \ref{tab:table3}. We note that 22\% of secular protostar variables, 23\% of secular disk variables, and 0.9\% of secular PMS+E variables are of combined type.
A representative example is presented in Figure \ref{fig:combined_ex}. 

\section{Discussion}\label{sec:discussion}

\subsection{Previous mid-IR variabilty studies of YSOs}

Careful analysis of the mid-IR variability of YSOs is becoming increasingly possible thanks to dedicated observations of star-forming regions by \textit{Spitzer} \citep{2004Werner} and \textit{WISE} \citep{wright10}. However, most previous studies have focused on searching for EXor/FUor accretion-related outbursts. For example, \citet{scholz13} and \citet{fischer19} compared two epochs of mid-IR photometry for known samples of YSOs (8000 and 319 sources, respectively)  from the \textit{Spitzer} and \textit{WISE} observations. 
From the detection of a handful of YSO outbursts with amplitudes larger than 1 mag over a baseline of $\sim$5 years, the frequency of FUor outbursts was estimated for the early stages of star formation.

\citet{2014Antoniucci} did a similar comparison but with a lower amplitude threshold in order to select and study the EXor type outbursts. More recently, \citet{lucas20} used \textit{WISE/NEOWISE} observations taken between 2010 and 2017 to search for high-amplitude variability in sources projected towards 7000 known Infrared dark clouds. They found 23 highly variable objects, one of which corresponds to a protostellar outburst with an amplitude of 8 mag at 4.6 $\mu$m. Similarly, \citet{uchiyama19} found five mid-IR variable candidates in NEOWISE monitoring of 331 massive protostars.  Finally, \citet{pena20} used \textit{WISE/NEOWISE} observations of sub-mm variables found in the JCMT transient survey observations \citep[e.g.][]{johnstone18}, revealing an observed correlation between the mid-IR and sub-mm variability, with implications for interpreting SEDs of outbursting protostars.

%The observed correlation between mid-IR and submm variability allows us to study the effect of outbursts on the SEDs of deeply embedded protostars.  
     
The YSOVAR program, on the other hand, provides dedicated observations at 3.6 and 4.5 $\mu$m for YSOs in 5 known star-forming regions \citep{morales11,wolk18}. YSOVAR is a high cadence survey over timescales of $\sim$40 days. The survey shows the complexity of mid-IR variability in YSOs, which might be associated with various physical mechanisms affecting the stellar photosphere and the inner disk. 
  
Our analysis in this paper enables an ensemble study for the overall YSO variability of a larger sample and over longer timescales than those covered by the YSOVAR program. In Section \ref{subsec:mech_var}, we discuss how the physical mechanisms studied in previous surveys of mid-IR variability in YSOs can apply to the different variability classes defined in Section \ref{sec:classification}. 

%We also compare the outburst frequency obtained from our sample against those discussed above in Section \ref{subsec:outburst_freq}. 

\subsection{Mechanisms for Variability}\label{subsec:mech_var}

The mechanisms that lead to variability in YSOs are associated with accretion processes, variable extinction, and changes in disk properties operating alone or in combination. These mechanisms lead to variability on a wide range of amplitudes and timescales \citep{morales11, 2014Cody, wolk18, pena20, lucas20,guo21}. 

Variable accretion in YSOs can be caused by a variety of different physical mechanisms, with perhaps a continuum of outbursting behaviour with a wide range of amplitudes (0.2-7 mag at optical wavelengths) and time-scales \citep[0.1 d to 100 yrs, e.g.][]{herbig77,2017Cody}. Outbursts lasting from 0.1 days to a few months are thought to be caused by viscous and magnetic instabilities at the boundary between the stellar magnetosphere and the accretion disk \citep{2008Kulkarni, 2012Dangelo,takasao19} while larger amplitude ($\Delta m>3$~mag), longer duration events (a few to up to 100 years) are linked to gravitational instabilities \cite[GIs,][]{zhu09,vorobyov10}, planet-induced thermal instabilities \citep{2004Lodato}, or binary interactions \citep{1992Bonnell}.

These type of events have been previously observed at optical, near-IR, and mid-IR wavelengths, i.e.~the short-term bursters that last for hours \citep{2013Findeisen, 2014Cody, 2014Stauffer}, EXor (months-long) or FUor type (decades-long) outbursts \citep{lorenzetti12, connelley18}, as well as outbursts with durations that are between those of EXors and FUors \citep[so-called MNors][]{contreras17}, such as the outbursts of V1647 Ori \citep{2007Acosta} and ASASSN-13db \citep{Holoien14,2017Sicilia}.

On the other hand, dips in the light curves of YSOs have usually been ascribed to variable extinction along the line of sight \citep{1994Herbst, carpenter01}. These events also occur with a variety of timescales and amplitudes. AA Tau-like objects display (quasi-)periodic obscuration events, with periods on the order of  a few days, that result from the obscuration of the central star by a warped inner disk \citep{2013Bouvier}. UXors show periodic dimming events (lasting days to weeks) due to dust clouds blocking the stellar light \citep{natta97,2007Rostopchina}. Finally years-long fading events have also been observed, for example, in RW Aur, AA Tau and V409 Tau \citep{2013Bouvier, 2016Bozhinova, 2015Rodriguez}. These long-duration dimming events are interpreted as obscuration from inhomogeneities located at large distances in the accretion disk or perhaps even a dusty wind. 

The models that describe variability in YSOs arise from observations mostly at optical and near-IR wavelengths. Contemporaneous observations at mid-IR and optical wavelengths show a complex behaviour that can challenge some of the known models \citep[e.g.][]{2014Cody}. For example, in variable extinction we expect to observe optical to infrared correlation of the variability, which simply reflects the wavelength dependence of extinction, diluted by any flux from the inner disk \citep{2014Cody}. However, some dippers in NGC 2264 show larger infrared than optical amplitudes. This could reflect a more unique geometry of the YSO system and might be the result of occultations of the disk by itself \citep{2014Cody}. The long-term optical to near-IR fading in AA Tau occurs as the mid-IR flux of the system increases. The anti-correlated variability might be explained by an increase in the scale height of the inner disk \citep{covey21}. 

Young stellar objects undergoing outbursts of accretion show correlated variability across the optical to mid-IR wavelengths. However, inclusion of longer wavelengths can help understand the way the outburst propagates through the disk. \citet{hillenbrand18} find that the outburst of FUor object Gaia17bpi started in the mid-IR at least a year earlier than the observed increase at optical wavelengths. This is explained as an outburst that starts at larger distances in the disk and then propagates inward.

For most protostars, the lack of contemporaneous photometry at shorter wavelengths and spectroscopic follow-up makes the task of associating our variability classes to these physical mechanisms difficult. Nevertheless, the observed NEOWISE light curves still allow us to obtain a rough understanding of the underlying mechanism driving the variability in our YSOs.
 
The observational cadence of NEOWISE does not allow us to study in detail the physical mechanisms that lead to variability with timescales of less than 6 months. These timescales are associated with processes affecting the stellar photosphere and the inner disk and include variable accretion (short-term bursters and EXors), variable extinction  (AA Tau-like objects and UXors) and quasi-periodic variability arising from hot spots of accretion \citep{morales11}. As noted in Section \ref{sec:wise}, however, there are roughly a dozen exposures at each epoch, spread over a day (see also Figure \ref{fig:short_var}). We will present results on this short-time variability analysis in a future publication.

Stochastic \textit{Burst} and {\textit Drop} variability is only observed at one to a few epochs in the light curves of YSOs falling in these classifications. These are likely short-timescale events associated with processes occuring close to the star. Changes in the extinction along the line of sight due to obscuration from inhomogeneities located close to central star are the most likely explanation for the variability of YSOs falling in the {\it Drop} classification. Bursters are probably related to the short-term bursters found by \citet{2013Findeisen} and \citet{2014Stauffer} or the longer-duration EXor outbursts\citep{lorenzetti12}, which are explained by viscous and magnetic instabilities at the boundary between the stellar magnetosphere and the accretion disk. We note that for both Class 0/I and Class II sources,  the fraction of \textit{Drop} variables compared with \textit{Burst} variables is relatively low, 0\% and 23\% respectively (Table \ref{tab:table2}). 
The lack of observed \textit{Drop} variables among the Class 0/I may be due to a selection bias against nearly edge-on disks in our mid-IR brightness limited sample.

We are not able to resolve the variability of repetitive short-term bursting events, (quasi-)periodic AA Tau-like variability or quasi-periodic variability arising from hot spots of accretion. It is likely that these mechanisms \citep[potentially in combination, similar to the stochastic variability class of ][]{2014Cody} are responsible for the observed variability in many members of the {\it Irregular} class.

Longer-term secular changes are likely induced by disk instabilities that lead to changes in the accretion rate (outbursts usually classified as FUors). The \textit{Linear} variability found in our analyses might be associated with accretion events whose outbursts duration are longer than the NEOWISE coverage.  If this is the case, \textit{Linear}($+$) objects are associated with the rise of the accretion rate, whilst \textit{Linear}($-$) YSOs are showing slow decay in the accretion rate.

The rarer \textit{Periodic}, and also some of \textit{Curved}, variability might be induced by regular/periodic dynamics such as binary interactions \citep[e.g.][]{2012Hodapp}. The detailed accretion process occurring in the disk might be hinted by the color variation along with the brightness variation with time as found for EC 53 (V371 Ser) by \citet{leeyh20}, although the associated separations (1-10 AU) may facilitate rapid disk depletion \citep[e.g.][]{kraus12}. We return to this idea in Section \ref{sec:discussion-color}. In addition, some objects showing periodic variability due to accretion changes might be classified as \textit{Irregular} due to the sparse sampling of NEOWISE light curves, as is found for the case for the known periodic variable YSO EC 53 \citep{leeyh20}.

The increase in the height of the inner disk could explain some of the {\it Curved} objects. For example YSO LkH$\alpha$ 337 is classified as {\it Curved} based on its mid-IR light curve. The increase at mid-IR flux does not correlate with shorter-wavelengths, which instead show repetitive dips in the light curve (see Section \ref{sec:long-term}).

For Class 0/I protostars we find many fewer \textit{Burst} variables compared with the longer-term secular variables (\textit{Linear, Curved}, and \textit{Periodic}), whereas for Class II sources the numbers are comparable (Table \ref{tab:table2}).

\begin{figure}[t] %ht
% \epsscale{2}
\centering
%\subfigure{
{
\includegraphics[trim={0.3cm 0.4cm 0.0cm 0.1cm},clip,width=0.98\columnwidth]{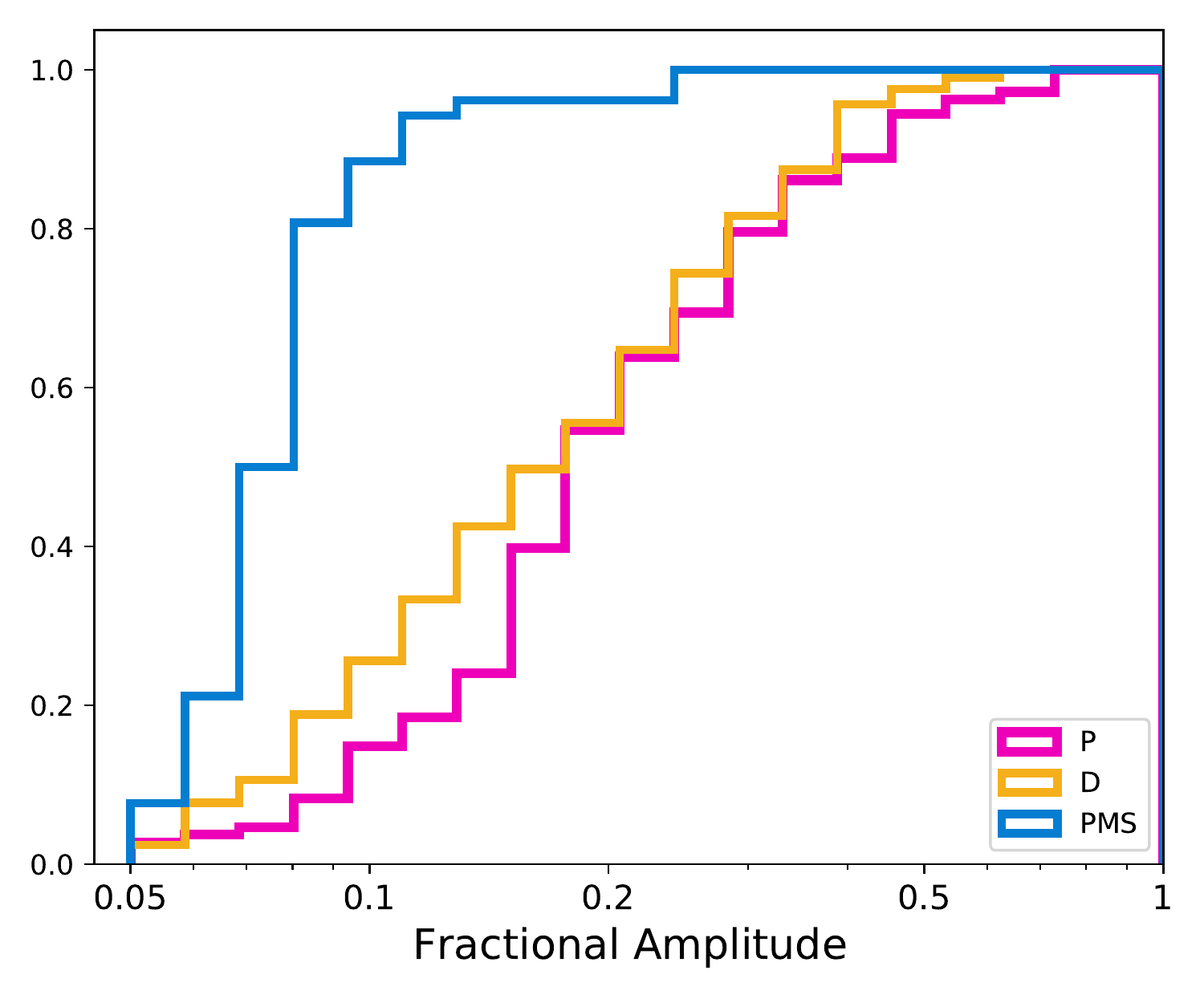}
}
\caption{The cumulative distribution of fractional amplitude of \textit{Curved}/\textit{Periodic} variables. Colors are the same as those in Figure \ref{fig:meanw2}: magenta for P, yellow for D, and blue for PMS. Note that the higher amplitude and shorter period PMS+E sources, which are believed to be AGB contaminants to the sample have been omitted (see text).
\label{fig:amp_cdf}}
\end{figure}

\subsection{Variability with Evolutionary Stage}\label{subsec:var_evol}

As presented in Figure \ref{fig:yso_sc_h}, the amplitude of variation is largest for the earliest evolutionary stage Class 0/I (P), except for the small bump for PMS+E at SD/$\sigma\,\sim\,3$ and $\Delta$W2$\sim$0.7. In addition, the fractional number of variables relative to all NEOWISE samples at a given evolutionary type decreases with increasing age (Table \ref{tab:table2}); about half of protostars (55\%) are variable in contrast with $\sim$33\% and $\sim$15\% of disks and PMS+E sources, respectively. The fractional number of PMS+E is an upper limit, given that many objects in the class are probable AGB contaminants (Appendix \ref{appendix:agb}). We further note that protostars show the largest fractions (or an equivalent fraction to disks) of variables for all variability types except for \textit{Periodic} (numbers in parentheses of Table \ref{tab:table2}).

For the \textit{Curved} and \textit{Periodic} variables, the periods are mostly longer than 1200 days for protostars and disks while the periods for $\sim$75\% of the PMS+E variables are shorter than 1200 days (Figure \ref{fig:period_cdf}). These \textit{Periodic} variables of PMS+E with the periods shorter than 1200 days are associated with the higher fractional amplitude peak in the left panel of Figure \ref{fig:lowfapamp} (the hatched blue histogram). Given the discussion in Appendix \ref{appendix:agb},  these shorter period and higher amplitude PMS+E variables are likely AGB stars and can be excluded. Figure \ref{fig:amp_cdf} presents the cumulative distribution function for the fractional amplitude of the fitted sinusoidal function \textit{after excluding the AGB candidates, i.e., the higher amplitude and shorter period PMS+E variables.} The figure clearly shows that the amplitude of variability is the largest in the earliest evolutionary stage Class 0/I (P) and is reduced greatly for the latest evolutionary stage (PMS). 

For \textit{Linear} variables, the distributions of slopes for different evolutionary stages are presented in Figure \ref{fig:slope}. 
The fraction of \textit{Linear} variables, as well as the degree of asymmetry in the source numbers between the positive and negative slopes, is much larger in the protostellar stage than the PMS stage. 

In summary, both the fractional number of variables and the amplitude of variability generally decrease with the evolutionary stage, as suggested by previous observational studies \citep{morales11,rebull15, pena17,wolk18} as well as theoretical investigations \citep{hartmann98,bae14,vorobyov15}. 

\begin{figure*}[t]
\centering
%\subfigure{
{
\includegraphics[trim={0.1cm 0.4cm 0.1cm 0.3cm},clip,width=0.98\linewidth]{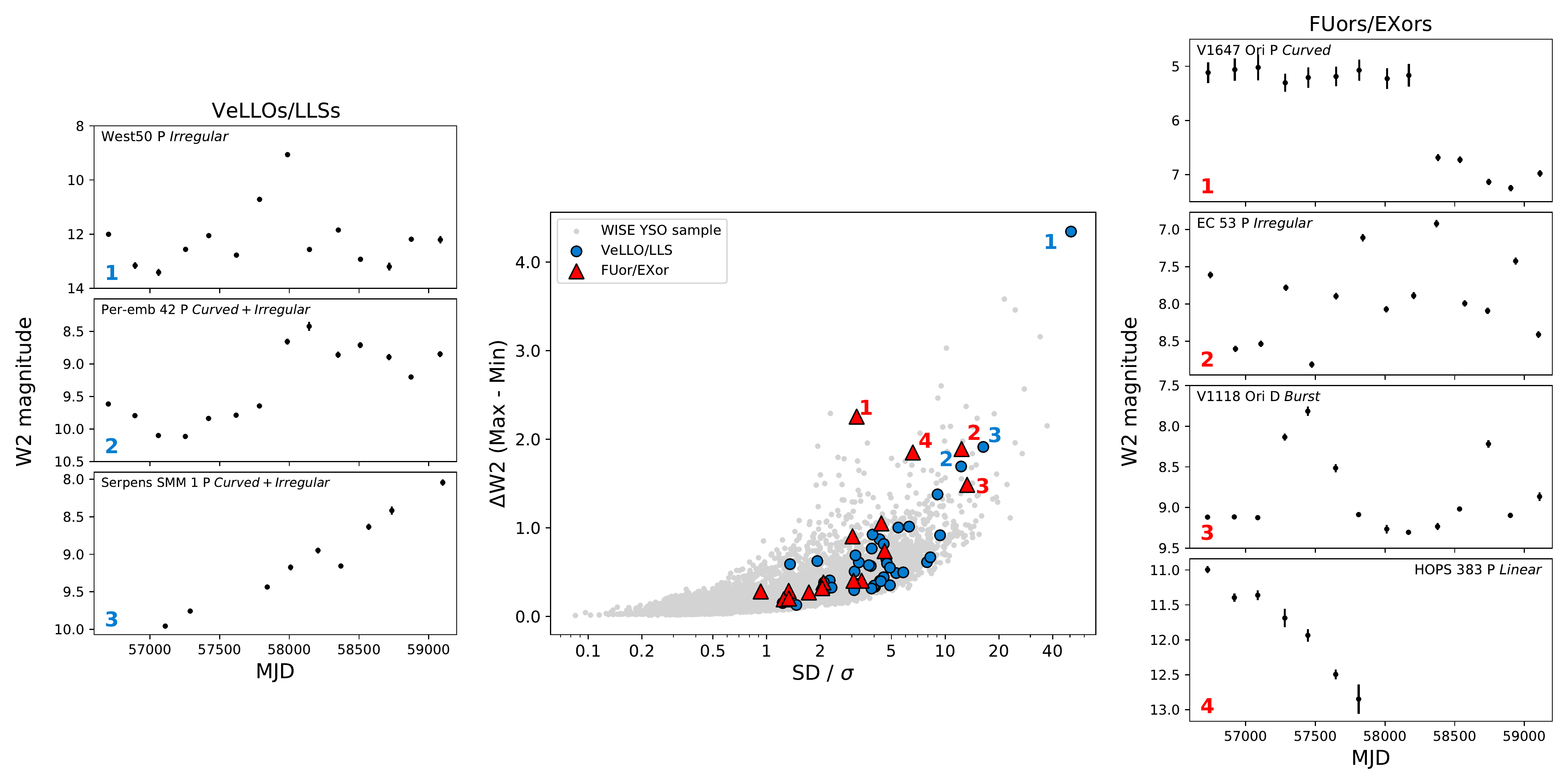}
}
\caption{Variable FUors/EXors, and VeLLOs/LLSs in the mid-IR. (Center) $\Delta$W2 vs SD/$\sigma$ for \textit{WISE} YSO samples (gray circles), variable VeLLOs/LLSs (blue circles) and variable FUors/EXors (red triangles). The left and right panels show the light curves of three VeLLOs/LLSs, which show recent brightening events, and four well-known FUors, respectively. The corresponding targets for the light curves are marked with numbers in the central panel.
%            }
\label{fig:vello_fuor}}
\end{figure*}

\begin{deluxetable}{ccc}

\tablecaption{Variability Types of variable FUors/EXors and VeLLOs/LLSs\label{tab:table4}}
\tablenum{4}
\tablehead{\colhead{Variability Type} & \colhead{FUors/EXors} & \colhead{VeLLOs/LLSs}}

\startdata
\textit{Linear} & 7 (25.9)\tablenotemark{a}  & 4 (5.5) \\
\textit{Curved}  & 4 (14.8) & 15 (20.5) \\%number
\textit{Periodic} & 0 (0)   & 0 (0) \\
\textit{Burst}  & 1 (3.7)   & 0 (0) \\
\textit{Drop} & 0 (0)  & 0 (0) \\
\textit{Irregular}& 3 (11.1) & 21 (28.8) \\
\hline
Total & 15 (55.6) & 40 (54.8)\\
\enddata
\tablenotetext{a}{Numbers in front are the counts of variables, while numbers in parentheses are the fractions (\%) of variables relative to the total FUors/EXors and VeLLOs/LLSs samples, 27 and 73, respectively.}
\end{deluxetable}

\begin{figure*}[t]
\centering
%\subfigure{
{
\includegraphics[width=.8\linewidth]{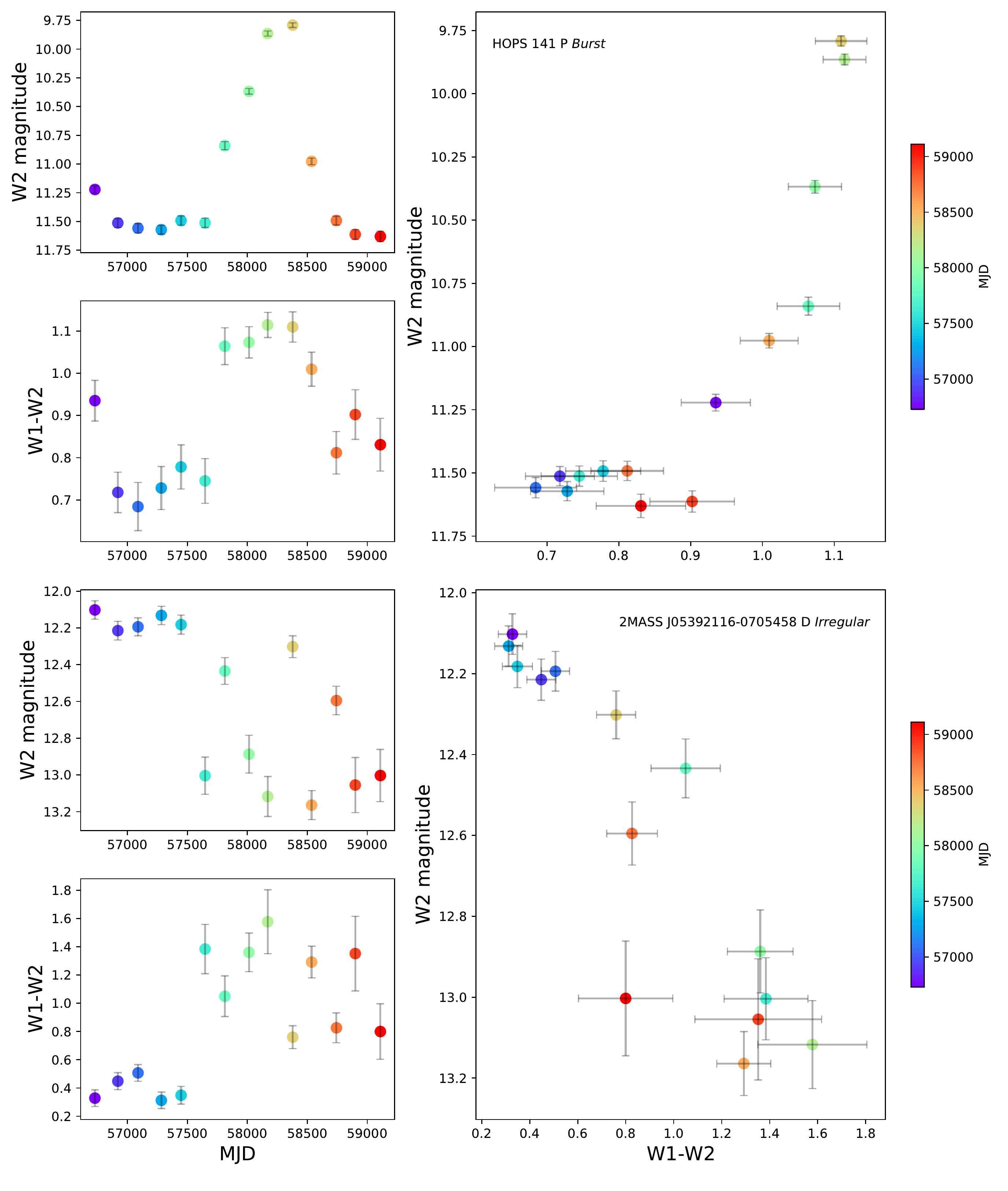}
}
\caption{Monotonic color variations. HOPS 141 is a \textit{Burst} protostar (upper panel), and 2MASS J05392116-0705458 is an \textit{Irregular} disk source (lower panel). The color variation relative to the brightness variation is much larger for 2MASS J05392116-0705458.
\label{fig:color_mono}}
\end{figure*}
    
\begin{figure*}[t]
\centering
%\subfigure{
{
\includegraphics[width=.8\linewidth]{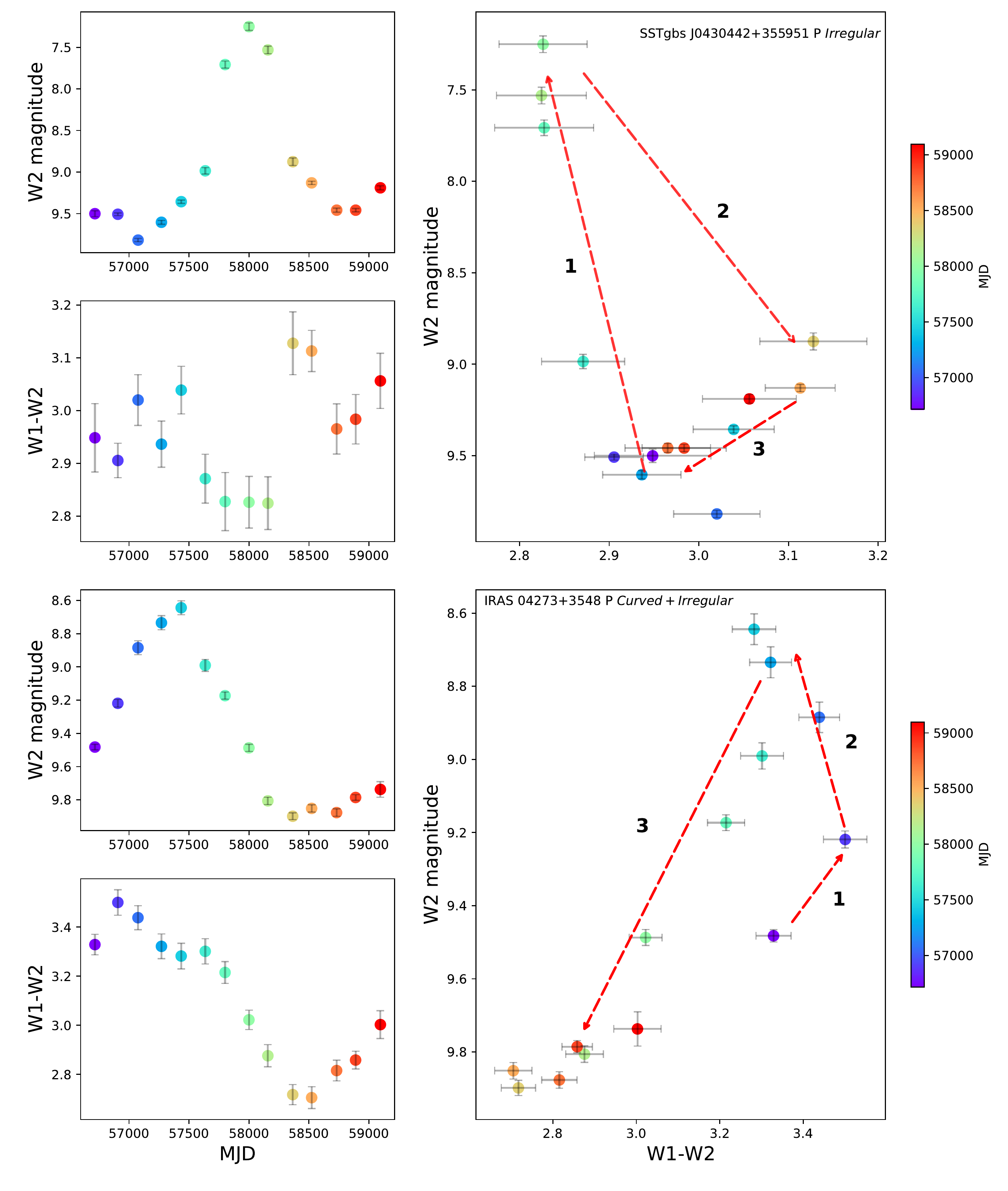}
}
\caption{Cyclic color variations. SSTgbs J0430442+355951 (\textit{Irregular}, P) shows a clockwise color variation with a two magnitude brightness change (upper panel). IRAS 04273+3548 (\textit{Curved}+\textit{Irregular}, P) has an anti-clockwise color variation with a one magnitude brightness change (lower panel). The color variation relative to the brightness variation is larger for IRAS 04273+3548 than for SSTgbs J0430442+355951.     
\label{fig:color_cyc}}
\end{figure*}

\subsection{Eruptive and subluminous YSOs}\label{subsec:fu_ex_vel}

The protostellar luminosity problem is recognized as an inconsistency between the protostellar luminosity function derived from observations and theoretical expectations \citep{dunham10}. Episodic accretion, consisting of quiescent-accretion phases interspersed with burst-accretion phases, has been suggested as a promising solution for the luminosity problem \citep[e.g.][]{audard14,dunham14}. 
In the episodic accretion model, stars build a significant fraction of their total mass during short outbursts of enhanced accretion. The largest accretion bursts have been detected as FU Orionis objects (FUors), which exhibit large-amplitude jumps in the optical ($\Delta m_V>4$~mag) and can last for decades. EXors, named after the prototype EX Lup, have lower amplitude outbursts every few years and stay bright for several months at a time.
In contrast, Very Low Luminosity Objects (VeLLOs, \citealp{young03,francesco07}), which are in the embedded stage, have a luminosity of lower than 0.1 L$_\odot$, possibly representing sources in the most quiescent phase of the episodic accretion process. 

Therefore, FUors/EXors and VeLLOs are considered as YSOs in the extrema of luminosity evolution at the observed moments.  Mass accretion rates vary during the YSO evolution and thus it should be natural for a YSO previously known as eruptive or subluminous to shift between categories during its evolution. Therefore, we do not identify these eruptive and subluminous phenomena as the cemented or intrinsic nature of the YSO. Instead, we examine whether there are clear differences in variability properties between the two types of phenomenon.

We cross-matched the variables identified in this study with the lists of known low luminosity sources ($<1$ L$_\odot$, LLSs) as well as VeLLOs \citep{dunham08,kimmr16,kimgj19} and FUors/EXors \citep{antoniucci13, audard14, pena14, pena17, connelley18}. We found 8 and 40 variables, respectively, out of 13 FUors/EXors and 73 VeLLOs/LLSs identified from the NEOWISE survey based on our criteria used in Section \ref{sec:wise}. 

In order to increase our sample of FUors/EXors, we extracted the photometric information for an additional 23 known FUors/EXors \citep{audard14, connelley18} from the NEOWISE archive. Twelve of these satisfy our criteria, in Section \ref{sec:wise}, for obtaining robust mid-IR light curves. Finally, we found 7 out of these 12 FUors/EXors are confirmed variables. As a result, a total of 15 FUors/EXors are variables out of 25 FUors/EXors in our NEOWISE samples. The fractional number of variables are similar in FUors/EXors (60\%) and VeLLOs/LLSs (55\%), indicative of the same nature of variability, over 6.5 years, both in eruptive and subluminous YSOs.

The variable types for FUors/EXors and VeLLOs/LLSs, however, are very distinct even though the overall fractions of variables are similar in the two types of objects, as listed in Table \ref{tab:table4}.
The variable FUors/EXors include 10 protostars (5 \textit{Linear}, 3 \textit{Curved}, and 2 \textit{Irregular}) and 5 disks (2 \textit{Linear}, 1 \textit{Curved}, and 1 \textit{Burst}, 1 \textit{Irregular}), while the variable VeLLOs/LLSs include 32 protostars (4 \textit{Linear}, 11 \textit{Curved}, 17 \textit{Irregular}) and 8 disks (4 \textit{Curved}, 4 \textit{Irregular}). 
Most FUor/EXor variables are secular with only 4 classified stochastic, while more than half of VeLLO/LLS variables are irregular. In addition, about half of FUor/EXor variables are {\it Linear} with all exhibiting declining light curves, {\it Linear}($-$), as listed in Table \ref{tab:tabled1}.

Figure \ref{fig:vello_fuor} (center) shows the distribution of $\Delta$W2 vs SD/$\sigma$ for variable VeLLOs/LLSs (blue circles) and FUors/EXors (red triangles). FUors/EXors typically show larger variations than VeLLOs/LLSs in $\Delta$W2 while VeLLOs/LLSs show larger SD/$\sigma$ compared to FUors/EXors. Three VeLLOs/LLSs, however, are located at large $\Delta$W2 and SD/$\sigma$ similar to those of four well-known FUors. The left and right panels of Figure \ref{fig:vello_fuor} present the light curves of these three VeLLOs/LLSs, which show recent brightening events, and these four FUors with large $\Delta$W2, respectively. 
This clearly demonstrates that the eruptive and subluminous classifications are not uniquely separable for YSOs, even over short timescales.
We summarize the information on all variable FUors/EXors and VeLLOs/LLSs, which have large values of $\Delta$W2 and SD/$\sigma$, in Tables \ref{tab:tabled1} and \ref{tab:tabled2}, respectively.

%{\bf (some discussions for special characteristics of variable FUors?)}
%12 variable FUors identified from our analysis are HOPS 383 (\textit{Linear}, P), V1647 Ori (??, P), V1118 Ori (\textit{Burst}, D), and IC 348 (\textit{Irregular}, D), ....

\subsection{Secular Color Variation}\label{sec:discussion-color}

The observed color variations of FUors and EXors are diverse. Some EXors and HBC 722, which is known as a FUor \citep{leeje15,connelley18}, become bluer around the burst and redder during the quiescent phase in near-IR observations \citep{kospal11,lorenzetti12}. On the other hand, V346 Nor becomes redder during its slow brightening \citep{kospal20}. The outburst Gaia 18bpi occurred initially in the mid-IR before moving to shorter wavelengths \citet{hillenbrand18}.

\cite{leeyh20} presented a detailed study of near-IR color variation for EC 53, which is classified as \textit{Irregular} in this study but is actually a quasi-periodic variable with a period of eighteen months. 
EC 53 quickly reddens just before the burst, likely due to the buildup of its inner disk mass leading to an increase in the geometric height of the disk, and thus, a largely increased extinction. Right after the burst event, likely caused by the draining of the inner disk material to the protostar, the color suddenly moves to the blue and very slowly becomes even bluer as both the extinction and brightness decrease. Then, a new buildup of the inner disk mass begins, following again the counterclockwise cyclic track with time in the color-magnitude diagram (see Figure 9 of \citealp{leeyh20}). 

We have analysed the secular W1-W2 color variations of our NEOWISE sample for similarly clear patterns. A caveat of our color analysis is that the NEOWISE light curves do not have sufficient time resolution to reveal the detailed color variation, as caught in EC 53. 
Nevertheless, we find diverse color variations in our samples. 
Here we present representative examples of color variation, and leave the more complete analysis to a future work.

The protostar HOPS 141 (P) clearly shows a red color at the brightest phase and becomes bluer monotonically as it becomes fainter while the disk source 2MASS J05392116-0705458 (D) presents the exactly opposite trend (Figure \ref{fig:color_mono}).
The color change in 2MASS J05392116-0705458 (D) is much larger than that in HOPS 141 (P) although the brightness change is greater in HOPS 141 (P) than 2MASS J05392116-0705458 (D). 
The color variation of HOPS 141 (P) is probably affected by the extinction predominantly while that of 2MASS J05392116-0705458 (D) appears to respond sensitively to the temperature change. 
In this respect, the color of 2MASS J05392116-0705458 behaves similarly to Gaia 19ajj \citep{hillenbrand19}; bluer when brighter. An extinction variation of about 100 magnitudes is required if the color change is solely attributed to extinction. Therefore, as suggested by \cite{hillenbrand19}, this large color variation is most likely intrinsic to the source (i.e., temperature variation). 

In our survey we find additional sources  with the cyclic color variations seen in EC 53. Figure \ref{fig:color_cyc} shows examples of clockwise and counterclockwise cyclic variations in two protostars.
Following the analysis presented for EC 53 by \cite{leeyh20}, this diversity of color variation is probably caused by the competitive interplay between the accretion luminosity (i.e., source temperature) and the extinction by material associated with the accretion process. For example, SSTgbs J0430442+355951 (upper panel in Figure \ref{fig:color_cyc}) follows a clockwise pattern, becoming slightly bluer ($\sim$0.1 mag) when brightened by 2 mag in W2 (track 1 in the cycle). We expect a stronger bluing given the large brightening, suggesting that extinction is also increasing at the same time.
Next, the source dims by 1.5 mag over $\sim$1 year and becomes redder by $\sim$0.25 mag (track 2), indicative of extinction. Finally, it becomes again blue close to the original color (track 3) despite only a small brightness change of $\sim$0.5 mag, probably due to the extinction clearing.

On the other hand, IRAS 04273+3548 (lower panel of Figure \ref{fig:color_cyc}) follows a similar color variation to that of EC 53. Along track 1, it becomes brighter, but the color gets redder. Then, on track 2 the color turns to the blue while the brightness continues to increase, resulting in a counterclockwise rotation in the color-magnitude diagram. As interpreted for EC 53 \citep{leeyh20}, extinction due to an enlarged geometric height of the inner disk during the initiation of a burst accretion can produce the red color (track 1), and  draining of the built-up inner disk material onto the protostar can reduce the extinction and slowly reveal the central source at a hotter stage (track 2). In track 3, the material is cleared quickly as the accretion rate greatly decreases.

\section{Long-Term Variability and the frequency of FUor outbursts}\label{sec:long-term}

The long-term, multi-wavelength photometric behaviour of variable objects in our sample provides insights into the physical mechanism driving the variability in the YSOs. This is particularly important if the variability is related to long-term fading due to structures in the disc or to long-duration (FUor-like) outbursts. Episodic accretion is likely to have an impact on both star and planet formation. Long-lasting outbursts alter the properties of the central star, such as luminosity and radius, and could explain the observed spread in the Hertzprung-Russell diagrams of pre-main-sequence clusters \citep[e.g.][]{2017Baraffe}.  Outbursts will alter the chemistry of protoplanetary disks \citep{2019Artur}, the location of the snowline of various ices \citep{2016Cieza,Lee19} and could affect orbital evolution of planets  \citep{2013Boss}. As well, the long quiescent-accretion phases could help to produce low-mass companions \citep{2012Stamatellos}. Determining the frequency of YSO outbursts is therefore an important input for models of star and planet formation.

We estimate the frequency of FUor outbursts \citep[the recurrence timescale, $\tau$, see][]{contreras19} following the analysis of \citet{scholz13} and \citet{fischer19}. To select YSO outbursts, \citet{scholz13} and \citet{fischer19} use an amplitude threshold of 1 to 1.6 mag in the mid-IR. Choosing this value should discard YSOs where variability is driven by common physical mechanisms such as hot-spots or inner-disk inhomogeneities. In our sample, the selection of YSOs that vary by more than 1 magnitude in $\Delta$W1 or $\Delta$ W2 yields 227 Objects, with 112 disks, 104 protostars and 11 PMS+E objects. The latter are all classified as periodic and are likely AGB contaminants (see Appendix \ref{appendix:agb}).

From this initial selection, we choose only the sources whose light curves resemble those of long-term outbursts (or FUors). Although short-term outbursts can still have an effect on processes of planet formation \citep{2019Abraham}, FUor-like outbursts are more likely to have a long-lasting impact on stellar/planet formation. We therefore inspected visually the mid-IR light curves of all 227 candidate YSOs. Where available, we also inspected the photometry arising from optical, near, and mid-IR surveys via public catalogues available from both Vizier \citep{2000Oschsenbein} and the NASA/IPAC Infrared Science Archive (IRSA). These catalogues include the \textit{Spitzer}/GLIMPSE surveys, 2MASS \citep{2006Skrutskie}, DENIS \citep{1994Epchtein}, UKIDSS GPS and GCS \citep{2007Lawrence,2008Lucas}, VISTA VHS \citep{2013Macmahon}, Carlsberg Meridian Catalogue \citep[CMC15][]{2014Muinos}, APASS \citep{2015Henden}, Pan-STARRS\citep{2016Chambers} and the Zwicky Transient Facility \citep[ZTF][]{2019Bellm}.

For all of the YSOs in our sample we are able to obtain photometric information from {\it Spitzer}, providing a long-baseline to study mid-IR variability. We do not have such complete coverage at shorter wavelengths as some objects were not covered by the surveys or are too faint at these shorter wavelengths. Nevertheless, we are able to provide a classification for the majority of the high-amplitude variable YSOs using the available data.

The classification of the 227 YSOs as candidate FUors is done based only on the visual inspection of the long-term light curves (i.e. including photometry taken before NEOWISE observations) and thus will not necessarily have a similar variability class to the one determined from NEOWISE data alone. In addition, the classification is done blindly without knowing if the YSOs have a previous classification as eruptive variables. This explains why the known eruptive variable star EC 53 is not included in the list of candidates, as the light curve of the object appears irregular and would not be classified as an FUor candidate from visual inspection alone.

\begin{figure}[t]
\centering
%\subfigure{
{
\includegraphics[width=1\linewidth]{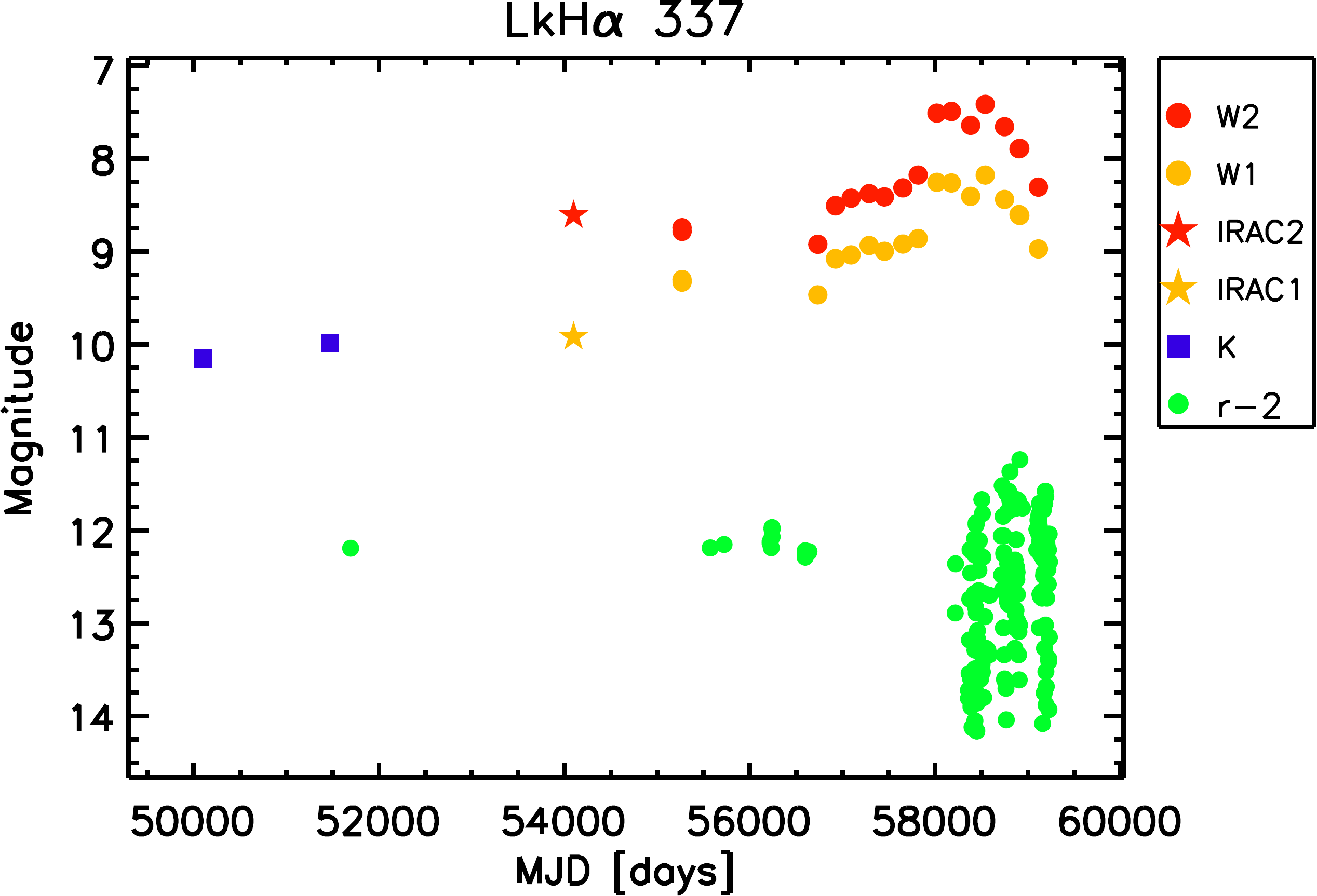}
}
%\vspace{-2cm}
\caption{Light curves for LkH$\alpha$ 337: optical sloan r (green circles), near-IR (blue squares), and mid-IR $3.4 \mu$m (W1; yellow circles) and $4.6 \mu$m (W2; red circles).  {\it Spitzer} IRAC1 and IRAC2 filters (colored stars) were converted to the associated {\it WISE} filters using the relations by \citet{2014Antoniucci}. Optical data arises from the APASS, CMC15, Pan-STARRS and ZTF surveys.      
\label{fig:long_var}
}
\end{figure}

Comparison with shorter wavelengths proves useful to understand the physical mechanism driving the observed variability.  For example Figure \ref{fig:long_var} shows the variability of LkH$\alpha$ 337, classified as \textit{Curved+Irregular} in our work. Along with its linear rise during NEOWISE monitoring, the source abruptly brightens by a magnitude in the mid-IR near MJD 58000, which resembles an outburst. The r-band photometry of the object shows sudden drops in flux that are coincident with this abrupt increase in the mid-IR. This large optical variability was detected by the All-Sky Automated Survey \citep[ASASSN-V J055437.95+012951.4,][]{2017Kochanek} and was also the subject of a \textit{Gaia} alert \citep[Gaia18beu,][]{2013Hodgkin}. It is possible that an increase in the scale height of the inner disk can lead to the inverse correlation between the optical and mid-IR variability of this YSO, similar to the observed behaviour in AA Tau \citep{covey21}.

%\subsection{Outburst frequency calculated from our sample}\label{subsec:outburst_freq}

\begin{figure}[t] %ht
% \epsscale{2}
\centering
%\subfigure{
{
\includegraphics[width=1\columnwidth]{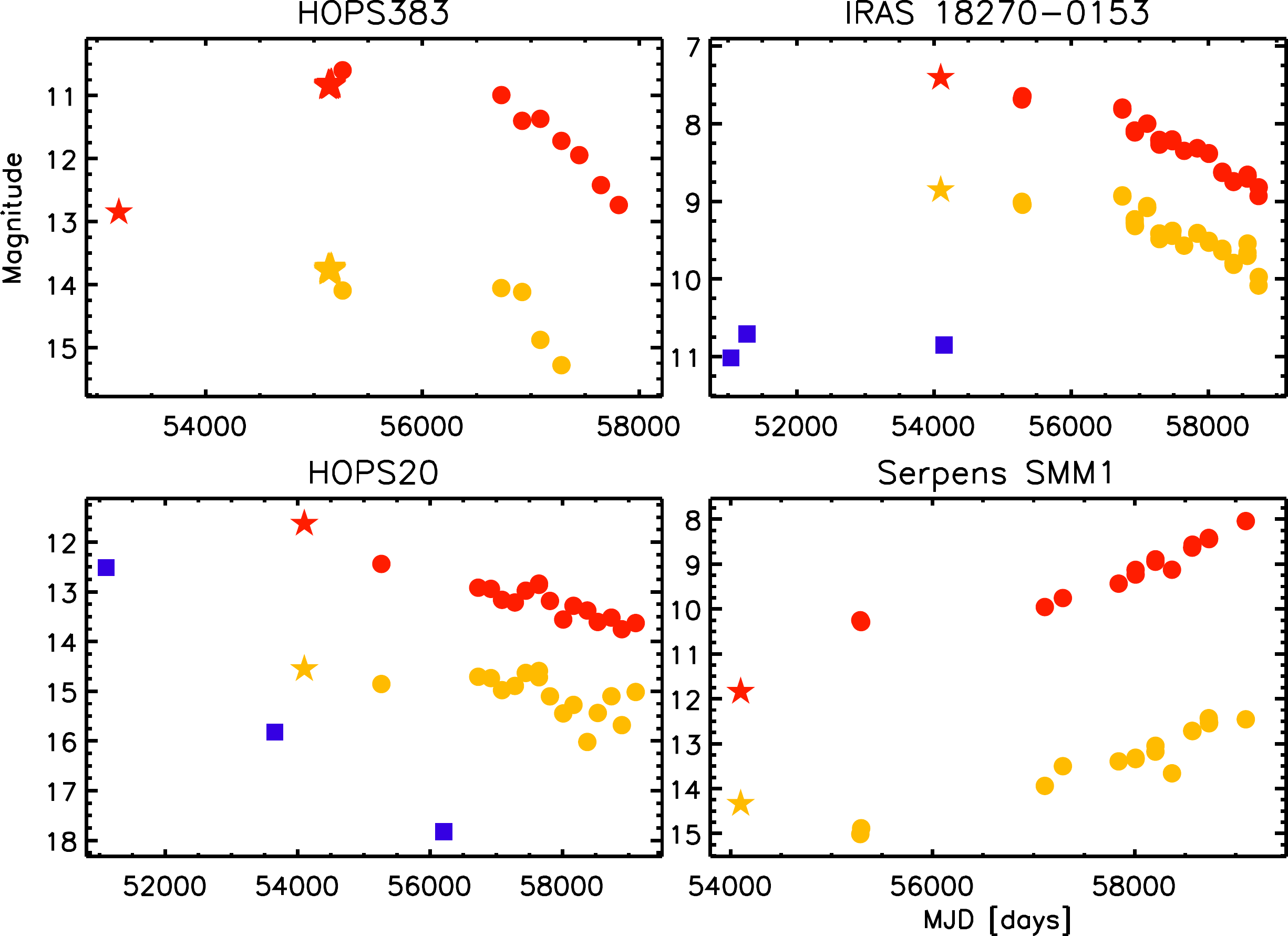}
}
\caption{Example light curves of protostars that are considered in the calculation of $\tau$. These include confirmed FUors HOPS383 (top left) and IRA18270-0153 (top, right). We also include HOPS20 (bottom left) and Serpens SMM1 (bottom right), two protostars that lack spectroscopic confirmation, but show photometric characteristics of FUor outbursts. Symbols are the same as in Figure \ref{fig:long_var}.
\label{fig:erupt_yso}}
\end{figure}

Our careful visual inspection shows 20 YSOs that could be classified as FUor outbursts from their light curves. These include four disks and sixteen protostars. The light curves of these YSOs are shown in Appendix \ref{App:out} (Figures \ref{fig:fuor_D} to \ref{fig:fuor_P_u}). Furthermore, examples of high-amplitude variable YSOs that are not selected as candidate FUors are shown in Figure \ref{fig:fuor_ns}.

The four YSOs classified as disks show high-amplitude fading events that on first impression resemble an FUor returning to quiescence. However, these YSOs, V1902 Ori, V409 Ori, 2MASS J18300168$+$0104430 and  2MASS J22350248$+$7517584, have observations that contradict the FUor interpretation. V1902 Ori and V409 Ori show H$\alpha$ emission \citep{2008Furesz, 2016Dario} with $\dot{M}\sim 10^{-7.6}$~M$_{\odot}$ yr$^{-1}$ and $\dot{M}\sim 10^{-11.5}$~M$_{\odot}$ yr$^{-1}$, respectively \citep[estimated from H$\alpha_{10\%}$ using the relation derived by][]{2004Natta}. The observations of \citet{2016Dario} and \citet{2008Furesz} were taken at a time when the objects were still in a bright state and thus the derived $\dot{M}$ are uncomfortably low for a typical FUor outburst. \citet{2009Kirk} classify 2MASS J22350248$+$7517584 as a Class II YSO with an IR luminosity of L$_{IR}=0.18$~L$_{\odot}$, whilst 2MASS J18300168$+$0104430 has a bolometric luminosity of L$_{bol}=0.1$ L$_{\odot}$ in \citet{evans09} . These low luminosities, derived from observations taken during bright states of the YSOs, also contradict the expected luminosity for an FUor outbursts. 

The lack FUor outburst detections in disk systems is consistent with the frequency of FUor outbursts at the Class II stage determined by \citet{contreras19}. If FUor outbursts occur every 112 kyr during the Class II stage, then we would not expect to observe such outbursts within a survey of 4477 Class II YSOs over a time baseline of 6.5 years.

Our inspection uncovers sixteen candidate long-term eruptive protostars (see Figure \ref{fig:erupt_yso} for a few examples), including six objects showing long-term fadings (V1647 Ori,  2MASS J03470544$+$3243084,  2MASS J05423314$-$1001197, HOPS20, HOPS297 and [TLL2016] Per-emb-40) and seven YSOs showing long-term rises (HOPS343, Serpens SMM 1, 2MASS J21013280$+$6811204, 2MASS J04283510$+$3625065, \linebreak HBC340, 2MASS J21533472$+$4720439 and SSTgbs \linebreak J21470601$+$4739394). YSO HOPS41 reveals a large amplitude ($\Delta$W2 $>4$ mag) variability and it is classified as {\it curved+irregular} with a period of $p=4800$d based on the NEOWISE data. The large period is consistent with the observed variability in the long-term data (See Figure \ref{fig:fuor_P_u}). Finally, IRAS 18270$-$0153 and HOPS383 show large amplitude changes during NEOWISE observations and are known eruptive YSOs.

The most probable frequency, or recurrence timescale $\tau$, for objects classified as protostars can be determined from equation 9 in \citet{contreras19}, whilst 90\% confidence intervals are estimated by integrating equation 8 in the same work. The choice of the total number of outbursts, $k$, requires some consideration given the properties of the 16 candidates.

%We assume $N=1059$ and $t=6.5$ yr, given by the total number of protostars in the adopted catalogues (Table \ref{tab:table1}) and the baseline of NEOWISE observations, respectively.  The choice of the total number of outbursts, $k$, requires some consideration given the properties of the 9 candidates.

%HOPS41 could also be classified as intermediate given the apparent periodicity of $\sim$10 years in the long-term light curve of the source

V1647 Ori is a known eruptive variable that has shown repetitive outbursts with each duration no longer than 5 years \citep{2013Ninan}. This YSO is usually classified as intermediate between EXors and FUors. \citet{dahm17} concludes that variable extinction is the most likely explanation for the observed brightening in HBC340. The light curve of HOPS343 seems to be decaying fast after reaching the peak and might also be an intermediate duration outburst. There is an apparent rise in the latest epochs of NEOWISE observations for 2MASS J03470544$+$3243084, which could indicate that the object is coming back to bright state. This might indicate a long-term extinction event rather than the YSO coming back to outburst. The low amplitude ($\Delta$W1 $\sim1$~mag and $\Delta$W2 $<1$~mag) of 2MASS J21533472$+$4720439 and lack of spectroscopic observations make it hard to confirm an FUor classification for the source. Finally, [TLL2016] Per-emb-40 and 2MASS J05423314$-$1001197 show long-term fading at mid-IR wavelengths that could point to an FUor coming back to quiescence (similar to IRAS 18270$-$0153, see below). However, the objects lack additional photometry at shorter wavelengths and/or spectroscopic data that could help to confirm an FUor classification and are therefore not included in our estimate of $\tau$.

HOPS383 is a known eruptive YSO with an outburst duration that is longer than 10 years (see Figure \ref{fig:erupt_yso}). IRAS 18270$-$0153 was classified as an FUor by \citet{2010Connelley} based on the similarity of its near-IR spectroscopic characteristics to known FUors (strong CO and H$_{2}$O absorption). The object has a second epoch of near-IR spectroscopic data where \citet{connelley18} notice a decrease in the strength of the CO absorption. The change in the spectra of the YSO is consistent with the long-term fading observed at mid-IR (see Figure \ref{fig:erupt_yso}). We note that seven other known FUor objects are part of the YSO sample described in Section \ref{sec:wise}. These correspond to Reipurth 50 N IRS 1, L1551 IRS5, V2775 Ori, V1735 Cyg, Haro 5a/6a, IRAS 05450$+$0019 and V883 Ori. These objects all went into outburst prior to NEOWISE observations and therefore are not detected in our analysis as they show low amplitude variability during the survey observations.

Additionally, objects  Serpens SMM 1, 2MASS J04283510+3625065, 2MASS J21013280$+$6811204 and  SSTgbs J21470601$+$4739394 show long-term rises that resemble those of known FUors. The large amplitude ($\Delta$W1, W2 $>3$~mag) and long-period of the variability of HOPS41 is very likely driven by changes in the accretion rate of the system. Finally, the long-term data of HOPS297 and HOPS20 show a steady decline for the past 15 years with a brightness change larger than 2 magnitudes in the mid-IR. The large changes are also seen at near-IR wavelengths. Such large amplitude decays in the mid-IR are expected after large accretion bursts \citep{scholz13}. Thus, these protostars are likely in the decaying phase after an outburst event.

Given all of the arguments above, seven protostars are not included in the final list of outbursts. We find two objects with confirmed classification as FUor outbursts (HOPS383 and IRAS18270$-$0153). In seven objects we lack spectroscopic confirmation, but have strong characteristics of FUor outbursts. 

To derive $\tau$ we consider the number of outbursts $k$ in our sample, to be in the range 2 to 9 YSOs. The value of $k$=6 is provided to take into account the possibility that not all the outbursts that lack spectroscopic confirmation are actual FUors. We also assume that the total number of protostars varies between $N=735$ (the number of protostars that satisfy our selection criteria) and $N=1059$ (the total number of protostars in the adopted catalogues Table \ref{tab:table1}). Finally, we assume $t=6.5$ yr, given by the baseline of NEOWISE observations. Table \ref{tab:table5} shows the value of $\tau$ for different values of $N$ and $k$.

\begin{deluxetable}{ccccc}

\tablecaption{Frequency of FUor outbursts\label{tab:table5}}
\tablenum{5}
\tablehead{\colhead{t} & \colhead{N} & \colhead{k} & \colhead{$\tau$ (yr)} & \colhead{90\% interval (yr)}}

\startdata
6.5 & 1059 & 9 & 688 &  403--1321\\
6.5 & 1059 & 6 & 983 &  511--2270\\
6.5 & 1059 & 2 & 2295 &  751--13519\\
6.5 & 735 & 9 & 478 &  281--919\\
6.5 & 735 & 6 & 683 &  356--1579\\
6.5 & 735 & 2 & 1593 &  521--9385\\
\enddata
%\tablenotetext{a}{Numbers in front are the counts of variables, while numbers in parentheses are the fractions (\%) of variables relative to the total FUors/EXors and VeLLOs/LLSs samples, 27 and 73, respectively.}
\end{deluxetable}

The results of Table \ref{tab:table5} show that for a given number of outbursts, varying the total number of protostars does not have a considerable effect on the most probable value of $\tau$, i.e. the values are contained within the 90\% confidence intervals. The biggest effect is given by changing the value of $k$. 

Our results are very similar to the work of \citet{fischer19}. The latter estimate the most probable value as $\tau=1000$~years based on comparison of {\it Spitzer} and {\it WISE} observations of 319 protostars selected from the Herschel Orion Protostars Survey \citep[HOPS,][]{2016Furlan}. These objects are contained within our larger sample of protostars. The results from our work show that the value of $\tau$ for protostars does not change significantly when increasing the number of SFRs surveyed. Our estimate also agrees extremely well with the time between ejection events of $\simeq1000$~years determined from the observation of gaps between H$_{2}$ knots \citep{2012Ioannidis, 2016Froebrich, 2018Makin}. The observation of emission knots in jets likely trace the accretion related events that occurred during the earlier stages of young stellar evolution \citep{2012Ioannidis}.  The values also agree with the interval between bursts of 2400 years derived from tracing the location CO and H$_{2}$O snowlines for the Class 0 stage in \citet{2019Hsieh}.\\

\section{Summary}
In this paper, we investigate the variability of known YSOs in twenty nearby low-mass star-forming regions, using 6.5-years of mid-IR NEOWISE photometric data. About 5400 sources out of $\sim$7000 known YSOs identified from the NEOWISE photometric data are analyzed for variability; $\sim$14\%, 64\%, and 22\% of the NEOWISE samples are protostars, disks, and PMS stars (including potential AGB contaminants), respectively.  

We develop a scheme to classify six types of YSO variability based on individual light curves: secular variability (\textit{Linear}, \textit{Curved}, \textit{Periodic}) and stochastic variability (\textit{Burst}, \textit{Drop}, \textit{Irregular}). Just under a third of all YSOs, $\sim$1700, are determined to be variable, with significant variation in the fraction by evolutionary class; $\sim$55\%, 33\%, and 15\% of protostars, disks, and PMS stars, respectively. Along with finding a higher fraction of variables at earlier evolutionary stage, our statistical results also reveal that the variability of YSOs in the earlier evolutionary stages is more secular and has higher amplitudes. Furthermore, the secular variability is associated with longer timescales (periods) at earlier evolutionary stages. Many objects are classified as non-variable despite some variability in their light curves, a consequence of our criteria developed to identify the most variable objects.

We calculated the recurrence timescale of FUor-type outbursts (with $\Delta$W1 or $\Delta$W2 $>1$ mag) from our sample of 735 protostars. Via visual inspection, 9 protostars are found to have FUor-type light curves with long timescales, concluding that a outbursting event occurs every $\sim$1000 years in the early protostellar evolutionary stage. The non-detection of FUor-type light curves in disk systems also agrees with previous estimates on the frequency of FUor outbursts during the Class II stage.

Combined, these variability phenomena suggest that the mass accretion process of YSOs is not continuous but episodic, and that YSOs in the early embedded stage acquire mass more violently and more frequently, with the individual events lasting longer.  

The episodic accretion process has been suggested to arbitrate the discrepancy between theory and observations for the protostellar luminosity function. The FUors/EXors outbursts are considered to be the most prominent and direct phenomenon of the episodic accretion model. Additionally, VeLLOs/LLSs have been revealed as YSOs in their most quiescent phase. 
%of the episodic accretion model. 
We extracted the NEOWISE light curves of 25 known FUors/EXors and 73 VeLLOs/LLSs to investigate their variability in the mid-IR. We find 60\% and 55\% of FUors/EXors and VeLLOs/LLSs to be variable, which are the same fraction as typical protostellar variables.
%VeLLOs/LLSs are somewhat less likely to be variable than typical protostars \textbf{($\sim$54\%)} while FUors/EXors have a much higher variability rate, indicative of relatively unstable dynamical state. 
Nevertheless, distinctively, the FUors/EXors variables are dominated by long-term secular variations (\textit{Linear} and \textit{Curved}) with few stochastic candidates.

Various mechanisms for producing variability are needed to interpret the diverse NEOWISE light curves. Time-dependent accretion rates, as predicted by the episodic accretion model, are an important mechanism of secular YSO variability. In addition to this intrinsic physical condition of YSOs, extinction changes due to inhomogeneous mass distributions within the disk or varying disk geometry can also cause variability, as dimming events. Hydromagnetic interactions between stellar surfaces and inner disk edges and reconnections within the stellar magnetosphere can produce short burst variability. Binary interaction will produce periodic variability with binary orbital motion timescales, which may result in irregular variability in the light curves of YSOs. Hot and cold spots on stellar surfaces can also lead to variability in the mid-IR with stellar rotation timescales. NEOWISE observations do not allow to resolve these short timescales, however YSOs where variability is driven by these mechanisms still may appear as \textit{Irregular} due to the NEOWISE sampling.

%Hot and cold spots on stellar surfaces or binary interaction will produce periodic variability with stellar rotation or binary orbital motion timescales, respectively. Any combination of these mechanisms may result in irregular variability in the light curves of YSOs. 

In addition to the mid-IR brightness variability, we also find diverse secular color variability; YSOs can become either bluer or redder as they brighten, and some YSOs show cyclic color variations in the color-magnitude diagram. This secular color variability can be interpreted as a competitive interplay between time-dependent accretion rates and extinction variations produced by the accreting material.

Our analysis in this paper mostly focuses on YSO variability over timescales of 0.5 to 6.5 years. However, we demonstrate that there is significant additional information on both shorter and longer mid-IR variability timescales. Short-term variability over 1-2 days can be also investigated by the NEOWISE data since each observing epoch of NEOWISE consists of 10-20 exposures with a cadence of $\sim$2 hours, while the study of longer-term variability, over $\sim$15 years, is also possible when the \textit{Spitzer} and \textit{WISE} data are combined with the NEOWISE data.

\section*{Acknowledgement}

The authors wish to acknowledge their valuable discussions on the time-variability of protostars with members of the JCMT Transient Team, in particular early discussions about NEOWISE with Aleks Scholz, discussions about statistics with Tim Naylor, and discussions on disk viewing angles with Wen-Ping Chen. The authors are also very grateful to Yong-Hee Lee for his significant support in setting up our periodogram analysis. 

This publication makes use of data products from the Near-Earth Object Wide-field Infrared Survey Explorer (NEOWISE), which is a project of the Jet Propulsion Laboratory/California Institute of Technology. NEOWISE is funded by the National Aeronautics and Space Administration. This research has made use of the NASA/IPAC Infrared Science Archive, which is operated by the Jet Propulsion Laboratory, California Institute of Technology, under contract with the National Aeronautics and Space Administration.

This work was supported by the National Research Foundation of Korea (NRF) grant funded by the Korea government (MSIT) (grant number 2021R1A2C1011718).
D.J. is supported by the National Research Council of Canada and by an NSERC Discovery Grant. 
G.J.H. is supported by general grant 11773002 awarded by the National Science Foundation of China.

\software{
Numpy \citep{numpy}, Scipy \citep{scipy}, Matplotlib \citep{matplotlib}, Pandas \citep{pandas}, Astropy \citep{astropy2013, astropy2018}}

%\clearpage
\bibliographystyle{aasjournal}
\bibliography{wise}

%% Appendix %%
% \pagebreak
\clearpage
\appendix

% \restartappendixnumbering

\section{NEOWISE multi-epoch photometry and statistic data}
\setcounter{figure}{0}
\renewcommand{\thefigure}{A.\arabic{figure}}

Here we present a portion of the NEOWISE multi-epoch photometry data set (Table \ref{tab:table6}) that is used as input to the analysis in this paper and whose calculations are described in Section \ref{sec:wise}. As well, we present a portion of the tabulation by source of derived statistics (Table \ref{tab:table7}) from Section \ref{sec:methods}. Full versions of these tables can be accessed in the online journal.

% \vspace{30mm}

% Table 1. Multi-epoch NEOWISE photometry

\begin{deluxetable*}{ccccccc}[h]
\tablecaption{Multi-epoch NEOWISE photometry\label{tab:table6}}
\tablenum{6}
%\tablenum{A.1}
\tablehead{\colhead{Index\tablenotemark{a}} & \colhead{MJD} & \colhead{Magnitude} & \colhead{Magnitude error} & \colhead{Band} }
%% All data must appear between the \startdata and \enddata commands
\startdata
%  & & & & & \\
M1  & 56729.827   &  8.78 & 0.069 & W1  \\
M1  & 56921.735   &  8.76 & 0.039 & W1  \\
\vdots   & \vdots   &  \vdots & \vdots &\vdots  \\
M1  & 56730.239   &  7.48 & 0.071 & W2  \\
% M1   & 56729.815   &  7.49 & 0.074 & W2  \\
\vdots   & \vdots   &  \vdots & \vdots &\vdots  \\
D1   &  56740.148  &  6.74 & 0.135 & W1  \\
\vdots   & \vdots   &  \vdots & \vdots &\vdots  \\
EL2   &  56708.812  &  12.10 & 0.035 & W1  \\
\vdots   & \vdots   &  \vdots & \vdots &\vdots  \\
\enddata
\tablenotetext{a}{M, D, and EL are used for the YSOs listed in \citet{megeath12}, \citet{dunham15}, and \citet{esplin19}, respectively. For M and D, source numbers are the same as those in their original catalogs. For EL, the source number is the same as the source order listed in Table 1 of \citet{esplin19}.}

%% Include any \tablenotetext{key}{text}, \tablerefs{ref list},
%% or \tablecomments{text} between the \enddata and 
%% \end{deluxetable} commands
%% No \tablecomments indicated
%% No \tablerefs indicated

\end{deluxetable*}
%%%%%%%%%%%%%%%%%%%%
\pagebreak

% \vspace{10mm}

% Table 2. Statistic data from NEOWISE photometry

\begin{deluxetable*}{cccccccccccccccc}
\rotate
\tablecaption{Statistics of NEOWISE light curves\label{tab:table7}}
\tablenum{7}
%\tablenum{A.2}
\tablehead{\colhead{Index\tablenotemark{a}} & \colhead{R.A. (deg)} & \colhead{Decl. (deg)} & \colhead{N$_{\rm W1}$\tablenotemark{b}} & \colhead{N$_{\rm W2}$\tablenotemark{b}} & \colhead{Class} & \colhead{SD/$\sigma$} & \colhead{$\Delta$W2 (mag)} & \colhead{FAP$_{\rm LSP}$} & \colhead{FAP$_{\rm Lin}$} & \colhead{Sec. Var\tablenotemark{c}} & \colhead{Stoch. Var\tablenotemark{d}} & \colhead{Slope (W2)\tablenotemark{e}} & \colhead{Period\tablenotemark{f}} & \colhead{Frac.amp\tablenotemark{g}} & \colhead{Cloud} }

%% All data must appear between the \startdata and \enddata commands
\startdata
%  & & & & & \\
M1  & 85.67783 & -10.41925 & 14 & 14 & P   & 5.15 & 0.83 & 2.726$\times10^{-2}$ & 4.717$\times10^{-2}$ & & \textit{Irregular} & & & & Orion \\
\vdots   & \vdots   &  \vdots & \vdots &\vdots & \vdots & \vdots & \vdots & \vdots & \vdots & \vdots & \vdots & \vdots & \vdots & \vdots & \vdots\\
% M1  & 56729.815   &  7.49 & 0.074 & W2 & P \\
% % M1   & 56729.815   &  7.49 & 0.074 & W2 & P \\
% \vdots   & \vdots   &  \vdots & \vdots &\vdots & \vdots \\
% D1   &  56729.661  &  12.46 & 0.035 & W1 & P \\
% \vdots   & \vdots   &  \vdots & \vdots &\vdots & \vdots \\
% EL1   &  56729.212  &  11.38 & 0.020 & W1 & D \\
% \vdots   & \vdots   &  \vdots & \vdots &\vdots & \vdots \\
\enddata
\tablenotetext{a}{Same as Table \ref{tab:table6}.}
\tablenotetext{b}{The number of observed epochs in NEOWISE W1/W2 bands to be found in Table \ref{tab:table6}.}
\tablenotetext{c}{Type of secular variability (\textit{Linear}, \textit{Curved}, and \textit{Periodic}), if any.}
\tablenotetext{d}{Type of stochastic variability (\textit{Burst}, \textit{Drop}, and \textit{Irregular}), if any. For secular variables, the stochasticity is determined after removal of secular trends.}
\tablenotetext{e}{Annual W2 flux change relative to the median value if a source is \textit{Linear} (See Section \ref{subsubsec:lin}).}
\tablenotetext{f}{Period by LSP if a source is \textit{Periodic} or \textit{Curved}.}
\tablenotetext{g}{Fractional amplitude (see Figure \ref{fig:lowfapamp}) if a source is \textit{Periodic} or \textit{Curved}.}

%% Include any \tablenotetext{key}{text}, \tablerefs{ref list},
%% or \tablecomments{text} between the \enddata and 
%% \end{deluxetable} commands
%% No \tablecomments indicated
%% No \tablerefs indicated

\end{deluxetable*}
% \end{rotatetable*}
%%%%%%%%%%%%%%%%%%%%
% \vspace{20mm}

\clearpage
\onecolumngrid

\section{Changes in YSO variability type after NEOWISE new data release}
\setcounter{figure}{0}
\renewcommand{\thefigure}{B.\arabic{figure}}

% \vspace{20mm}

Two additional epochs of NEOWISE photometric measurements are released every year, extending the mid-IR monitoring time-line and potentially modifying the results of our variability analysis and classification. Here we describe the degree to which the addition of one year's worth of measurements, from 5.5 to 6.5 years total coverage (typically going from 12 to 14 epochs per source), modifies our results.
%After submission to the journal for review, an additional year of NEOWISE photometric measurements was released to the public, extending the mid-IR monitoring time-line from 5.5 to 6.5\,yrs. We have taken the opportunity to recalculate our entire variability analysis for this extended data set and here we describe the degree to which the addition of two more epochs - typically going from 12 to 14 per source - modifies our results.

\begin{figure}[h]
\centering
%\subfigure{
{
\includegraphics[trim={0.3cm 0.1cm 0.7cm 0.0cm},clip,width=\columnwidth]{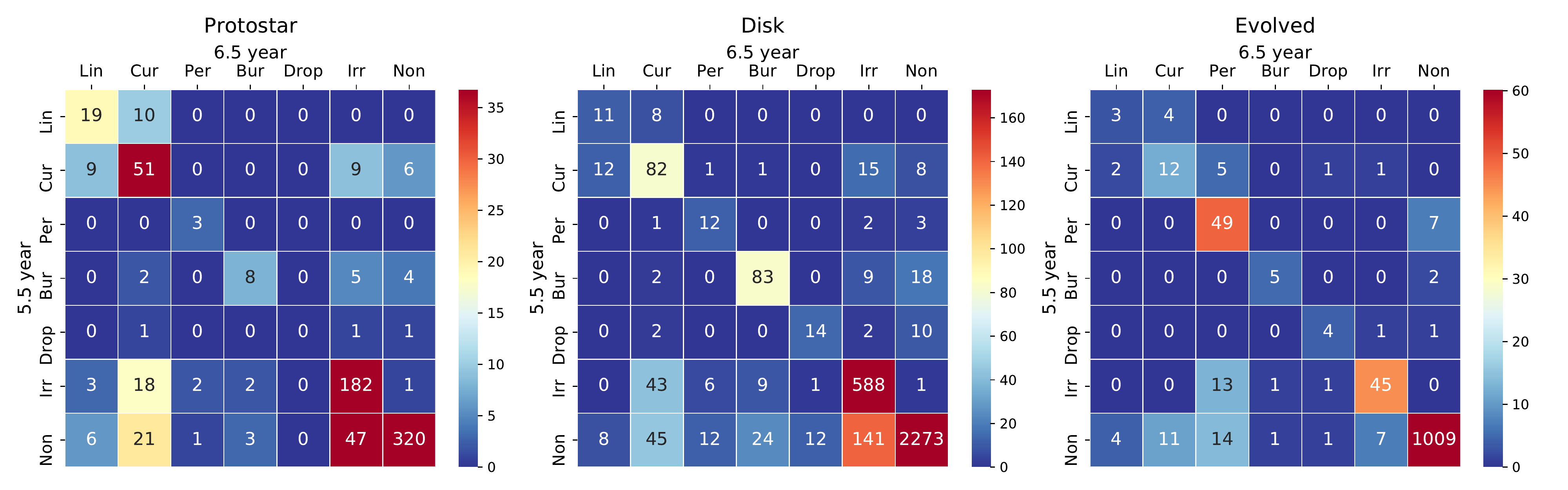}
}
%\vspace{-2cm}
\caption{Matrix heatmaps showing changing variability type between the 5.5\,year and 6.5\,year analyses for the different evolutionary stages of YSOs. Defined variability types are (\textit{Linear, Curved, Periodic, Burst, Drop, Irregular}, and Non-Varying). Rows show the 5.5\,year variability type while columns show the 6.5\,year variability type. The number in each box represents the sources that belong to that joint type. \label{fig:heatmap_num}
}
\end{figure}

\begin{figure}[h]
\centering
%\subfigure{
{
\includegraphics[trim={0.3cm 0.1cm 0.7cm 0.0cm},clip,width=\columnwidth]{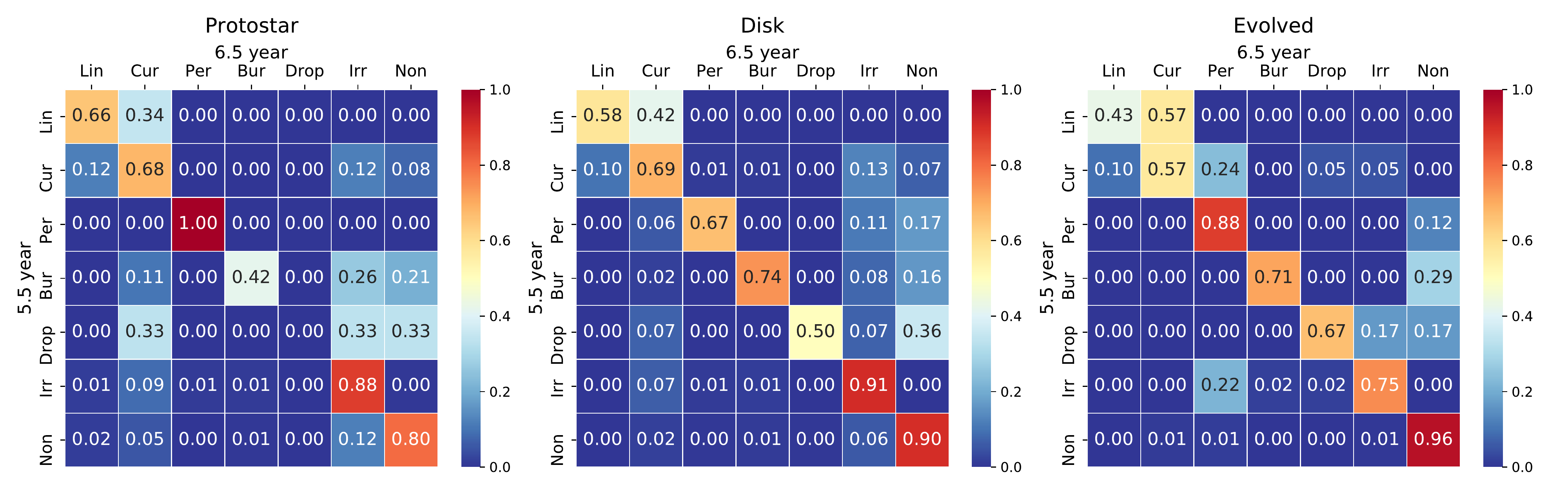}
}
%\vspace{-2cm}
\caption{Same as Fig.\ref{fig:heatmap_num}, but expressed as fractions of the 5.5\,year analysis variable type such that rows sum to 1. \label{fig:heatmap_frac}
}
\end{figure}

Figure \ref{fig:heatmap_num} presents heatmap matrices by source type, protostar, disk, and PMS+E, where each column represents the variability type found after 6.5\, years and each row represents the variability type found after 5.5\,years. If none of the sources changed type between these analyses then only the diagonal boxes would be non-zero. Similar information is presented in Figure \ref{fig:heatmap_frac} except that the raw counts for the 5.5\,year analysis are converted to fractions by type such that each row sums to unity.

\begin{figure}[h]
\centering
%\subfigure{
{
\includegraphics[trim={0.0cm 0.1cm 0.0cm 0.0cm},clip,width=0.7\columnwidth]{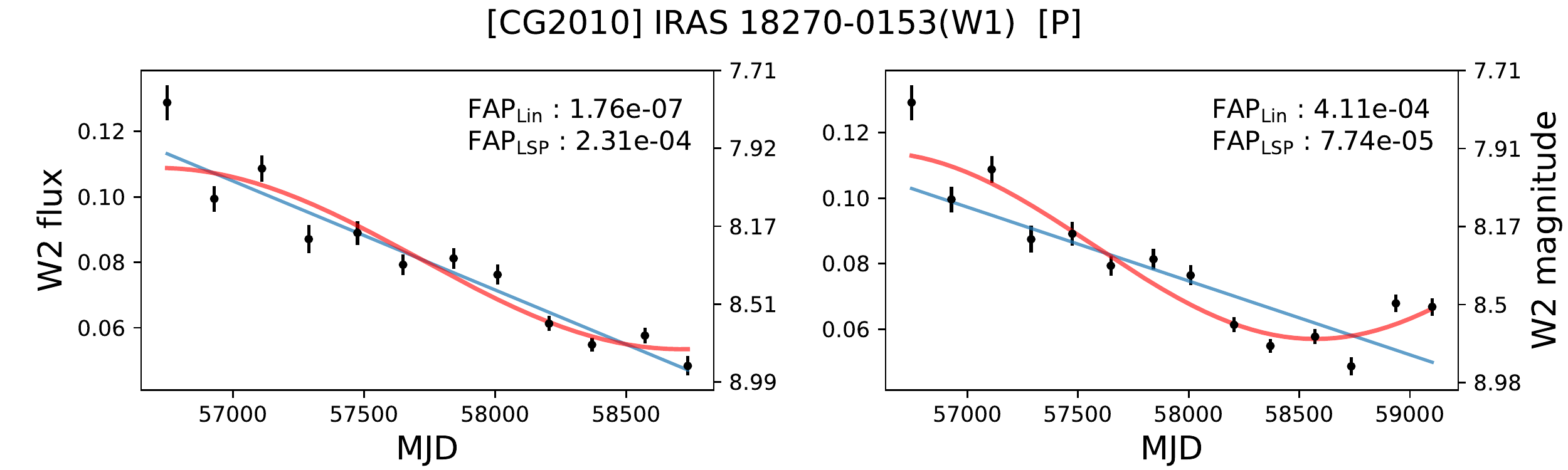}
\includegraphics[trim={0.0cm 0.1cm 0.0cm 0.0cm},clip,width=0.7\columnwidth]{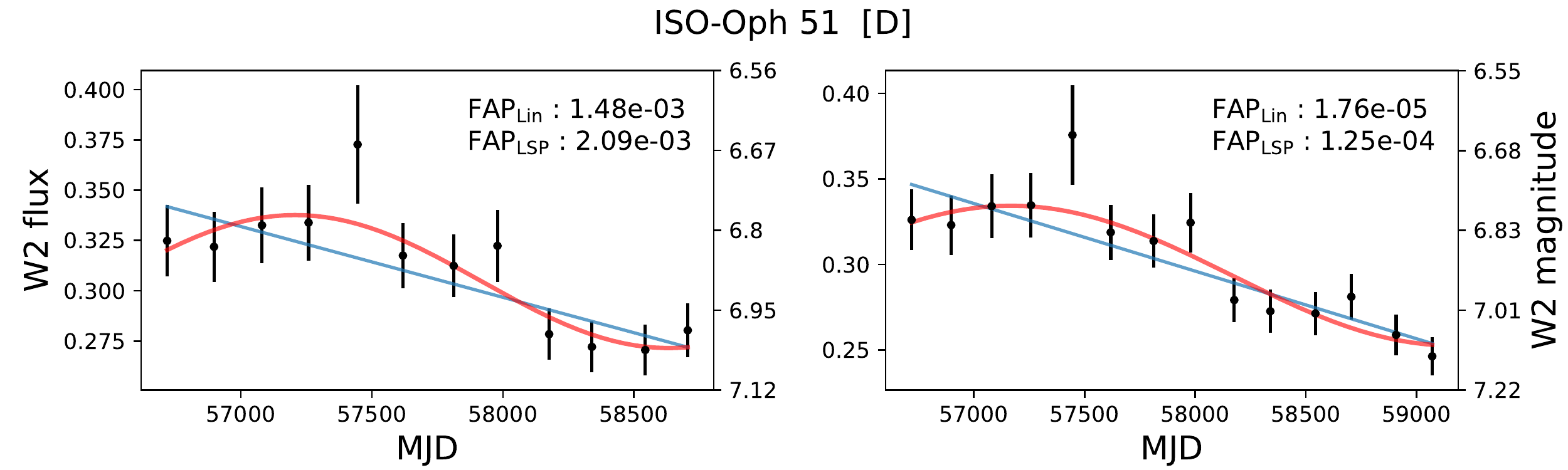}
}
%\vspace{-2cm}
\caption{Examples of changing secular variability type. Top and bottom rows show example sources with their names and evolutionary stages indicated. Figures on the left show 12 epochs, while 14 epochs are shown on the right. Color-lines are the same as in Figure \ref{fig:secular_ex} and the false alarm probabilities are provided in the top right corner of each light curve. [CG2010] IRAS 18270-0153(W1) changes variability type from \textit{Linear} to \textit{Curved}. ISO-Oph 51 changes variability type from \textit{Curved} to \textit{Linear}.  \label{fig:chvar_sec_tosec}
}
\end{figure}

Considering either figure, it is immediately obvious that the majority of secular variables, \textit{Periodic, Curved, and Linear}, remain secular but occasionally change secular type, especially between \textit{Curved and Linear}. This is not unexpected with the addition of epochs, as can be readily seen by considering a couple of specific cases (see Figure \ref{fig:chvar_sec_tosec}) where the new data points provide additional leverage on the best-fit long-timescale secular solution.

% changed variability lightcurve examples

\begin{figure}[h]
\centering
%\subfigure{
{
\includegraphics[trim={0.0cm 0.1cm 0.0cm 0.0cm},clip,width=0.7\columnwidth]{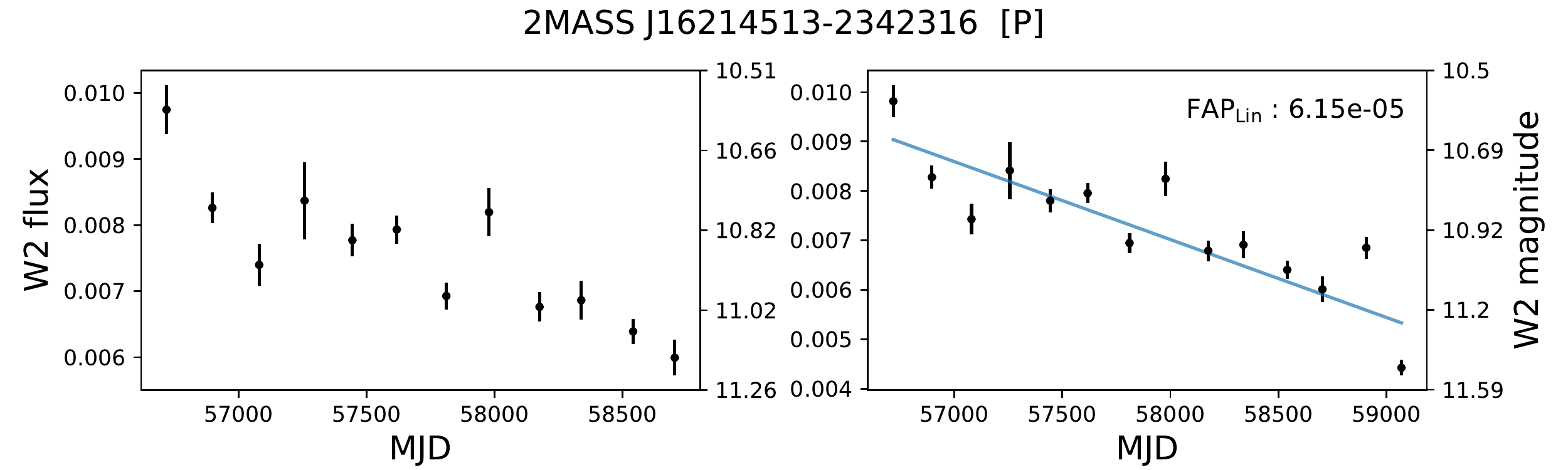}
\includegraphics[trim={0.0cm 0.1cm 0.0cm 0.0cm},clip,width=0.7\columnwidth]{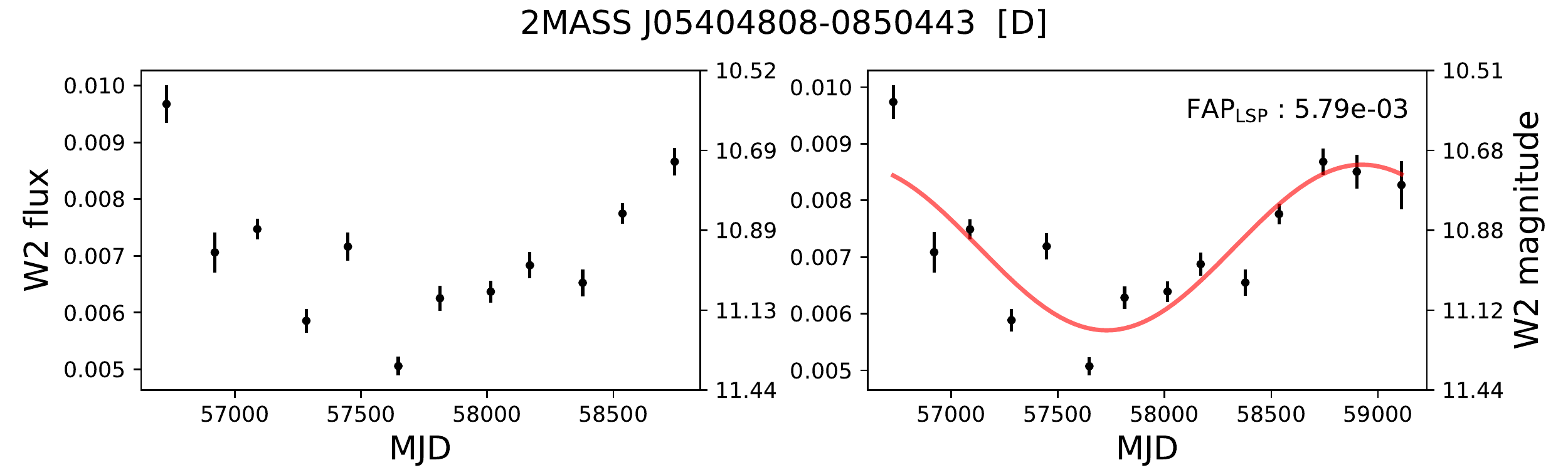}
\includegraphics[trim={0.0cm 0.1cm 0.0cm 0.0cm},clip,width=0.7\columnwidth]{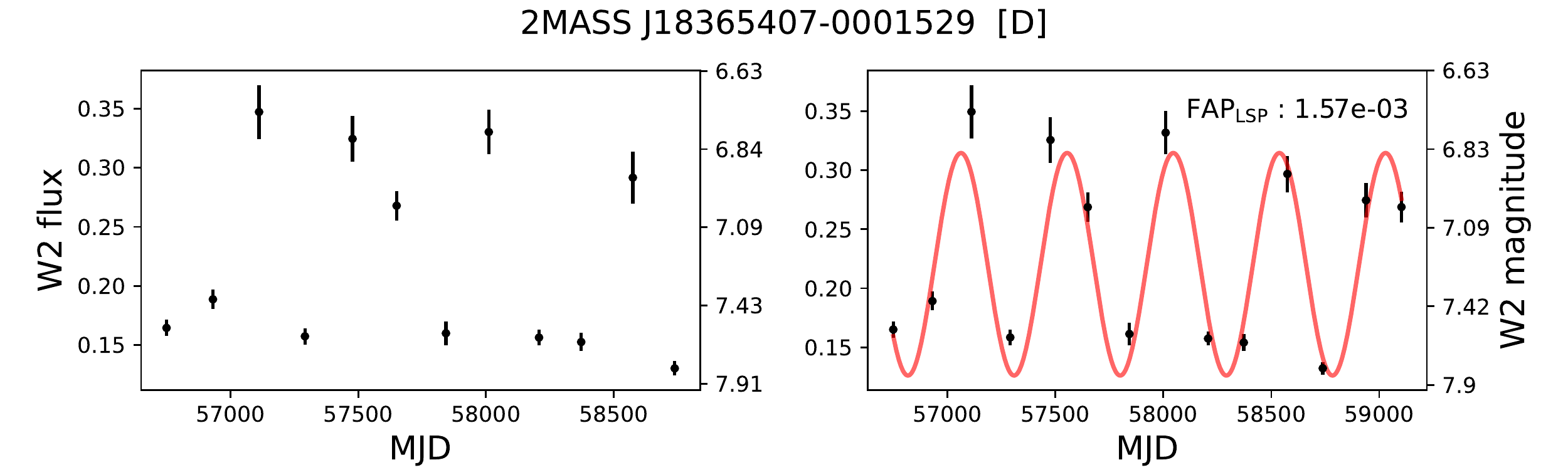}
}
%\vspace{-2cm}
\caption{Similar to Figure \ref{fig:chvar_sec_tosec}, except showing examples that change variability type from \textit{Irregular} to \textit{Linear}, \textit{Curved} and \textit{Periodic}, by row. 
\label{fig:chvar_irr_to_secular}
}
\end{figure}

Furthermore, for the protostars and disks, about 10\% of the \textit{Curved} type become each of \textit{Irregular} and non-varying with the additional epochs (Figure \ref{fig:heatmap_frac}), while by number more than twice as many (Figure \ref{fig:heatmap_num}) convert from both \textit{Irregular} and non-varying to \textit{Curved}. Alternatively, for the PMS+E case, virtually no sources move from \textit{Curved} to \textit{Irregular} or 
non-varying while a non-negligible addition move from 
%\textit{Irregular} and 
non-varying to \textit{Curved}.  For examples of these types of sources, see Figure \ref{fig:chvar_irr_to_secular}. For both disks and PMS+E, there are a non-negligible fraction of \textit{Periodic} sources that become non-varying, likely due to the modest false alarm threshold used for this type. It is worth noting, however, that many more sources move from \textit{Irregular} and non-varying to \textit{Periodic} than the other way.

\begin{figure}[h]
\centering
%\subfigure{
{

\includegraphics[trim={0.0cm 0.1cm 0.0cm 0.0cm},clip,width=0.7\columnwidth]{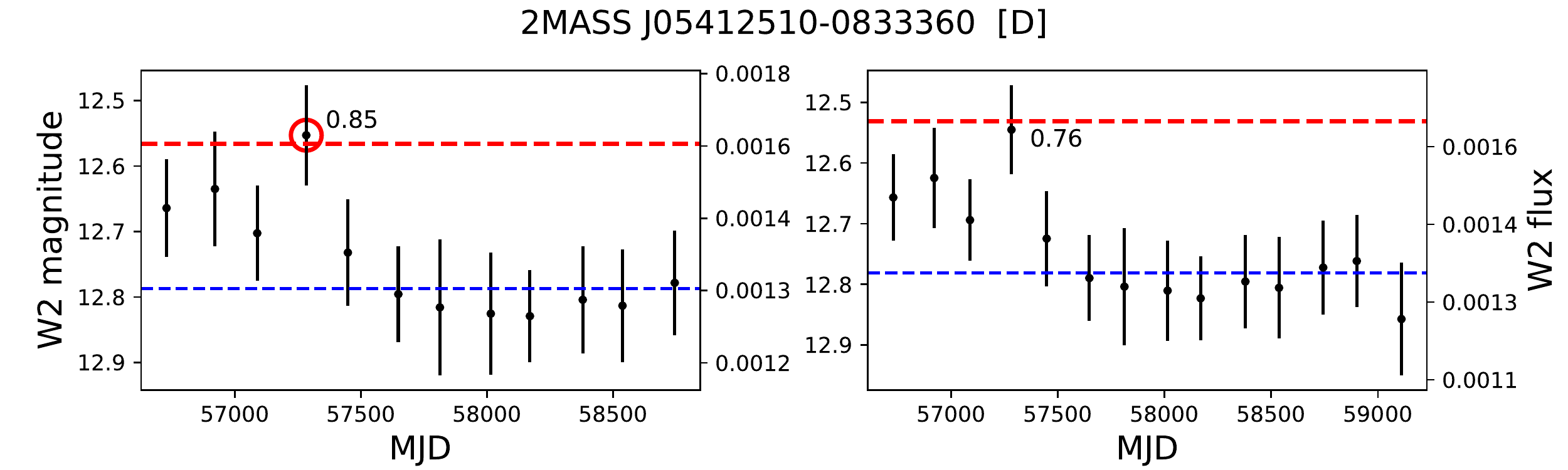}
\includegraphics[trim={0.0cm 0.1cm 0.0cm 0.0cm},clip,width=0.7\columnwidth]{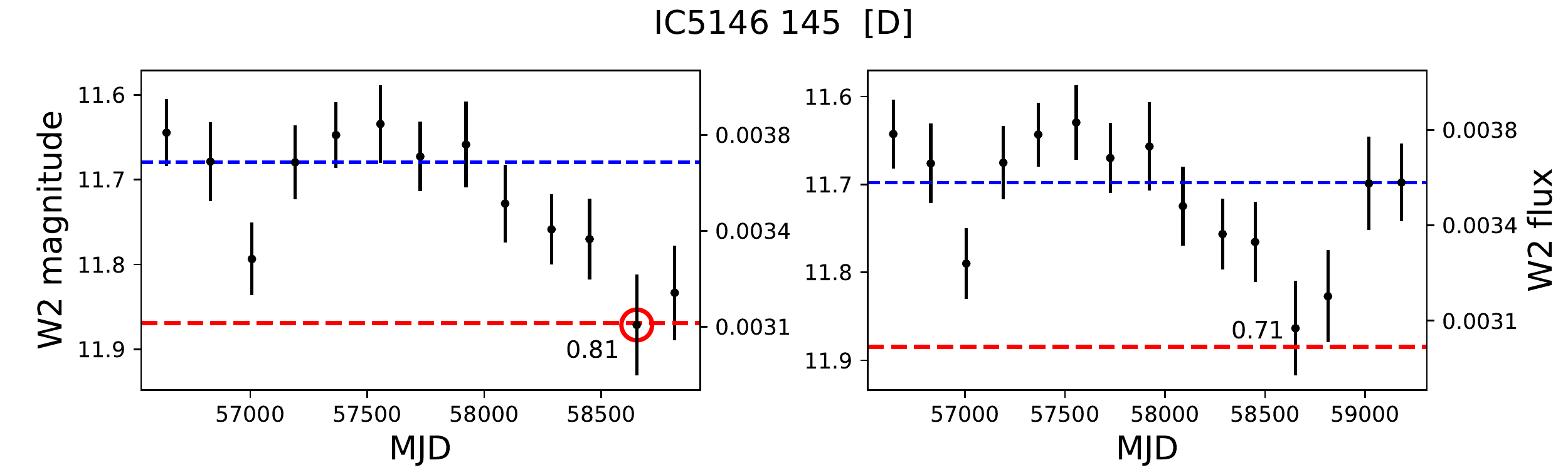}
}
\includegraphics[trim={0.0cm 0.1cm 0.0cm 0.0cm},clip,width=0.7\columnwidth]{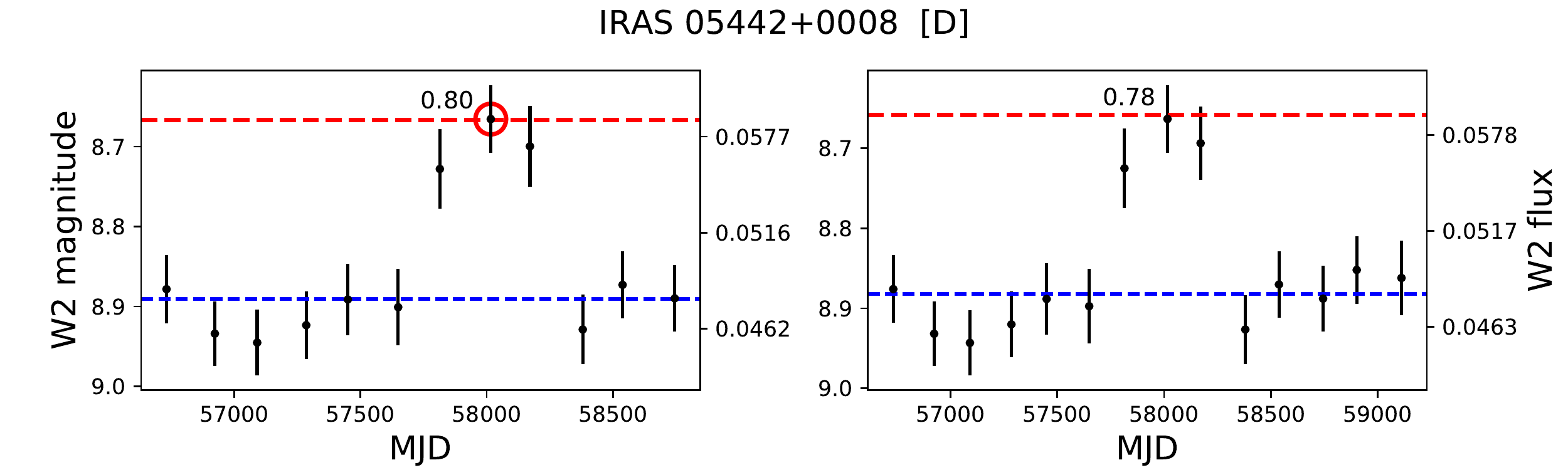}
\includegraphics[trim={0.0cm 0.1cm 0.0cm 0.0cm},clip,width=0.7\columnwidth]{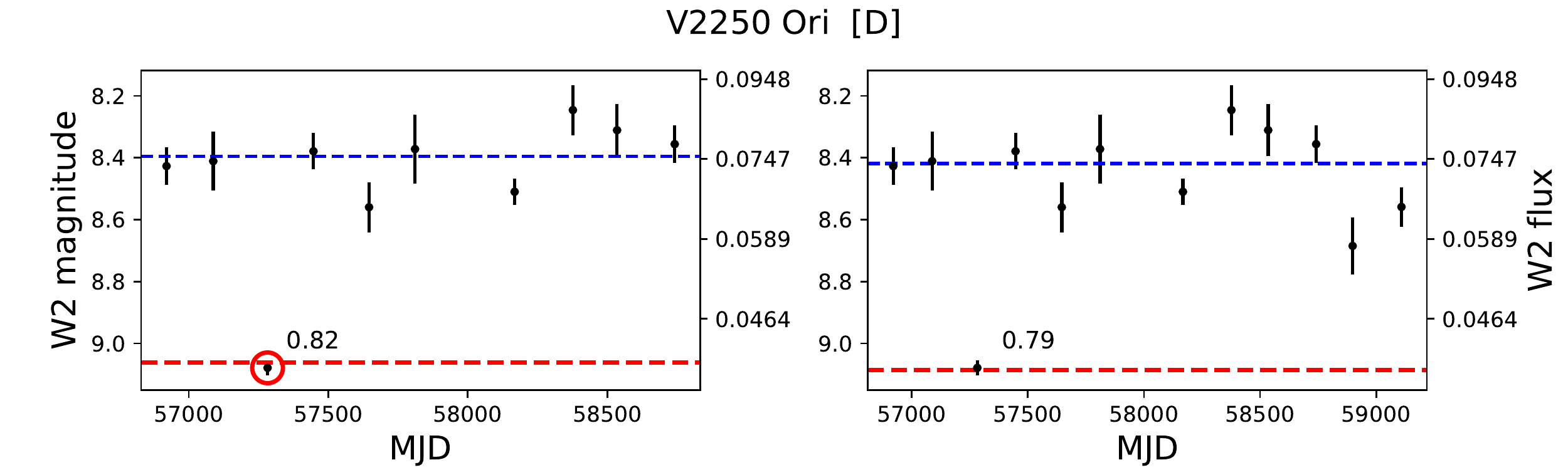}
%\vspace{-2cm}
\caption{Similar to Figure \ref{fig:chvar_sec_tosec} except showing changes in variability type from \textit{Burst/Drop} to non-variable. The dashed blue line indicates the median magnitude. The criteria for \textit{Burst} and \textit{Drop} are indicated by dashed red lines, while the burst and drop events are marked by red circles (see also Figure \ref{fig:stoch_ex}). In each panel the numbers indicate the fraction of $\Delta$W2 that the extrema is from the median. 2MASS J05412510-0833360, IC5146 145, IRAS 05442+0008 and V2250 Ori have SD/$\sigma$ of 1.13, 1.51, 2.10, and 2.50 respectively, from their 6.5-year light curve.}
\label{fig:burst_drop}

\end{figure}

The largest variation in identified types across the two analyses is found for \textit{Burst} and \textit{Drop} sources.  We note that the definition of \textit{Burst} and \textit{Drop}, see Section \ref{sec:methods}, requires a computational mechanism to distinguish a significant brightness change over a single epoch while simultaneously discounting sources displaying \textit{Irregular} stochasticity. Thus, it is not a surprise that these types present some ambiguity. Furthermore, as we want to identify only robust bursts/drops, we choose a high threshold, requiring the difference in magnitudes between the median and the extrema to be at least 0.8 times the difference between the maximum and minimum magnitudes (Section \ref{sec:methods}). By adding additional epochs, which slightly moves the median, some sources will cross this threshold in both directions. Furthermore, aside from the extrema event itself, many of these light curves have low variability measures, the standard deviation about the mean is less than three times the estimated measurement uncertainty, and thus the sources are not considered irregular if they fail to make the burst/drop cut. Figure \ref{fig:burst_drop} shows examples of bursts and drops that fail the cut after the inclusion of additional epochs.  We note that the light curves of both these sources continue to support the notion of an extrema event - suggesting that our high threshold hides many potential bursts and drop sources.

\begin{figure}[h]
\centering
%\subfigure{
{

\includegraphics[trim={0.0cm 0.1cm 0.0cm 0.0cm},clip,width=0.7\columnwidth]{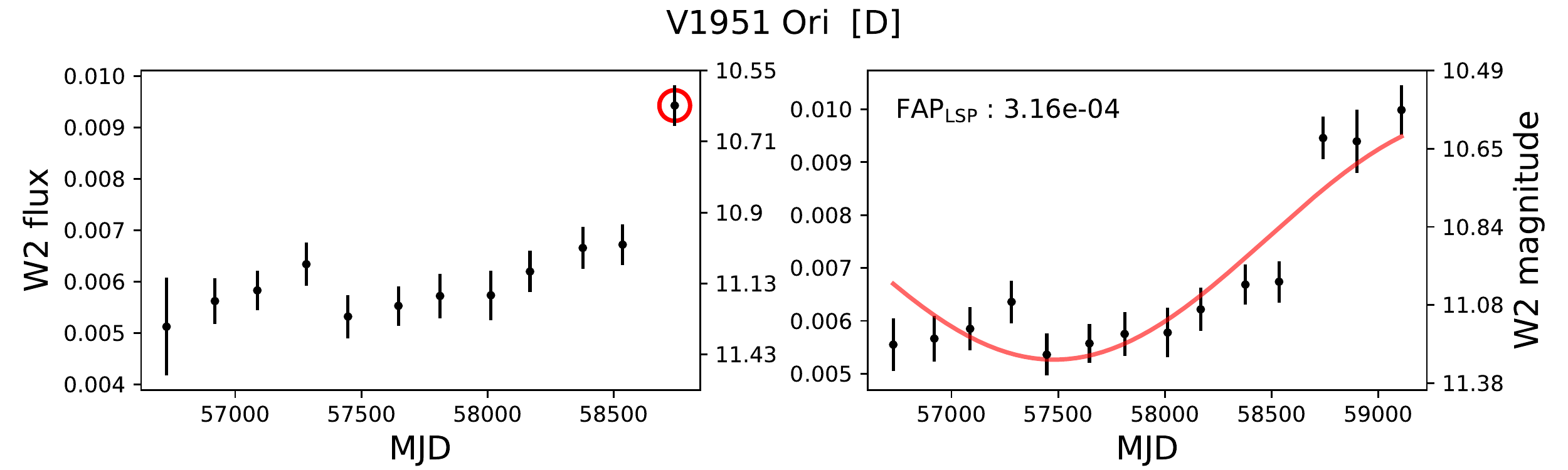}
\includegraphics[trim={0.0cm 0.1cm 0.0cm 0.0cm},clip,width=0.7\columnwidth]{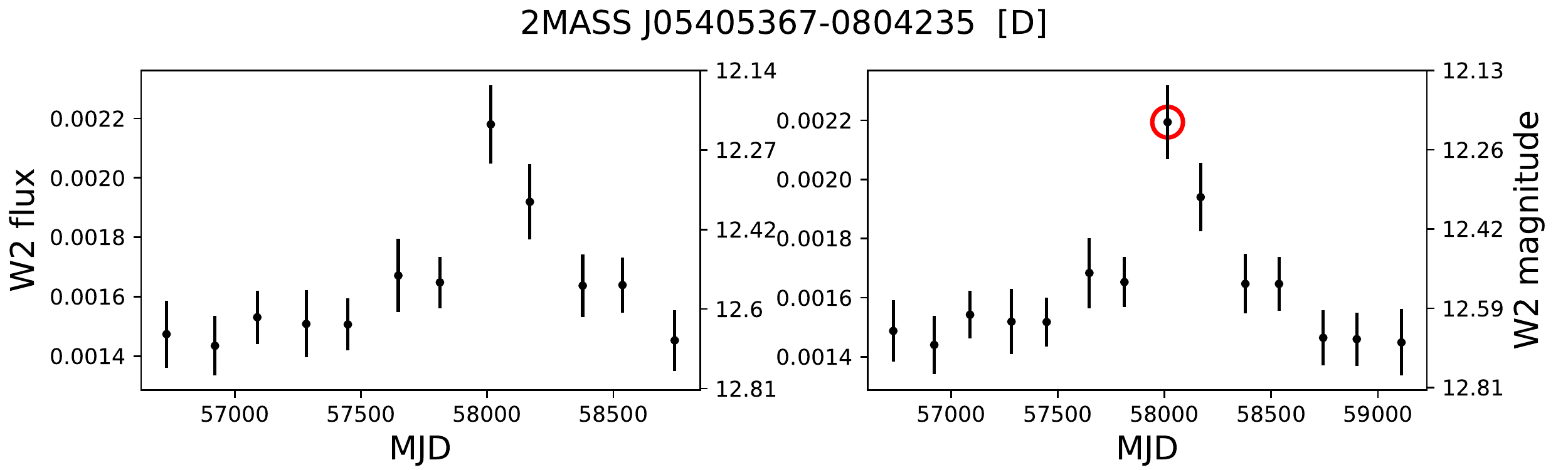}
\includegraphics[trim={0.0cm 0.1cm 0.0cm 0.0cm},clip,width=0.7\columnwidth]{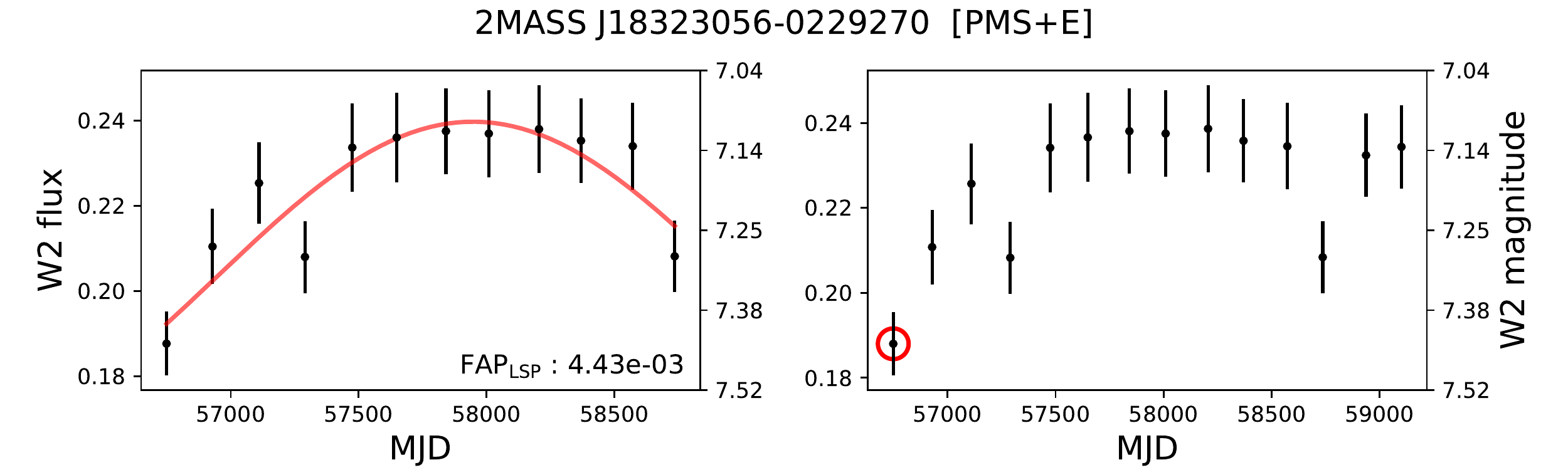}
}
%\vspace{-2cm}
\caption{Similar to Figure \ref{fig:chvar_sec_tosec} except showing changes in variability type from or to \textit{Burst/Drop}. Red lines are as in Figure \ref{fig:secular_ex} while red circles are as in Figure \ref{fig:stoch_ex}. V1951 Ori changes type from \textit{Burst} to \textit{Curved}. 2MASS J05405367-0804235 was classified as non-variable, but changes to \textit{Burst} with two  additional epochs. 2MASS J18323056-0229270 was \textit{Curved}, and changes variability type to \textit{Drop}. \label{fig:chvar_burst_drop}
}
\end{figure}

Examining by type, only 40\% of identified protostar \textit{Bursts} after 5.5\,years remain \textit{Bursts} after 6.5\,years, with  those changing type becoming \textit{Irregular} or non-varying. A similar number of sources change from either \textit{Irregular} or non-varying to become \textit{Burst}.  Furthermore, no \textit{Drop} sources remain after the 6.5\,year analysis, with the three previously identified \textit{Drop} type moving to \textit{Curved}, \textit{Irregular}, and non-varying. For the disk sources, a smaller fraction of \textit{Bursts} convert to \textit{Irregular} or non-varying while an even larger number change from \textit{Irregular} or non-varying to \textit{Burst}. For examples of these types of source changes, see Figure \ref{fig:chvar_burst_drop}.

%The largest variation in identified types across the two analyses is found for \textit{Burst} and \textit{Drop} sources, primarily protostars. Only 40\% of identified Protostar \textit{Bursts} after 5.5\,years remain \textit{Bursts} after 6.5\,years, with  those changing type becoming \textit{Irregular} or non-varying. A similar number of sources change from either \textit{Irregular} or non-varying to become \textit{Burst}. Furthermore, no \textit{Drop} sources remain after the 6.5\,year analysis, with the three previously identified \textit{Drop} type moving to \textit{Curved}, \textit{Irregular}, and non-varying. For the disk sources, a smaller fraction of \textit{Bursts} convert to \textit{Irregular} or non-varying while an even larger number change from \textit{Irregular} or non-varying to \textit{Burst}. For examples of these types of source changes, see Figure \ref{fig:chvar_burst_drop}. We note that the definition of \textit{Burst} and \textit{Drop}, see Section \ref{sec:methods}, requires a computational mechanism to distinguish a significant brightness change over a single epoch while simultaneously discounting sources displaying \textit{Irregular} stochasticity. Thus, it is not a surprise that these two types present some ambiguity.

Finally, we note that  12\% of the 5.5\,year-identified non-varying protostars are reclassified as \textit{Irregulars} after 6.5\,years and 5\% become \textit{Curved} type. The fraction of non-varying sources changing type is much lower for the disks, 10\%, and PMS+E, 4\%.

Putting together a taxonomy of light curves is fraught with uncertainty, especially when it is unknown if the underlying signal should contain regularity and if so what form the regularity might take. The classification system developed in this paper was designed to discriminate between secular and stochastic light curves, with three types in each category. The definitions were optimized for the original 5.5\,year analysis. The investigation undertaken in this appendix, comparing the classification results after the addition of almost 20\% in time coverage, and over 15\% in epochs, provides a strong degree of confidence in the robustness of our classification scheme.

\vspace{20mm}

%\onecolumngrid
\clearpage
\setcounter{figure}{0}
\renewcommand{\thefigure}{C.\arabic{figure}}

\section{Contamination from AGBs and AGNs} \label{appendix:agb}

Both forming and dying stars are bright in IR wavelengths since their effective temperatures are %{\sout as low as} 
2000 to 4000 K, and they are often enshrouded by cold and dense circumstellar material. As a result, YSOs as well as evolved stars are commonly identified via their location in IR color-magnitude or color-color diagrams \citep{suh11,tu13,koenig14}. However, the colors of YSOs and evolved stars, especially AGBs, overlap significantly in those diagrams resulting in contamination of source identifications \citep[e.g.][]{2008Robitaille}. In addition, AGB stars display large amplitude periodic variability \citep{2008Whitelock}, including in the mid-IR \citep{Karambelkar19}, that could also be misidentified as arising from YSOs \citep{pena17}.  

\begin{figure}[h] %ht
\epsscale{2}
\centering
{\includegraphics[trim={0cm 0.3cm 0cm 0.0cm},clip,width=0.5\columnwidth]{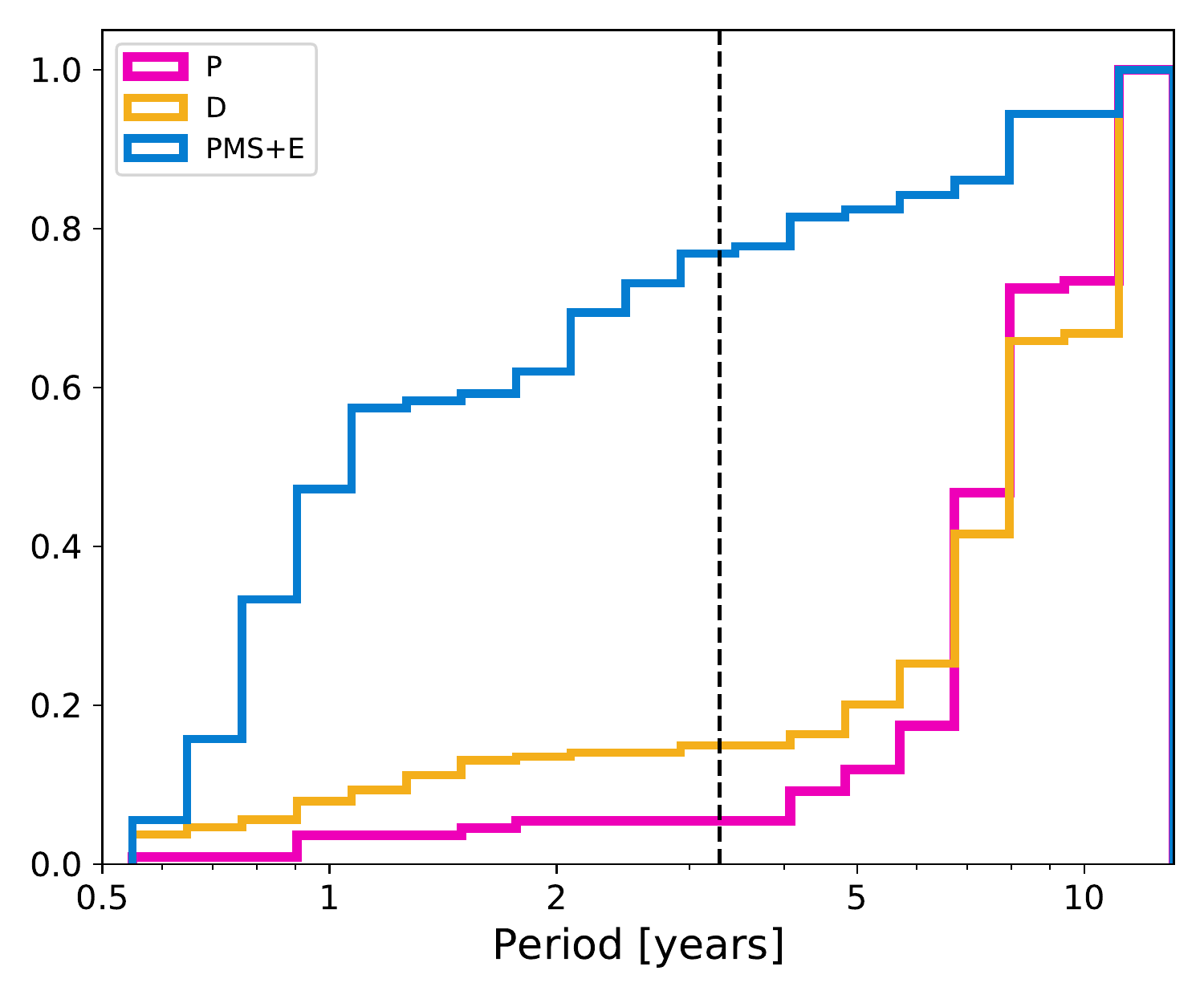}         
}
 
\caption{The cumulative  distribution function of the best-fit periods of \textit{Curved} and \textit{Periodic} variables. Colors are the same as those in Figure \ref{fig:meanw2}. The vertical dashed line indicates the period of 1200 days.            
\label{fig:period_cdf}}
\end{figure}

\begin{figure}[t] %ht
% \epsscale{2}
\centering
%\subfigure{
{
\includegraphics[trim={0.3cm 0.3cm 0.3cm 0.1cm},clip,width=0.5\columnwidth]{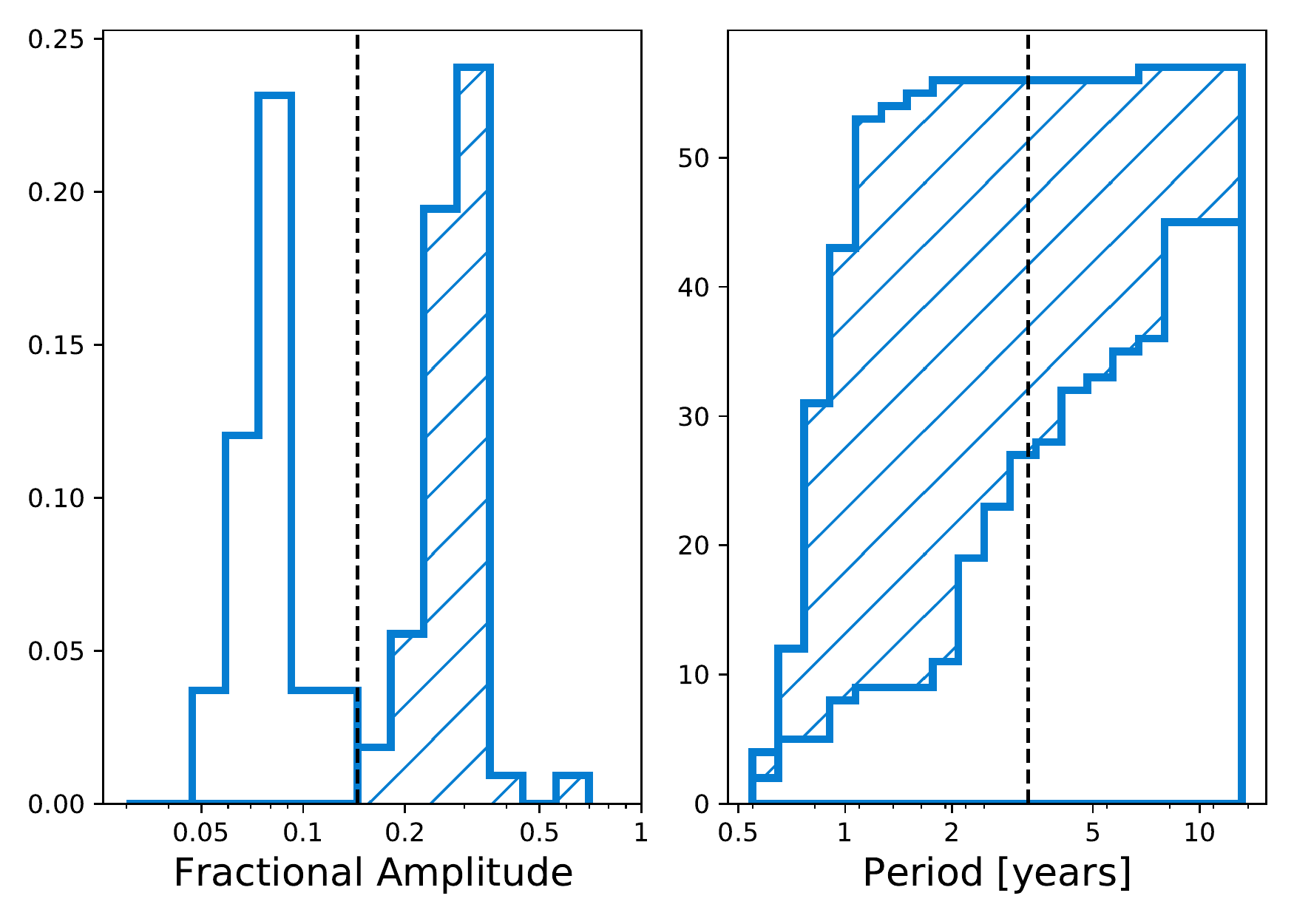}
}
\caption{ (Left) Distribution of fractional amplitudes of \textit{Curved} and \textit{Periodic} PMS+E variables. For each best-fit sinusoid, the amplitude is divided by the mean flux to derive the fractional amplitude.  The variables are clearly distinguishable as two types bounded by the fractional amplitude of 0.15 (dashed vertical line): low amplitude variables (non-filled histogram) and high amplitude variables (hatched histogram). (Right) The cumulative  histogram of the best-fit periods of \textit{Curved} and \textit{Periodic} PMS+E variables. The hatched histograms indicate PMS+E variables with high fractional amplitude ($>0.15$). The vertical dashed line indicates the period of 1200 days.     
\label{fig:lowfapamp}}
\end{figure}

A third of our PMS+E variables are \textit{Periodic} (Table \ref{tab:table2}) with well defined sinusoidal light curves having periods of a few hundred days and high fractional amplitudes (Figure \ref{fig:period_cdf} and Figure \ref{fig:lowfapamp}). Such regular variability with short periods and large amplitudes is unexpected for true Class III YSOs.
Therefore, these variables with nice sinusoidal light curves, periods shorter than 1200 days, and high amplitudes (hatched regions of Figure \ref{fig:lowfapamp}) are likely AGB \textit{interlopers}, which are mis-classified as YSOs. Further confirmation that these are contaminating AGB stars comes from the fact that no periodic PMS+E sources are found in Taurus through our analysis. In Taurus the YSO sample has been well classified, with no confusion due to background AGBs. Analyses of {\it Gaia} astrometry confirms that background AGB stars are a significant source of contamination in these catalogs \citep{manara18,herczeg19}.

\begin{figure}[t]
\centering
{\includegraphics[trim={0.5cm 0.3cm 1.0cm 0.3cm},clip,width=0.5\columnwidth]
{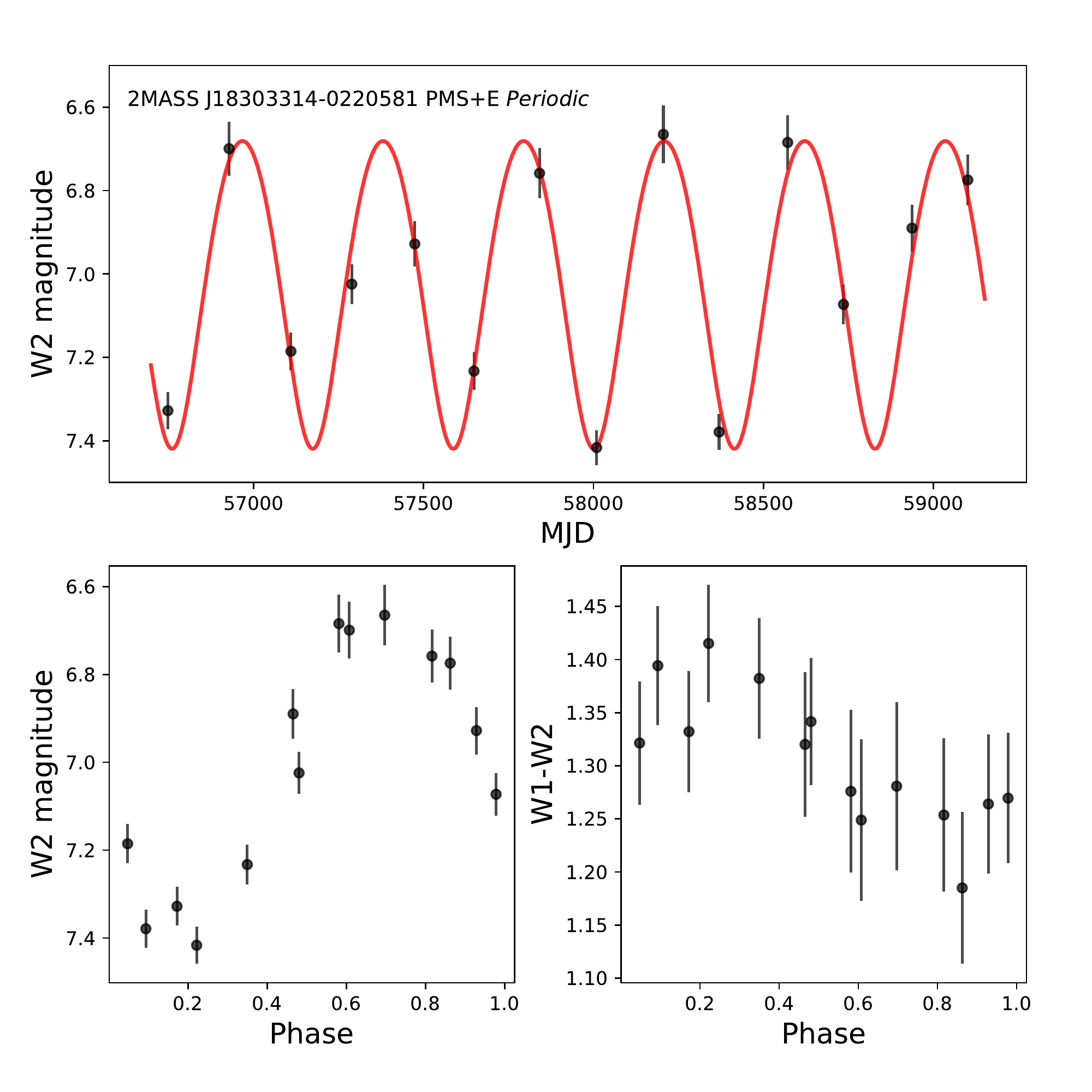}
}
\caption{The NEOWISE light curve of an AGB candidate (upper), the phase diagram of the W2 magnitude (bottom left), and the phase diagram of the W1-W2 color (bottom right). The red line overlaid in the light curve shows the best-fit sinusoidal function with the period of 413 days. 
\label{fig:agb}
}
\end{figure}

Figure \ref{fig:agb} shows an example light curve of an AGB candidate along with the phase diagrams of magnitude and color. Most of our AGB candidates are bluer when brighter, suggesting that the temperature change, caused by pulsations, results in the luminosity variation.
More detailed analyses for these AGB candidates are presented in a separate paper \citep{lee2021agb}. 

In addition to this AGB contamination, Active Galactic Nuclei (AGNs) can also contaminate the YSO catalogs. We cross-matched our NEOWISE YSO sample with the AGN catalog by \citet{Shu19} and found 21 overlapped sources, out of which 7 sources are variables (1 {\it Curved}+{\it Irregular} and 6 {\it Irregular}) in our analysis. These overlapped sources could be extragalactic; however, the majority of AGNs in the catalog are located at high galactic latitude, and the large visual extinction through the galactic plane potentially blocks most extra-galactic sources. Therefore, these overlapped sources are more likely YSO contamination within the AGN catalog, although decisive confirmation by spectral observations is required.

\clearpage
\section{Variable FUors/EXors and VeLLOs/LLSs}
\setcounter{figure}{0}
\renewcommand{\thefigure}{D.\arabic{figure}}

Table \ref{tab:tabled1} presents the name, evolutionary stage, variability type, and literature reference for all the known eruptive YSOs found to be variable in the mid-IR (see Section \ref{subsec:fu_ex_vel}).

Table \ref{tab:tabled2} presents the name, evolutionary stage, variability type, and literature reference for all the known  subluminous YSOs found to be variable in the mid-IR (see Section \ref{subsec:fu_ex_vel}).

\begin{deluxetable}{cccccc}[h]
\tablecaption{Variable FUors/EXors\label{tab:tabled1}}
\tablenum{8}
\tablehead{\colhead{Source} & \colhead{Known name} & \colhead{Stage} & \colhead{Variability Type} & \colhead{Region} & \colhead{Reference}}
\startdata
FUor   & Parsamian 21 (IRAS 19266+0932)    & D                  & \textit{Linear}($-$)            & Aquila  & \cite{connelley18}   \\
FUor   & V582 Aur                          & P                  & \textit{Linear}($-$)+\textit{Drop}          & Auriga  & \cite{connelley18}   \\
% FUor   & \textcolor{red}{V2775 Aur}                          &                   & \textit{Linear}($+$)            &   &    \\
% FUor   & \textcolor{red}{V900 Mon}                          &                   & \textit{Linear}($+$)            &   &    \\
FUor   & V733 Cep                          & D                  & \textit{Linear}($-$)            & Cepheus & \cite{connelley18}   \\
FUor   & HH354 IRS (IRAS 22051+5848)       & P                  & \textit{Linear}($-$)           & Cepheus & \cite{connelley18}   \\
FUor   & V1515 Cyg                          &  D       & \textit{Curved}            & Cygnus  &  \cite{connelley18}  \\
FUor   & V2495 Cyg                          &          P    & \textit{Curved}            & Cygnus  &  \cite{connelley18}  \\
FUor   & Haro 5a IRS                       & P          & \textit{Linear}($-$)            & Orion & \cite{connelley18}  \\
FUor   & HOPS 383                          & P                  & \textit{Linear}($-$)            & Orion   & \cite{pena17} \\
% FUor   & \textcolor{red}{IRAS 05450+0019}                          &                   & \textit{Curved}            &   &    \\
FUor   & BBW 76 (V646 Pup)                 & P                  & \textit{Linear}($-$)            & Puppis  & \cite{connelley18}   \\
FUor   & EC 53 (V371 Ser)                   & P                  & \textit{Irregular}        & Serpens & \cite{connelley18} \\
FUor   & V370 Ser                          & P                  & \textit{Irregular}          & Serpens & \cite{samus17}          \\
EXor   & V1647 Ori                         & P                  & \textit{Curved}            & Orion   & \cite{connelley18}   \\
EXor   & V1118 Ori                         & D                  & \textit{Burst}            & Orion   & \cite{audard14}          \\
EXor   & HBC 340 (2MASS J03284325+3117330) & P                  & \textit{Curved}            & Perseus & \cite{dahm17}   \\
%FUor   & IC 348 LRL 31                           & D                  & \textit{Irregular}          & Perseus &    \\
EXor   & VY Tau                         & D           & \textit{Irregular}            & Taurus &  \cite{audard14}  \\
\enddata
\end{deluxetable}

\begin{deluxetable}{cccccc}[h]

\tablecaption{Variable VeLLOs/LLSs\label{tab:tabled2}}
\tablenum{9}

\tablehead{\colhead{Source} & \colhead{Known name} & \colhead{Stage} & \colhead{Variability Type} & \colhead{Region} & \colhead{Reference}}
\startdata
LLS             & IRAS 18265-0148          & P              &  \textit{Irregular}          & Aquila          & \cite{kimmr16}, \cite{kimgj19}           \\
LLS             & IRAS 18277-0154          & P              &  \textit{Curved}             & Aquila          & \cite{kimmr16}, \cite{kimgj19}           \\
LLS             & IRAS 21017+6742          & P              &  \textit{Curved}             & Cepheus         & \cite{kimmr16}, \cite{kimgj19}          \\
LLS             & IRAS 22290+7458          & P              &  \textit{Irregular}          & Cepheus         & \begin{tabular}[c]{@{}c@{}}\cite{dunham08}; \cite{kimmr16}; \\ \cite{kimgj19}\end{tabular} \\
LLS             & SSTc2d J223846.1+751132 & P              &  \textit{Irregular}          & Cepheus   & \cite{dunham08}                 \\
LLS             & 2MASS J12534285-7715114 & P              &  \textit{Irregular}          & Chamaeleon II   & \cite{dunham08}                 \\
LLS             & 2MASS J16270524-2436297  & P              &  \textit{Curved}             & Ophiuchus       & \cite{dunham08}                 \\
LLS             & IRAS 16544-1604          & P              &  \textit{Irregular}          & Ophiuchus       & \cite{dunham08}                 \\
LLS             & {[}EES2009{]} Per-emb 42   & P              &  \textit{Curved}+\textit{Irregular}             & Perseus         & \cite{dunham08,enoch09}                  \\
LLS             & IRAS 03256+3055          & P              &  \textit{Curved}           & Perseus         & \cite{dunham08}                 \\
LLS             & 2MASS J03285630+3122279  & D              &  \textit{Irregular}          & Perseus         & \cite{dunham08}                 \\
LLS             & IRAS 03262+3123          & P              &  \textit{Linear}($-$)             & Perseus         &\cite{dunham08}                 \\
LLS             & IRAS 03293+3052          & P              &  \textit{Irregular}          & Perseus         & \cite{dunham08}                 \\
LLS             & {[}SDA2014{]} West50     & P              &  \textit{Irregular}          & Perseus         & \cite{dunham08,sadavoy14}               \\
LLS             & Cl* IC 348 LRL 1889      & P              &  \textit{Irregular}          & Perseus         & \cite{dunham08}                 \\
LLS             & SSTc2d J182844.0+005337  & P              &  \textit{Irregular}          & Serpens         & \cite{dunham08}                 \\
LLS             & 2MASS J18284503+0052028          & P              &  \textit{Curved}+\textit{Irregular}            & Serpens         & \cite{dunham08}                 \\
LLS             & Serpens SMM 1                & P              &  \textit{Curved}+\textit{Irregular}          & Serpens         & \cite{dunham08}                \\
LLS             & IRAS 18273+0034          & D              &  \textit{Curved}+\textit{Irregular}          & Serpens         & \cite{dunham08}                 \\
LLS             & 2MASS J18295434+0036014          & D              &  \textit{Irregular}          & Serpens         & \cite{dunham08}                 \\
VeLLO           & 2MASS J18285582-0137346  & P              &  \textit{Irregular}          & Aquila          &\cite{kimmr16}, \cite{kimgj19}            \\
VeLLO           & SSTgbs J1829054-034245   & P              &  \textit{Irregular}          & Aquila          &\cite{kimmr16}, \cite{kimgj19}            \\
VeLLO           & IRAS 18267-0139          & P              &  \textit{Irregular}          & Aquila          &\cite{kimmr16}, \cite{kimgj19}            \\
VeLLO           &2MASS J18292510-0147382  & P              &  \textit{Curved}             & Aquila          &\cite{kimmr16}, \cite{kimgj19}            \\
VeLLO           & 2MASS J18293368-0145103  & D              &  \textit{Irregular}          & Aquila          &\cite{kimmr16}, \cite{kimgj19}            \\
VeLLO           & SSTgbs J1839298+003740   & P              &  \textit{Curved}             & Aquila          &\cite{kimmr16}, \cite{kimgj19}            \\
VeLLO           & SSTgbs J0430149+360008   & P              &  \textit{Irregular}          & Auriga/CMC      &\cite{kimmr16}, \cite{kimgj19}            \\
VeLLO           & 2MASS J20405664+6723047  & P              &  \textit{Curved}             & Cepheus         & \begin{tabular}[c]{@{}c@{}}\cite{dunham08}; \cite{kimmr16}; \\ \cite{kimgj19}\end{tabular} \\
VeLLO           & SSTc2d J222933.4+751316  & P              &  \textit{Irregular}          & Cepheus         & \cite{dunham08}, \cite{kimgj19}         \\
VeLLO           & 2MASS J21470308+4733147  & P              &  \textit{Irregular}          & IC5146          &\cite{kimgj19}                     \\
VeLLO           & SSTgbs J21470601+4739394 & P              &  \textit{Linear}($+$)             & IC5146          &\cite{kimmr16}, \cite{kimgj19}            \\
VeLLO           & SSTgbs J21475567+4737113 & P              &  \textit{Curved}             & IC5146          & \cite{kimgj19}                     \\
VeLLO           & V1192 Sco                & D              &  \textit{Irregular}          & Lupus III       & \cite{dunham08}                 \\
VeLLO           & {[}SSG2006{]} MMS 126    & P              &  \textit{Irregular}          & Ophiuchus       & \begin{tabular}[c]{@{}c@{}}\cite{dunham08,stanke06};\\ \cite{kimmr16}; \cite{kimgj19}\end{tabular} \\
VeLLO           & 2MASS J03283258+3111040  & P              &  \textit{Irregular}          & Perseus         & \begin{tabular}[c]{@{}c@{}}\cite{dunham08}; \cite{kimmr16}; \\ \cite{kimgj19}\end{tabular} \\
VeLLO           & SSTc2d J032856.6+310737  & P              &  \textit{Linear}($-$)             & Perseus         & \cite{dunham08}                 \\
VeLLO           & IRAS 4B1 South           & P              & \textit{Linear}($-$)            & Perseus         & \cite{dunham08}                 \\
VeLLO           & {[}DAB2006{]} NOT- 239   & D              &  \textit{Curved}          & Serpens         & \cite{dunham08,djupvik06}                 \\
VeLLO           &IRAS F04110+2800         & D              &  \textit{Curved}             & Taurus          & \cite{kimgj19}                   \\
VeLLO           & IRAS 04381+2540          & D              &  \textit{Curved}             & Taurus          & \cite{dunham08}                
\enddata

\end{deluxetable}

\clearpage
\onecolumngrid
\section{Light curves of candidate FUors}\label{App:out}
\setcounter{figure}{0}
\renewcommand{\thefigure}{E.\arabic{figure}}

\begin{figure}[h]
\centering
{\includegraphics[width=0.7\columnwidth]
{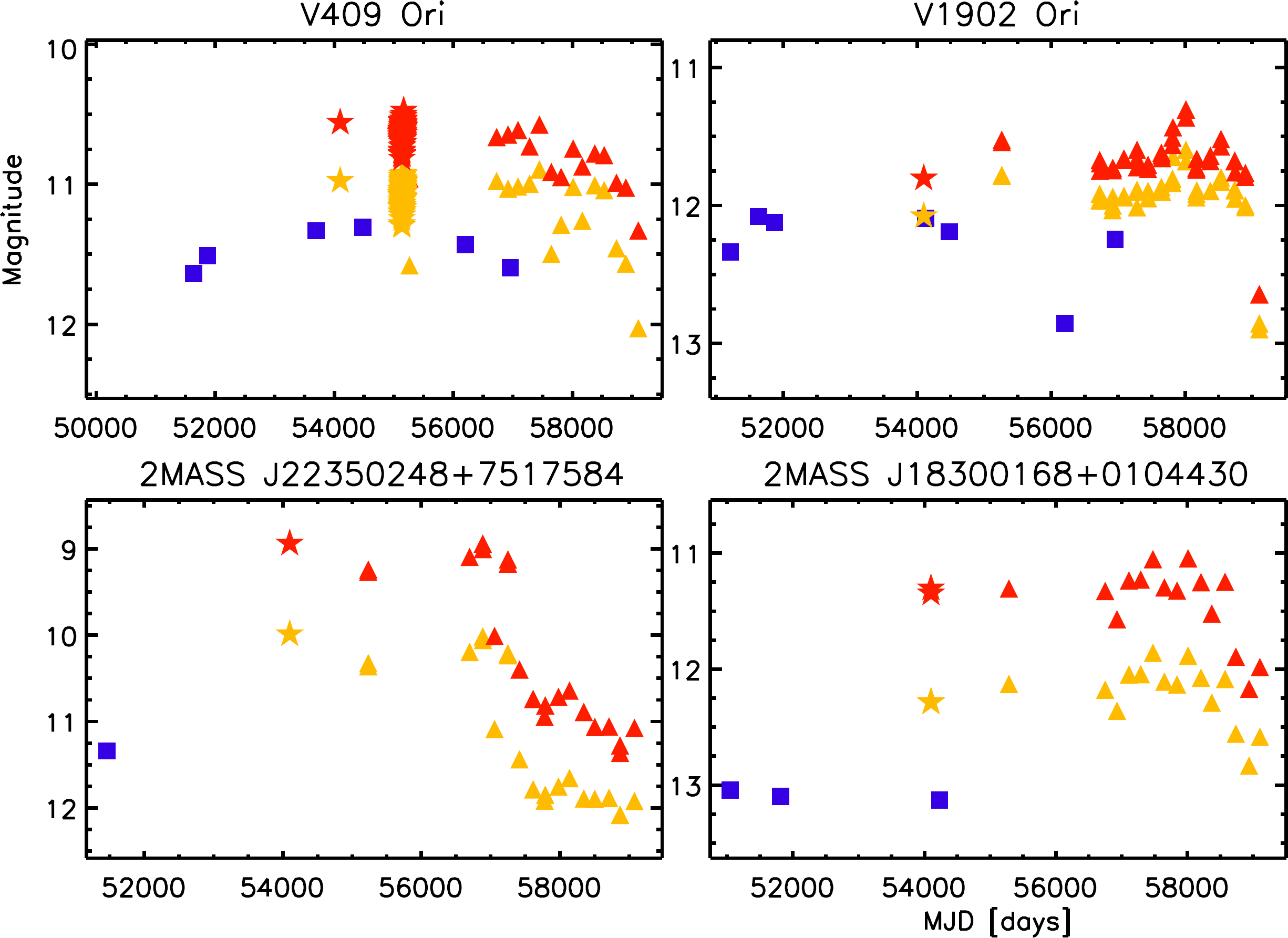}
}
\caption{Light curves for candidate FUors that are classified as disks: Near-IR $2.2 \mu$m (blue squares), and mid-IR $3.4 \mu$m (W1; yellow circles) and $4.6 \mu$m (W2; red circles).  {\it Spitzer} IRAC1 and IRAC2 filters (colored stars) were converted to the associated {\it WISE} filters using the relations by \citet{2014Antoniucci}.
\label{fig:fuor_D}
}
\end{figure}

\begin{figure}[h]
\centering
\includegraphics[width=0.7\columnwidth]{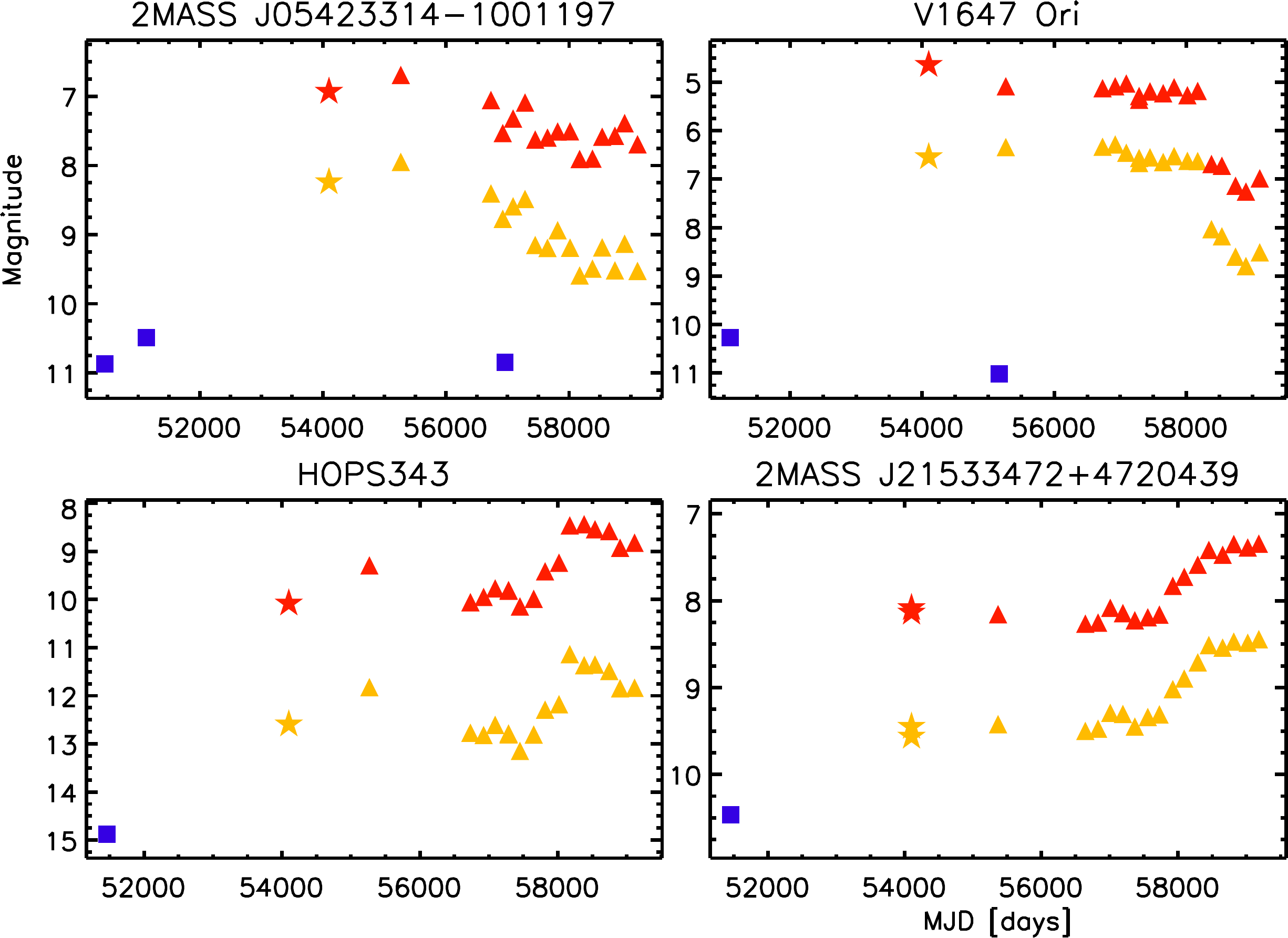}\\
\includegraphics[width=0.7\columnwidth]{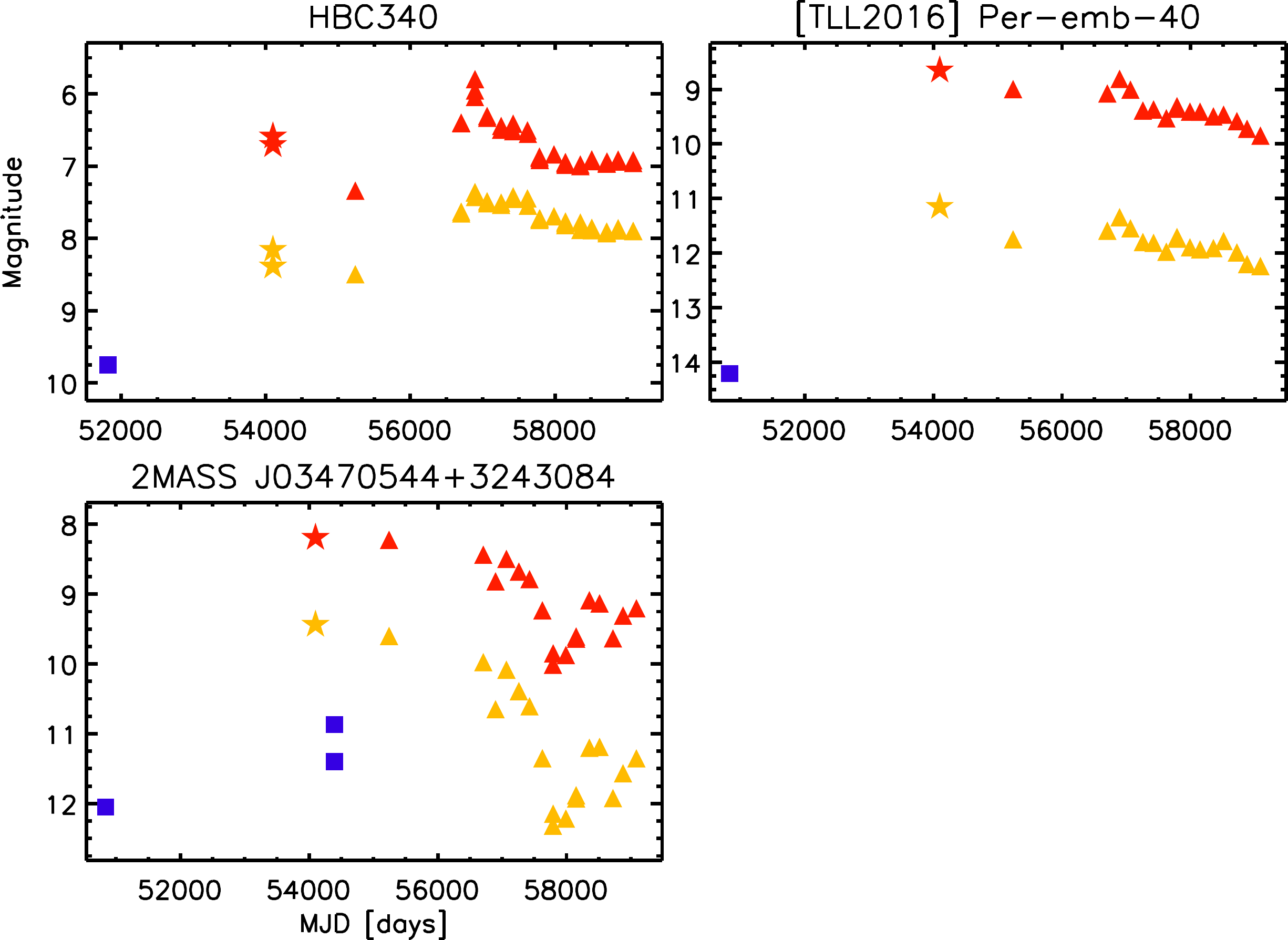}
\caption{Light curves for candidate FUors that are classified as protostars, but that are not used to determine $\tau$ in Section \ref{sec:long-term}. Symbols are the same as in Figure \ref{fig:fuor_D}.
\label{fig:fuor_P_nu}
}
\end{figure}

\begin{figure}[t]
\centering
\includegraphics[width=0.7\columnwidth]{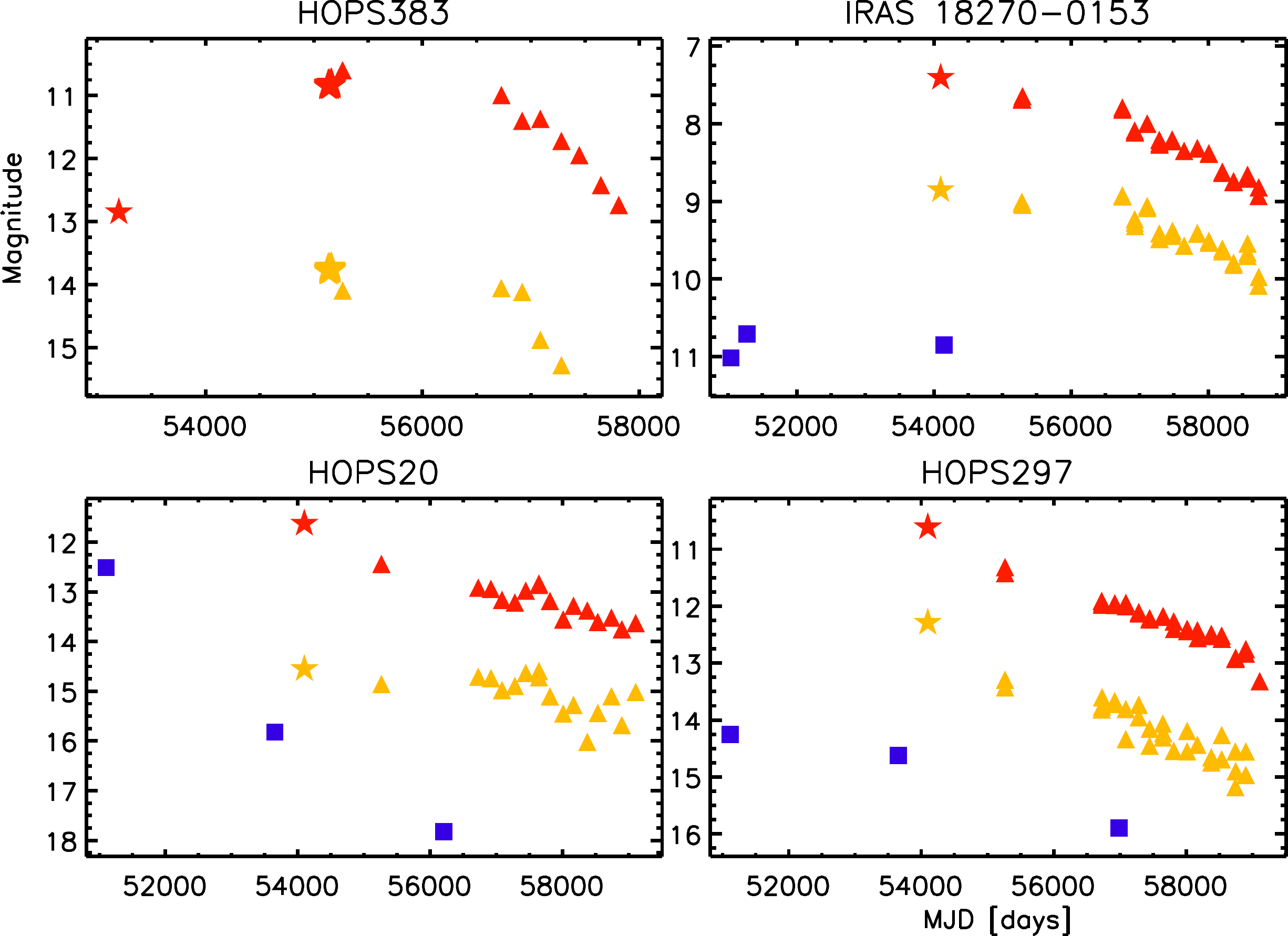}\\
\includegraphics[width=0.7\columnwidth]{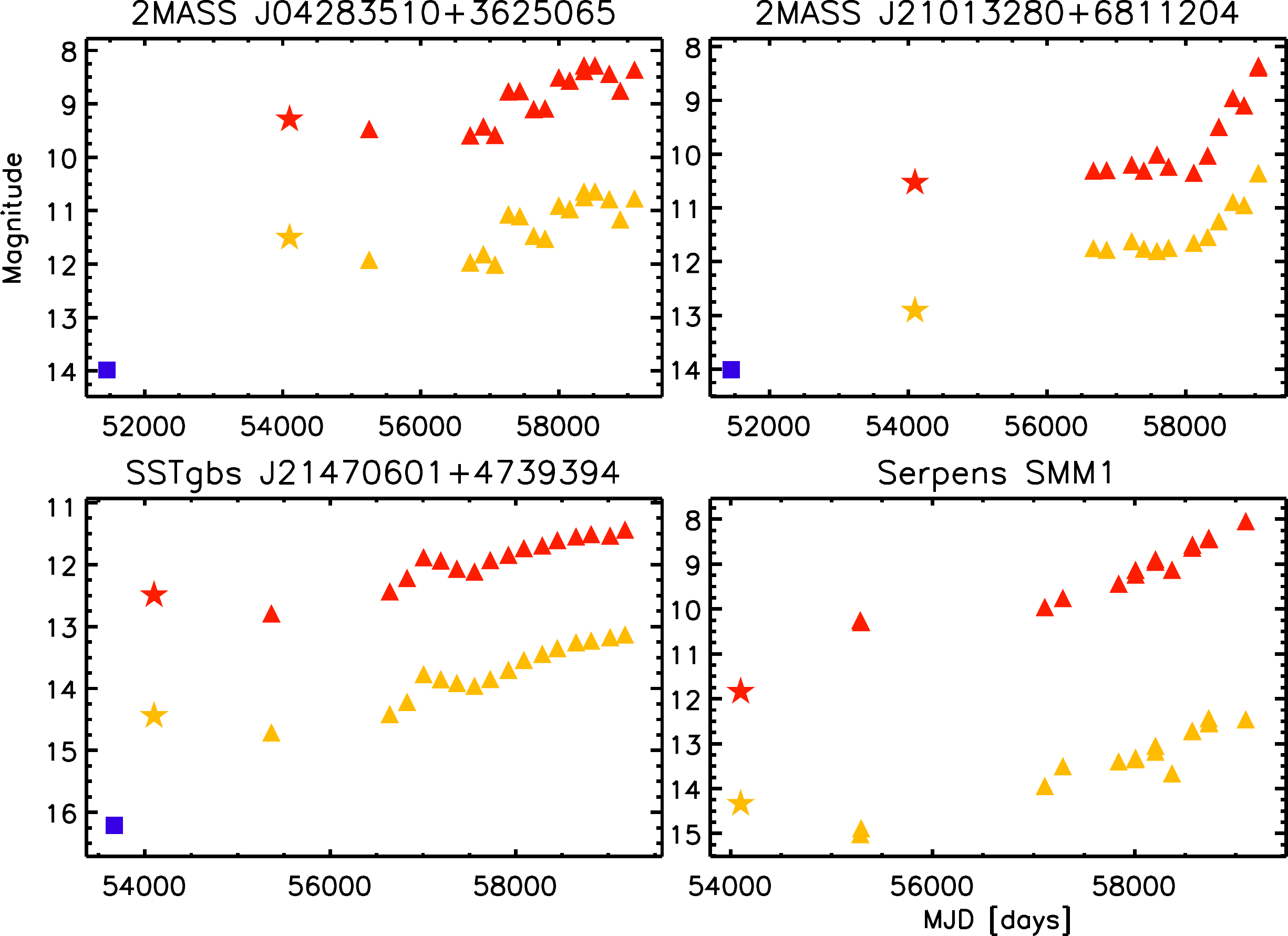}\\
\includegraphics[width=0.35\columnwidth]{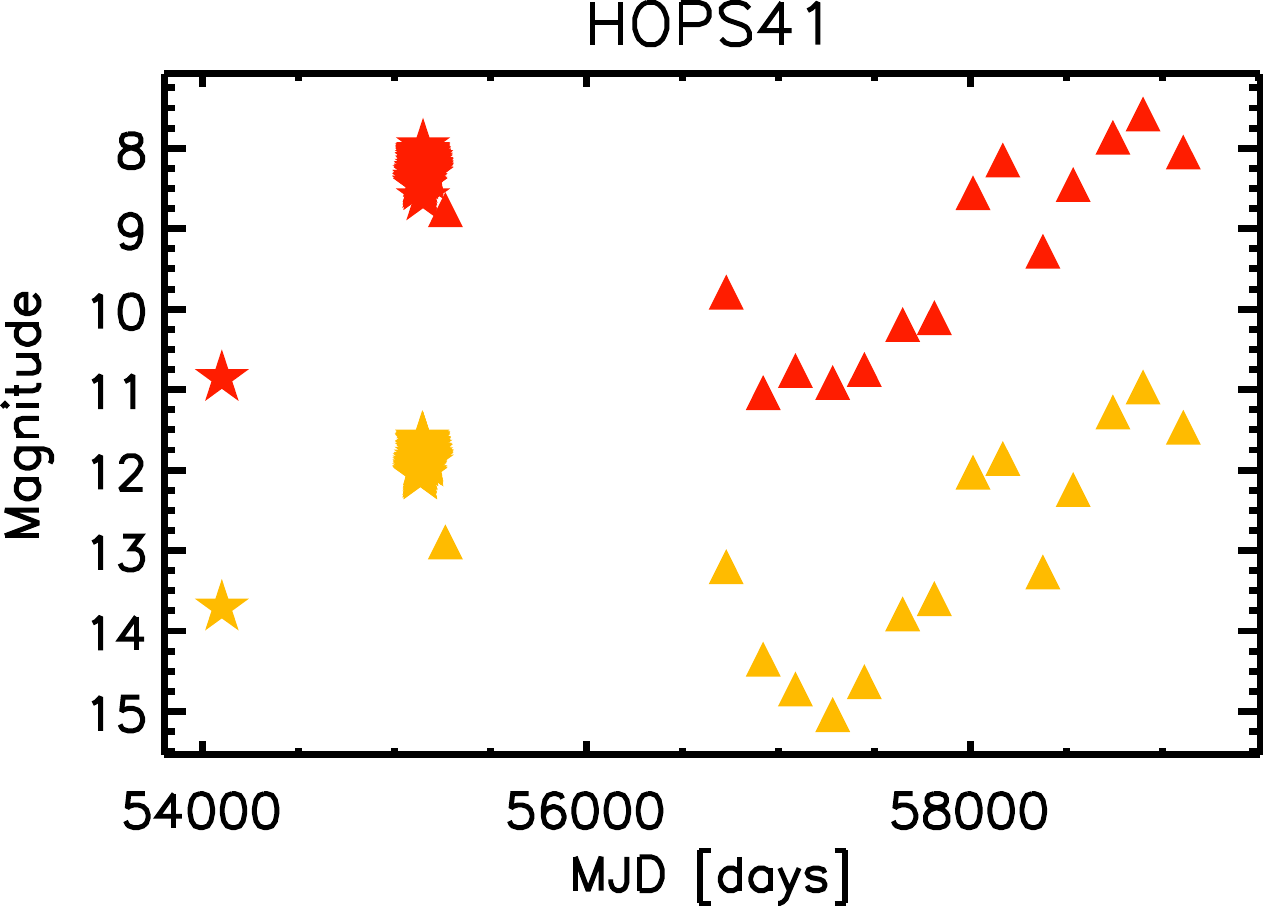}
\caption{Light curves for likely FUors that are classified as protostars. Symbols are the same as in Figure \ref{fig:fuor_D}.
\label{fig:fuor_P_u}
}
\end{figure}

\begin{figure}[h]
\centering
\includegraphics[width=0.7\columnwidth]{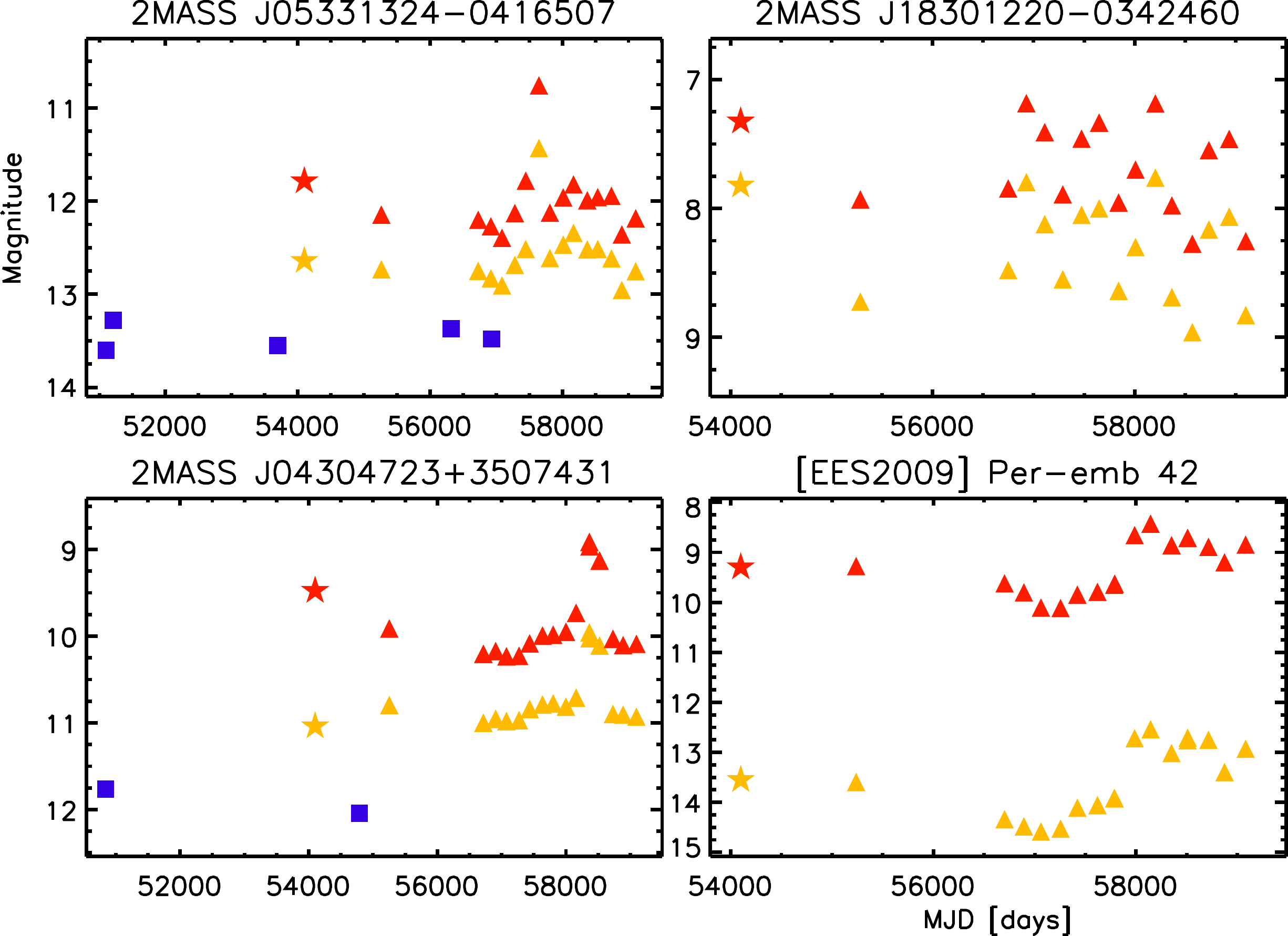}\\
\includegraphics[width=0.7\columnwidth]{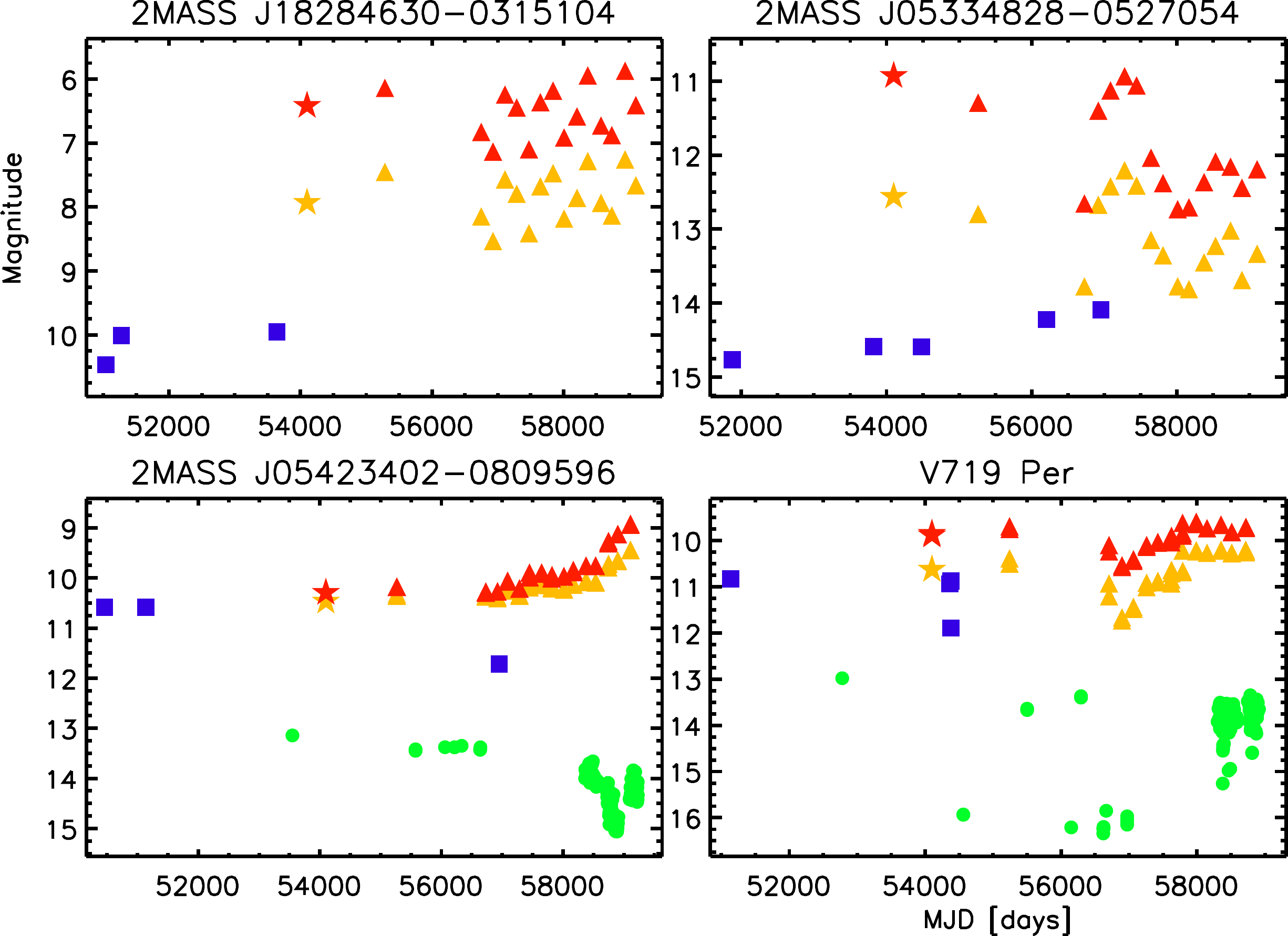}
\caption{Example light curves for high-amplitude variable YSOs, but that are not selected as candidate FUors in Section \ref{sec:long-term}. Symbols are the same as in Figure \ref{fig:fuor_D}. For YSOs  2MASS J05423402$-$0809596 and V719 Per, we also show optical photometry ($r-3$) as green circles.
\label{fig:fuor_ns}
}
\end{figure}

In this section we present the light curves of 20 YSOs that are considered candidate FUors and are discussed in more detail in Section \ref{sec:long-term} (Figures \ref{fig:fuor_D} to \ref{fig:fuor_P_u}). In addition, in Figure \ref{fig:fuor_ns} we present eight example light curves of YSOs that show high-amplitude variability, but are not selected as candidate FUors.

\clearpage
\onecolumngrid

\clearpage
\onecolumngrid

\end{document}